\documentclass[useAMS,usenatbib]{mnras}
\usepackage{aas_macros}
\usepackage{graphicx}
\usepackage{amsmath,amssymb}
\usepackage[T1]{fontenc}
\usepackage{aecompl}
\usepackage{times}
\title[Global Simulations of Galactic Discs]{Global Simulations of Galactic Discs: Violent Feedback from Clustered Supernovae during Bursts of Star Formation}
\author[D. Martizzi]{\parbox[t]{\textwidth}{Davide Martizzi$^{1,2}$\thanks{E-mail: davide.martizzi@nbi.ku.dk}} \\ \\
$^{1}$DARK, Niels Bohr Institute, University of Copenhagen, 2100 Copenhagen, Denmark \\
$^{2}$Department of Astronomy and Astrophysics, University of California, Santa Cruz, CA 95064, USA \\
}

\begin{document}

\maketitle

\begin{abstract}
A suite of idealised, global, gravitationally unstable, star-forming galactic disc simulations with 2 pc spatial resolution, performed with the adaptive mesh refinement code {\sc ramses} is used in this paper to predict the emergent effects of supernova feedback. The simulations include a simplified prescriptions for formation of single stellar populations of mass $\sim 100 \, M_{\odot}$, radiative cooling, photoelectric heating, an external gravitational potential for a dark matter halo and an old stellar disc, self-gravity, and a novel implementation of supernova feedback. The results of these simulations show that gravitationally unstable discs can generate violent supersonic winds with mass loading factors $\eta \gtrsim 10$, followed by a galactic fountain phase. These violent winds are generated by highly clustered supernovae exploding in dense environments created by gravitational instability, and they are not produced in simulation without self-gravity. The violent winds significantly perturb the vertical structure of the disc, which is later re-established during the galactic fountain phase. Gas resettles into a quasi-steady, highly turbulent disc with volume-weighted velocity dispersion $\sigma > 50 \, {\rm km/s}$. The new configuration drives weaker galactic winds with mass loading factor $\eta \leq 0.1$. The whole cycle takes place in $\leq 10$ dynamical times. Such high time variability needs to be taken into account when interpreting observations of galactic winds from starburst and post-starburst galaxies. 
\end{abstract}


\section{Introduction} \label{sec:intro}

Core-collapse supernovae (CC SNe) are ubiquitous in star forming galaxies. CC SNe arise from the rapid collapse of massive stars followed by an explosion that injects $\sim 10^{51} \, {\rm erg}$ of energy into the circumstellar and interstellar medium (ISM) in form a few solar masses of material ejected from the progenitor star. On average, CC SNe explode $\gtrsim 10 \, {\rm Myr}$ after their progenitor star is formed. For this reason, as long as a galaxy keeps forming stars at a given rate, a steady rate of CC SNe is expected to explode in that galaxy. Due to the high-energy-per-event and their high rates, CC SNe can completely change the outlook of a galaxy throughout its history. The collection of phenomena related to SNe that influences the evolution of a galaxy is referred to as supernova feedback. 

On the observational side, evidence has been accumulated that SN ejecta interact with the ISM by sweeping gas and heating it to X-ray emitting temperatures \citep{2012A&ARv..20...49V}. The hot, low density cavities of size $\lesssim 100 \, {\rm pc}$ inflated by supernova remnants (SNRs) constitute one of the main features of the ISM of galaxies. The overlap of multiple SNRs is also observed to give rise to the formation of even bigger hot, low density regions called superbubbles \citep{2004ARA&A..42..211E}, which are also interpreted as one of the driving phenomena of the galactic outflows (or winds) observed around star forming galaxies \citep{1999ApJ...513..156M,2005ARA&A..43..769V,2005ApJ...621..227M,2009ApJ...697.2030S}.

On the theoretical side, the role of supernovae in setting the phase structure and porosity of the ISM has been recognized for decades \citep{1977ApJ...218..148M}. The idea that supernovae can inflate hot superbubbles expanding in a disc, eventually blowing out and launching a galactic wind has also been studied early on \citep{CC85}. The feedback of SNe on the physical state of star forming gas has also been appreciated. In fact, the normalization of the observed star formation law \citep{1998ApJ...498..541K} implies that star formation in galaxies is  inefficient. Most star forming galaxies appear to be able to convert only 1-5\% of their gas into stars each free fall time. Theoretical models have been successful at explaining the observed star formation efficiency by assuming that star formation is regulated by turbulence in molecular clouds  \citep{2004RvMP...76..125M,2005ApJ...630..250K}, and complementary models have established the connection between SN feedback and turbulence in the ISM \citep{2010ApJ...721..975O,2011MNRAS.417..950H,2013MNRAS.433.1970F,2017MNRAS.465.1682H,2019arXiv190300962D}, or the direct connection between the regulation of star formation and SN feedback \citep{1997ApJ...481..703S,2019arXiv190300962D}. 

The effects of SN feedback postulated or predicted by theoretical models have been extensively tested with numerical simulations. In practice, all state-of-the-art  simulations of large cosmological volumes (box size $L \gtrsim 100 \, {\rm Mpc}$) require sub-resolution models for SN feedback to regulate star formation rates and prevent the formation of overly massive galaxies \citep{2014MNRAS.444.1453D,2016MNRAS.462.3265D,2014MNRAS.444.1518V,2015MNRAS.446..521S,2018MNRAS.473.4077P,2019MNRAS.486.2827D}. Adoption of the zoom-in technique has recently allowed computational astrophysicists to perform re-simulations of individual galaxies with extremely high spatial ($5-50 \, {\rm pc}$) and mass resolution ($10^3-10^4 \, M_{\odot}$), which is sufficient to (marginally) resolve the small scale structure of the ISM and capture explicitly the effects of stellar feedback \citep{2014MNRAS.445..581H,2015MNRAS.451.2900K,2015ApJ...804...18A, 2016MNRAS.460.2731C}. These achievements stimulated the community to further investigate the physics of SN feedback with direct numerical simulations, in order to check the robustness of the theoretical assumptions made in feedback models for cosmological zoom-ins, and highlight underappreciated aspects of this process. Several efforts focused on quantifying the evolution of SNRs in a turbulent and inhomogeneous ISM \citep{2015MNRAS.450..504M,2015ApJ...815...67K,2015MNRAS.454..238W,2015ApJ...809...69S}. These results were later used to design simulation of emergent effects such as the generation of a multi-phase ISM \citep{2015MNRAS.449.1057G,2015ApJ...814....4L,2017ApJ...846..133K}, the regulation of star formation, the driving of SN-driven turbulence, and the launching of galactic winds \citep{2014A&A...570A..81H,2015ApJ...815...67K,2016MNRAS.456.3432G,2016MNRAS.459.2311M}. 

In particular, \cite{2016MNRAS.459.2311M} and \cite{2017MNRAS.470L..39F} demonstrated that supersonic, highly mass-loaded galactic winds cannot be produced by SN feedback in locally stratified simulations of galactic disc patches, but only in global simulation of galactic discs. This line of research identified another important requirement for launching galactic winds similar to those observed in real galaxies: the non-linear interaction among multiple SNe exploding in the same cloud/star cluster \citep{2014MNRAS.443.3463S,2017MNRAS.465.1720Y,2017ApJ...834...25K,2017MNRAS.470L..39F,2018MNRAS.481.3325F,2019arXiv190209547E}. Isolated SNRs in high density regions can only efficiently inject momentum into the ISM, whereas their thermal energy is lost via radiative cooling in less than 1 Myr, preventing the maintenance of a hot phase of the ISM. On the other hand, clustered SNRs are able to better retain their thermal energy and inject more momentum-per-supernova, provided that each SN explodes within approximately one cooling radius and one cooling time from the previous one in the cluster. \cite{2017MNRAS.465.2471G} and \cite{2019MNRAS.483.3647G} discussed some of the numerical requirements that hydrodynamical and magneto-hydrodynamical simulations should satisfy in order to properly capture the momentum and thermal energy budget of clustered SNRs, suggesting that current 3D simulations might be underestimating the effect of clustered SNe. 

Most of the studies of clustered SN feedback so far have focused on test scenarios in which the spacing and timing of clustered SNe was manually controlled (see references above). Although these kinds of tests are crucial to identify the parameters that control the efficiency of SN feedback, a systematic study of how clustered SN feedback arises in rapidly star forming discs is currently missing. In order to fill this gap, idealised adaptive mesh refinement (AMR) simulations of isolated, gravitationally unstable galactic discs are performed and analysed in this paper. Due to their parsec-scale resolution, the inclusion of radiative cooling, photoelectric heating, self-gravity, formation of individual $100 \, M_{\odot}$ stellar populations and a novel model for SN feedback, these simulations are able to capture SN clustering in action, while being simple enough to be easily analysed and interpreted. 

The paper is structured as follows. Section~\ref{sec:methods} describes the numerical setup and the initial conditions. Section~\ref{sec:results} reports the main predictions of the simulations. Section~\ref{sec:conclusions} summarises the main findings of the paper and discusses them in the context of other work.
 
\section{Numerical Methods} \label{sec:methods}

With the goal of running parsec-resolution simulations of isolated galaxies, a modified version of the {\sc ramses} code \citep{Teyssier:2002p451} was developed. This version of the code can be found in the public online repository \href{https://bitbucket.org/dmartizzi/ramses}{https://bitbucket.org/dmartizzi/ramses}. {\sc ramses} uses adaptive mesh refinement (AMR) methods to solve for the dynamics of dark matter, stars and gas. The code also includes a solver for the self-gravity of the simulated collisional (gas) and collision-less (dark matter, stars, black holes, etc.) fluids. Finally, the code is equipped with several modules for star formation, radiative transfer and galactic feedback mechanisms.

Modifications to the code implemented for this paper include: 
\begin{enumerate}
\item Initial condition generator.
\item Treatment of the gravitational field.
\item Calculation of the optimal time step.
\item Inclusion of a novel supernova feedback scheme. 
\item Inclusion of a simplified prescription photoelectric heating.
\end{enumerate}
The details of the modified version of the code developed for this paper are discussed in the following subsections. 

\subsection{Simulation Setup}

A box size $L = 4.096 \, {\rm kpc}$ with standard outflow boundary conditions (zero gradients) was adopted for the fiducial runs and the AMR functionality of {\sc ramses} is fully used. The adaptive mesh uses a base Cartesian grid of size $256^3$ (level $\ell = 8$) with 3 additional levels of refinement (up to level $\ell = 11$, effective resolution $2048^3$) activated on-the-fly according to user-defined criteria. The base grid has a cell size $\Delta x_{\rm coarse} = 16 \, {\rm pc}$, whereas the highest resolution regions have a cell size $\Delta x_{\rm fine} = 2 \, {\rm pc}$, which is suitable to resolve the sites of star formation and marginally resolve the internal structure of star-forming clouds in a galactic disc. Cells are triggered for refinement if their total mass mass is $m_{\rm cell} \geq 8 \, m_{\rm sph}$, where $m_{\rm sph} = 125 \, {\rm M_{\odot}}$ is a user-defined mass resolution element. To assess the effect of the box size on the solutions, tests have also been performed with a larger box size $L = 8.192 \, {\rm kpc}$ with the same spatial and mass resolution, but no significant differences in the results were found. 

Gas hydrodynamics is solved using the second-order Godunov methods implemented in {\sc ramses}. Piecewise linear reconstruction with the MinMod slope limiter was used to interpolate cell-centered variables to cell faces, whereas the Harten-Lax-van Leer (HLL) Riemann solver was adopted for all the runs. The effect of magnetic fields was not included. To avoid numerical issues a density floor $n_{\rm H, floor} =10^{-10} \, {\rm cm^{-3}}$ and a temperature floor $T_{\rm floor} = 50 \, {\rm K}$ were enforced. The time step for the evolution of the system is principally determined by gas hydrodynamics using a Courant-Friedrichs-Lewy (CFL) condition, but the time step is restricted in such a way that fast radiative cooling of the gas can be resolved with appropriate time resolutions (see Subsection\ref{sec:cooling}).

The contribution to the gravitational potential from the dark matter halo hosting the simulated galaxy is included as a static external potential (Subsection~\ref{sec:ics1}). Although this choice does not allow the simulations to capture the effect of galactic feedback on dark matter dynamics \citep{2010Natur.463..203G,2013MNRAS.429.3068T,2016MNRAS.456.3542T}, this approach offers the opportunity of cutting completely the computational expenses associated to solving for the evolution of several million dark matter particles, while still allowing the code to follow realistic orbits for stars and gas. 

Besides the dark matter halo potential, the code also allows to include additional external gravitational potentials associated with a bulge and a disc, respectively (Subsections~\ref{sec:ics1} and \ref{sec:ics2}). These components are intended to model the gravity of stars formed before the start of the simulation or the gravity of a non self-gravitating gaseous disc. 

{The code can also compute the gravitational potential and acceleration generated by the gas and by any massive particle present in the simulation. This functionality is particularly useful for cases in which gas self-gravity is important, and it is achieved via a multi-grid implementation of the Gauss-Seidel method with Red-Black
Ordering and Successive Over Relaxation \citep[see][for reference]{Teyssier:2002p451}. In this method, an approximate solution to the Poisson equation is iteratively found for each level of refinement, while keeping the appropriate boundary conditions for each region into account. In the simulations considered in this paper, it was chosen to solve the Poisson equation only down to $\ell_{\rm grav} = 10$, a level coarser than the maximum level of refinement used for the hydrodynamics solver $\ell = 11$. For cells at the highest level of refinement, the gravitational forces are interpolated from $\ell_{\rm grav} = 10$ to $\ell = 11$. Several tests were performed in order to choose the fiducial value for $\ell_{\rm grav}$. It was found that setting $\ell_{\rm grav} = \ell = 11$ in these simulations causes spurious numerical effects in highly self-gravitating regions where the gravitational potential and the local hydrodynamic variables change on similar time scales. These spurious effects are absent if the gravitational potential is made sufficiently smooth in the regions with highest level of refinement, e.g. by setting $\ell_{\rm grav}=10$.
} 

All the simulations in this paper include gas dynamics, radiative cooling, photoelectric heating, star formation and supernova feedback. A subset of the simulations was run with self-gravity and was labeled as WSG. Another subset of the simulations was run without self-gravity and was labeled as NOSG. With the intention of observing the consequences of star formation bursts of various intensities, discs with different gas surface densities were considered. Since the presence of a bulge stabilises the gas against gravitational instability, only the contribution to the potential from a disc was included. Galaxies with this setup may settle to conditions similar to those of gas-rich, high-redshift galactic discs that do not yet have a bulge, or to those in the nuclei of post-merger high-redshift galaxies \citep{2014MNRAS.441..389W}.

\subsection{Initial Conditions in Simulations without Self-gravity}\label{sec:ics1}

The initial conditions for the simulated galaxies assume vertical hydrostatic equilibrium and centrifugal equilibrium in the external gravitational potential specified by the user, which is given by:
\begin{equation}
\Phi_{\rm ext}({\bf r}) = \Phi_{\rm halo}({\bf r}) + \Phi_{\rm bulge}({\bf r}) + \Phi_{\rm disc}({\bf r}),
\end{equation}
where ${\bf r}$ is the position vector, $\Phi_{\rm halo}(\bf{r})$ is the contribution from the dark matter halo, $\Phi_{\rm bulge}(\bf{r})$ is the contribution from the bulge, and $\Phi_{\rm disc}(\bf{r})$ is the contribution from the disc. 

The gravitational potential generated by the dark matter halo is given by the Navarro-Frenk-White (NFW, \citealt{1997ApJ...490..493N}) model:
\begin{equation}\label{eq:pot}
\Phi_{\rm halo}(r)=4\pi G r_{\rm s}^2 \rho_{\rm s} \frac{\ln (1+\frac{r}{r_{\rm s}})}{\frac{r}{r_{\rm s}}}
\end{equation}
where $r$ is the distance from the halo centre, $\rho_{\rm s}$ is a characteristic density of the halo, $r_{\rm s}$ is the halo scale radius. 

Let $r_{\rm 200}$ be the radius within which the mean density is 200 times the critical density of the universe, then we define the concentration
\begin{equation}
 c_{\rm 200}=\frac{r_{\rm 200}}{r_{\rm s}}.
\end{equation}
We label the mass within $r_{\rm 200}$ as $M_{\rm 200}$. Numerical N-body simulations have shown that a mass-concentration relation exists. We adopt the mass-concentration relation at redshift $z=0$ from \cite{2014MNRAS.441.3359D}:
\begin{equation}
 c_{\rm 200} = 8.03\times\left(\frac{M_{\rm 200}}{10^{12} \hbox{ M}_{\odot}/h}\right)^{-0.101}.
\end{equation}
We set $h = 0.7$ for the Hubble parameter. Once the mass-concentration relation is set, the NFW potential model only depends on the choice of $M_{\rm 200}$. For our simulations we assume $M_{\rm 200} = 1.5\times 10^{11} \, {\rm M_{\odot}}$. 

The gravitational potential associated with a central bulge is modeled using a Plummer model:
\begin{equation}
\Phi_{\rm bulge}(r)=-\frac{GM_{\rm bulge}}{\sqrt{r^2+h_{\rm bulge}^2}},
\end{equation}
where $M_{\rm bulge}$ is the mass of the bulge and $h_{\rm bulge}$ is the characteristic size of the bulge. The bulge potential is fully implemented in the code, but we do not consider the contribution from a fully formed bulge, as discussed in the previous Subsection. For numerical reasons, we set the mass of the bulge and its scale radius to small numbers, $M_{\rm bulge}=10^{-3}\, {\rm M_{\odot}}$ and $h_{\rm bulge} = 10^{-6} \, {\rm kpc}$.

The gravitational potential associated with the disc is given by a Miyamoto \& Nagai model:
\begin{equation}\label{eq:pot_disc}
\Phi_{\rm disc}(R,z)=-\frac{GM_{\rm disc}}{\sqrt{R^2 + \left[ h_{\rm R} + (h_{z}^2+z^2)^{1/2}\right]^2}},
\end{equation}
where $R$ and $z$ are cylindrical coordinates centered at the galaxy centre, and $M_{\rm disc}$, $h_{\rm R}$ and $h_{z}$ are the mass, scale radius and scale height of the disc, respectively. 

In NOSG simulations that do not include self-gravity, it is assumed that the disc external potential is generated by the combination of an old stellar disc  formed before the beginning of the simulation and a non-self-gravitating gaseous disc, both with the same scale height and scale radius, but with different masses. In practice, for NOSG runs the disc mass is set to $M_{\rm disc}=M_{\rm gas}+M_{\rm star,old}$, with a fixed old stellar mass $M_{\rm star,old}=5\times 10^8\,{\rm M_{\odot}}$ and $M_{\rm gas}$ depending on the considered initial condition model (see Subsection~\ref{sec:sim_catalog}). In reality, such disc would be self-gravitating, but the NOSG runs have to be considered only as test cases in which the supernova rate is proportional to the gas surface density, but supernovae do not cluster as strongly as in the case with self-gravity. The scale radius and height are set to $h_{\rm R} = 0.4\, {\rm kpc}$ and $h_{\rm z} = 0.2\, {\rm kpc}$, respectively.

Once the external potential is set, then the initial condition for the gas density distribution in vertical hydrostatic equilibrium is computed. Vertical hydrostatic equilibrium with respect to the external gravitational potential requires:
\begin{flalign}\label{eq:vertical_balance}
&\frac{\partial P_{\rm gas}(R,z)}{\partial z}= \frac{k_{\rm B}T_0}{\mu m_{\rm p}}\frac{\partial\rho_{\rm gas}(R,z)}{\partial z} = \nonumber \\ 
&=-\rho_{\rm gas}(R,z)\frac{\partial\Phi_{\rm ext}(R,z)}{\partial z},
\end{flalign}
where $P_{\rm gas}=\rho_{\rm gas}k_{\rm B}T_0/(\mu m_{\rm p})$ is the gas pressure, and where the gas is assumed to initially be isothermal, with temperature $T_0/\mu=1850 \, {\rm K}$ and mean molecular weight $\mu\approx 1.3$. We adopt a particular solution to equation~\ref{eq:vertical_balance}: 
\begin{flalign} \label{eq:density}
&\rho_{\rm gas}(R,z) = \nonumber \\ 
&=\rho_0\exp\left\{-\frac{R^2}{h_{\rm R}^2}-\frac{\mu m_{\rm p}}{k_{\rm B}T_0} \left[\Phi_{\rm ext}(R,z)-\Phi_{\rm ext}(R,0)\right]\right\},
\end{flalign}
where the central density $\rho_0$ is set in such a way that the following normalisation condition is satisfied: 
\begin{equation}\label{eq:norm}
M_{\rm gas} = \int_0^{+\infty} 2\pi R dR \int_{-\infty}^{+\infty} dz\rho_{\rm gas}(R,z).
\end{equation}

Finally, the velocity of the gas is computed by assuming centrifugal equilibrium in the external potential. The gas is assumed to move in circular orbits with vertical and radial velocities equal zero, $v_{\rm z}=v_{\rm R}=0$. The only non-vanishing component of the velocity field is the azimuthal component $v_{\rm \vartheta}$, which is the solution of the following radial balance equation:
\begin{equation}
-\rho_{\rm gas}(R,z)\frac{v_{\rm \vartheta}^2}{R} = -\frac{\partial P_{\rm gas}(R,z)}{\partial R}-\rho_{\rm gas}(R,z)\frac{\partial\Phi_{\rm ext}(R,z)}{\partial R}. 
\end{equation}
Using the fact that $P_{\rm gas}=\rho_{\rm gas}k_{\rm B}T_0/(\mu m_{\rm p})$, the solution for the azimuthal velocity is found to be:
\begin{equation}\label{eq:vrot}
v_{\rm \vartheta}=\sqrt{R\left( \frac{\partial \Phi_{\rm ext}(R,0)}{\partial R} -2\frac{k_{\rm B}T_0}{\mu m_{\rm p}}\frac{R}{h_{\rm R}^2}\right)}.
\end{equation}

Finally, the gas pressure is set $P_{\rm gas}=\rho_{\rm gas}k_{\rm B}T_0/(\mu m_{\rm p})$, whereas the gas metallicity is set to the solar value. 

\subsection{Initial Conditions in Simulations with Self-gravity}\label{sec:ics2}

In the WSG simulations which include the self-gravity of the gas and the newly formed stars, the external potential is used to only model the dark matter halo and the old stellar disc potential, but not the gas and the newly formed stars. For this reason, the mass of the disc associated to the external potential of equation~\ref{eq:pot_disc} is set to $M_{\rm disc}=M_{\rm star,old}=5\times 10^8\,{\rm M_{\odot}}$, with scale radius and height are set to $h_{\rm R} = 0.4\, {\rm kpc}$ and $h_{\rm z} = 0.2\, {\rm kpc}$, respectively. An isothermal rotating gaseous disc is initialised following equations~\ref{eq:density} and \ref{eq:vrot}, subject to the normalisation condition of equation~\ref{eq:norm}. Under these assumptions, the gravitational potential in the WSG runs is given by 
\begin{equation}\label{eq:pot_tot}
\Phi_{\rm tot}(R,z,t) = \Phi_{\rm ext}(R,z) + \Phi_{\rm SG}(R,z,t),
\end{equation}
where $\Phi_{\rm SG}(R,z,t)$ is the gravitational potential generated by the dynamically evolving gaseous disc and by the stellar particle that form during the evolution of the system. 

In practice, the self-gravity term $\Phi_{\rm SG}$ in equation~\ref{eq:pot_tot} is the solution to the Poisson equation
\begin{equation}\label{eq:poisson}
\nabla^2\Phi_{\rm SG}(R,z,t) = 4\pi G \left[\rho_{\rm gas}(R,z,t)+\rho_{\rm star,new}(R,z,t)\right], 
\end{equation}
where $\rho_{\rm gas}(R,z,t)$ and $\rho_{\rm star,new}(R,z,t)$ are the gas density and the density of stellar particles at time $t$, respectively. The {\sc ramses} code numerically solves equation~\ref{eq:poisson} at each time step. To summarise, the total potential in the WSG runs is the sum of an analytical static external potential and a time-varying potential numerically computed by the code. Notice that because of this choice the initial conditions for the WSG runs are generally prone to development of the gravitational instability (see Subsection~\ref{sec:sim_catalog}). 

\subsection{Radiative Cooling, Heating and the Time Step}\label{sec:cooling}

The global disc simulations considered in this paper include radiative cooling based on the metal-dependent cooling curves from \cite{1993ApJS...88..253S}. For simplicity, and to isolate processes internal to the simulated galaxies, heating from an external UV background is not included. However, photo-electric heating from dust grains by stellar UV radiation has been shown to strongly influence the conditions of the ISM \citep[e.g.][]{2015ApJ...814....4L}. Due to the lack of explicit dust and radiation modeling in our simulations, photoelectric heating is included in an approximate way following \cite{2011piim.book.....D}. The photoelectric heating rate in each gas cell is computed as:
\begin{equation}
    \Gamma_{\rm pe} = 1.4\times 10^{-26}\, {\rm erg/s} \times n_{\rm H},
\end{equation}
where $n_{\rm H}$ is the hydrogen number density in the cell. 

Radiative cooling and heating are computed at each hydrodynamical time step. If the cooling time in a  cell is too short compared to the hydrodynamical time step, the temperature of the gas is adjusted by  performing multiple cooling/heating sub-cycles within the time step. 

After performing tests, it was found that cooling sub-cycling alone is often unable to prevent gas cells to artificially cool within a hydrodynamic time step. To prevent numerical inaccuracies caused by this issue, we also restrict the time step so that it is shorter than the cooling time:
\begin{equation}
 \Delta t = \min(\Delta t_{\rm hydro},C\times t_{\rm cool}),
\end{equation}
where $C<1$ is a dimensionless factor. This choice allows to accurately track cooling and heating in the simulations at the price of significantly reduced time steps.

\subsection{Star Formation Scheme and Supernova Seeding}\label{sec:star-formation}

One of the main features of these simulations is the natural inclusion of the effects of clustered supernovae, achieved by explicitly resolving the sites where star clusters and supernova progenitors form. To keep the focus of the study on supernova feedback, rather than on the physics of star formation, a simplified approach to the latter was followed. It was assumed that stars forms from gas in clusters of a given mass, and that one core-collapse supernova explodes every $100 \, M_{\odot}$ of newly formed stars. {Supernovae Ia were not included, in order to isolate the effect of prompt core-collapse supernova feedback, which is more sensitive to star formation clustering.}

In practice, star formation is included following the standard recipes implemented in {\sc ramses}, but the mass of stellar particles created from the gas is kept under strict control. Stellar particles of mass $m_{\rm star} = 100 \, {\rm M_{\odot}}$ are created in cells with density $n_{\rm H}\geq 100 \, {\rm cm^{-3}}$ with an efficiency-per-free-fall-time $\epsilon =0.01$. Stellar particles inherit the metallicity of the gas from which they formed, and their trajectories are computed by taking into account the gravity of the whole system. The mass of each stellar particle is chosen to represent the typical mass of a single stellar population. Multiple particles can be thought as the assembling components of a star cluster. After a time $\Delta t_{\rm delay} = 10 \, {\rm Myr}$ each stellar particle spawns a supernova, which injects momentum, thermal energy and metals following the scheme described in subsection~\ref{sec:SNscheme}. 

{Since the stellar particle mass is comparable to the mass of the most massive stars, each stellar particles should be formally considered as a random sub-sample of stars extracted from a larger population characterised by a given stellar initial mass function (IMF). Since a given stellar particle mass can be obtained by summing the mass of multiple low-mass stars sampled from the IMF, it is not guaranteed that each stellar particle will seed a core-collapse supernova. An accurate sub-grid treatment of star formation in this regime should include explicit stochastic sampling of the stellar IMF. With this approach, each stellar particle could seed a different number of supernovae. However, given that the simulations in this paper focus on highly star-forming discs with frequent supernovae, even if each star particle seeded a different number of supernovae, one should expect the results to converge to those of a scenario in which each stellar particle seeds the same average number of supernovae, which is the assumption made in this paper. }

More sophisticated models for star and star cluster formation were used by other groups \citep[e.g.]{2017ApJ...834...69L,2018MNRAS.480.1666S,2019MNRAS.486.5482G}, but their adoption adds complexity to simulations, which may complicate the interpretation of the results. A study of the effects of more sophisticated models for star formation on supernova feedback is deferred to future work. 

\subsection{Supernova Feedback Scheme}\label{sec:SNscheme}

In a previous paper, \cite{2015MNRAS.450..504M} ran a series of sub-pc resolution simulations of supernova remnants and characterised the momentum and thermal energy injected by them into the ISM. The results of the simulations where fitted with piece-wise power-law functions that capture the phenomenology of the interaction between SN ejecta and the ISM. As the ejecta travel through the inhomogeneous ISM, the latter is shocked and accelerated in a quasi-spherical fashion. The evolution of the system is initially energy-conserving and dominated by thermal energy, until radiative cooling modifies its dynamics, leading to a momentum conserving phase with little thermal energy left in the remnant. The formulae from \cite{2015MNRAS.450..504M} capture the functional dependence of SNR radial momentum $P_{\rm r}(R, \rho, Z)$ and thermal energy $E_{\rm th}(R, \rho, Z)$, where $R$ is the distance of the shock from the SN progenitor, $\rho$ is the local average ISM density, and $Z$ is the local average ISM metallicity. 

\begin{figure*}
\begin{center}
 \includegraphics[width=0.99\textwidth]{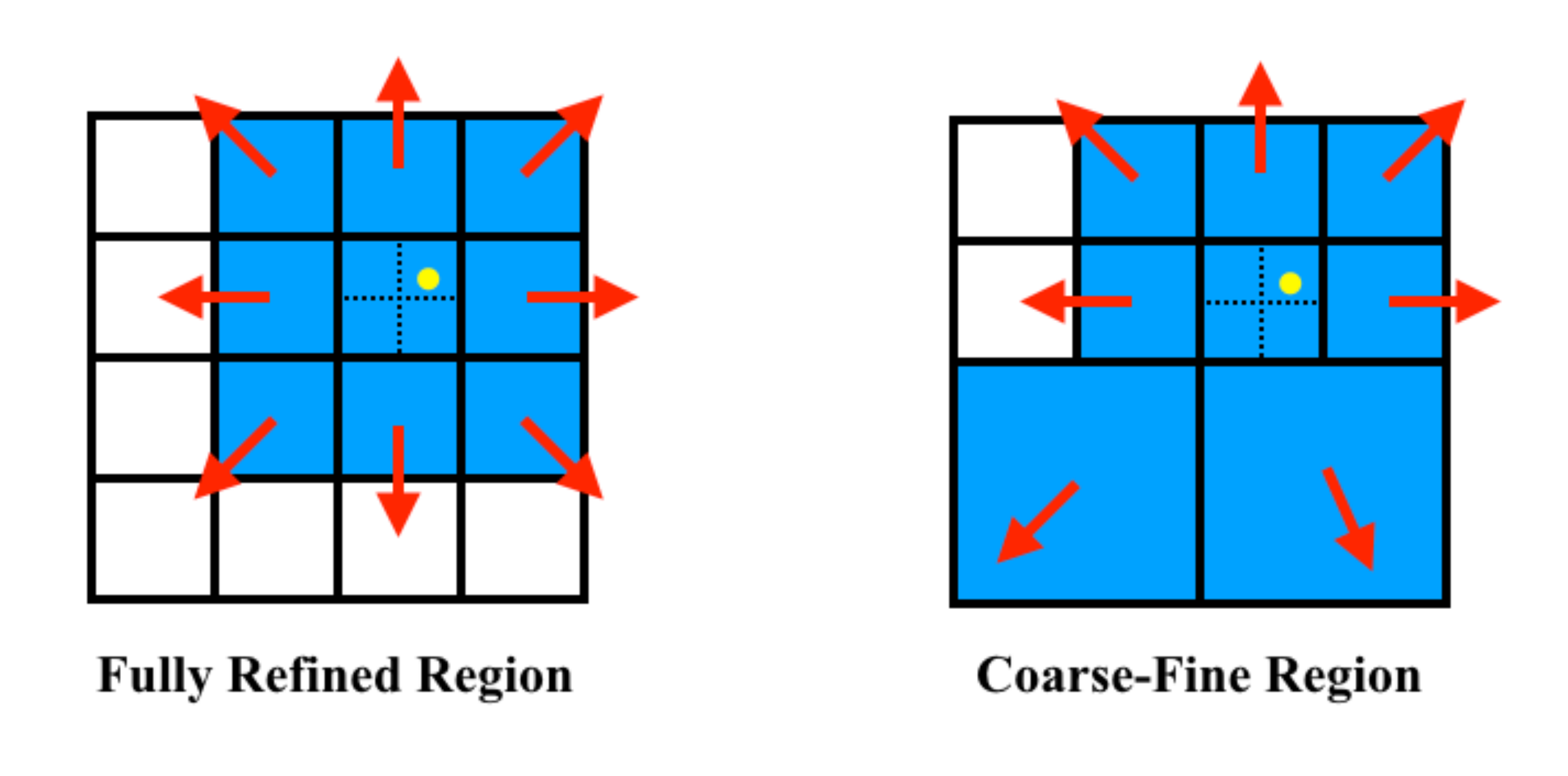}
\end{center}
\caption{Schematic representation of the supernova momentum deposition scheme. The scheme is fully implemented in 3-d, but the figures above only show the 2-d case, for simplicity. The yellow circle represents the location of the supernova progenitor. The cyan-shaded region represents the region where thermal energy is injected, whereas the red arrows represent the cells that receive momentum. Left: a stellar particle spawns a SN deep into a fully refined region, and momentum is deposited in the surrounding $3^3$ cells. Right: a stellar particle spawns a SN near a coarse-fine region; in this case, momentum is deposited within the fine level, when possible, and in the neighbouring cells at the coarse level, when fine cells are not available. }\label{fig:fb_scheme}
\end{figure*}

For densities typical of the ISM of star forming galaxies $n_{\rm H} > 1 \, {\rm cm^{-3}}$, the full evolution of individual radiating SNRs happens at sup-pc scales and cannot be tracked explicitly with parsec-resolution simulations like the ones performed in this paper. Formulae for the momentum and thermal energy of SN remnants are particularly useful, because they allow to include physics that cannot be explicitly resolved. For this paper, the analytical formulae from \citep{2015MNRAS.450..504M} were used to design an adaptive SN feedback scheme based on the physics of individual SN remnants that was fully implemented for the adaptive mesh in {\sc ramses}. {These formulae quantify the thermal energy $E_{\rm th}(R,\rho,Z)$ and radial momentum $P_{\rm r}(R,\rho,Z)$ deposited within a sphere of radius $R$ by a SN exploding in a patch of the ISM with local average density $\rho$ and metallicity $Z$: }
\begin{eqnarray}\label{eq:formulae}
E_{\rm th}(R,\rho,Z)&=& E_{\rm th,0}\theta(R_{\rm c}-R) \nonumber \\
&+&E_{\rm th,0}\left(\frac{R}{R_{\rm c}}\right)^{\alpha}\theta[(R-R_{\rm c})(R_{\rm r}-R)] \nonumber \\
&+&E_{\rm th,0}\left(\frac{R_{\rm r}}{R_{\rm c}}\right)^{\alpha}\theta(R-R_{\rm r}) \\
P_{\rm r}(R,\rho,Z)&=& P_0\left(\frac{R}{R_0}\right)^{1.5}\theta(R_{\rm b}-R)\nonumber \\
&+&P_0\left(\frac{R_{\rm b}}{R_0}\right)^{1.5}\theta(R-R_{\rm b}),
\end{eqnarray}
{where $\theta$ is the step function, $E_{\rm th,0}= 7.10\times 10^{50} \, {\rm erg}$, $P_0= 3.48\times 10^{42} \, {\rm g cm s^{-1}}$ and the model parameters are given by:}
\begin{eqnarray}
&\alpha=&-7.8\left(\frac{Z}{Z_{\odot}}\right)^{0.050}\left(\frac{n_{\rm H}}{100 {\rm~cm}^{-3}}\right)^{0.030} \nonumber \\
&R_{\rm c}=&3.0 \hbox{\rm~pc} \left(\frac{Z}{Z_{\odot}}\right)^{-0.082}\left(\frac{n_{\rm H}}{100 {\rm~cm}^{-3}}\right)^{-0.42} \nonumber \\
&R_{\rm r}=&5.5 \hbox{\rm~pc} \left(\frac{Z}{Z_{\odot}}\right)^{-0.074}\left(\frac{n_{\rm H}}{100 {\rm~cm}^{-3}}\right)^{-0.40} \nonumber \\
&R_{0}=&0.97 \hbox{\rm~pc} \left(\frac{Z}{Z_{\odot}}\right)^{0.046}\left(\frac{n_{\rm H}}{100 {\rm~cm}^{-3}}\right)^{-0.33} \nonumber \\
&R_{\rm b}=&4.0 \hbox{\rm~pc} \left(\frac{Z}{Z_{\odot}}\right)^{-0.077}\left(\frac{n_{\rm H}}{100 {\rm~cm}^{-3}}\right)^{-0.43}. 
\label{eq:fits-h}
\end{eqnarray}

After a stellar particle is marked for spawning a SN, 10 Myr after the star formation event, the following algorithm is followed to inject momentum and thermal energy:
\begin{itemize}
    \item Determine an effective injection radius $R_{\rm fb}=2.05\times \Delta x$ for the SN that depends on the local cell size given by the adaptive mesh $\Delta x$. {The factor 2.05 was chosen so that at least 2 fine cells are always included in the injection region along each Cartesian axis.}
    \item Measure the local ISM density $\rho$ and metallicity $Z$ in the region within the injection radius. 
    \item Use the analytical formulae to predict the SN remnant radial momentum $P_{\rm r}(R_{\rm fb}, \rho, Z)$ and thermal energy $E_{\rm th}(R_{\rm fb}, \rho, Z)$. 
    \item Inject radial momentum and thermal energy in a volume-weighted fashion in cells progenitor star particle. 
\end{itemize}

For the feedback scheme to be fully functional, several choices need to be made to inject momentum and thermal energy from a SN remnant on the {\sc ramses} adaptive mesh. Particular care needs to be placed in considering whether a SN goes off in a fully refined region or in a region at the coarse-fine boundary between two refinement levels (see Figure~\ref{fig:fb_scheme}). If the cell where the SN is spawned is entirely surrounded by cells at the same level of refinement, the volume where the momentum and thermal energy are injected includes this cell, plus the $3^3-1$ neighbours. If the cell where the SN is spawned is near a coarse-fine boundary, it is likely that some of the neighbouring cells will not be refined. In this case, a search for neighbouring cells at the coarse level is performed. In both cases, thermal energy and radial momentum are assigned to each cell in the injection volume in a volume-weighted fashion. 

Momentum injection is performed taking into account (I) that each progenitor stellar particle moves with respect to the mesh, (II) that each SN remnant ejects a mass $M_{\rm ej} = 3 \, M_{\odot}$, and (III) that each SN remnant is spherically symmetric in the frame of reference of the progenitor star. For simplicity, the centre of each SN remnant is taken as the location of the stellar particle that spawns it. All cells in the injection region receive a contribution from the mass of the ejecta. In practice, the mass in each cell in the injection region is updated as:
\begin{equation}
    \rho_{\rm new,cell} = \rho_{\rm old,cell} + \frac{M_{\rm ej}}{V_{\rm fb}},
\end{equation}
where $V_{\rm fb}$ is the volume of the injection region, i.e. the sum of the volumes of all cells in the injection region. Each cell in the search region receives momentum associated to the deposition of the ejecta from a moving progenitor star. Finally, radial momentum from the SN remnant evolution is also added to the cells in the injection region. In summary, the momentum density in each cell in the injection region is updated as:
\begin{equation}
    {\bf p}_{\rm new,cell} = {\bf p}_{\rm old,cell} + \frac{M_{\rm ej}}{V_{\rm fb}}{\bf v}_{\rm p} + \frac{P_{\rm r}(R_{\rm fb}, \rho, Z)}{V_{\rm fb}}\hat{\bf r},
\end{equation}
where ${\bf v}_{\rm p}$ is the velocity of the progenitor stellar particle, and $\hat{\bf r}$ is a unit vector pointing from the particle position towards the centre of the cell where momentum is injected. 

The thermal energy density in each cell of the injection region is updated as:
\begin{equation}
    \epsilon_{\rm th,cell} = E_{\rm th}(R_{\rm inj}, \rho, Z)/V_{\rm fb},
\end{equation}
The total energy density of the system is also updated by taking into account the changes to the thermal and kinetic energy of the ISM.

\subsection{Catalogue of Simulations}\label{sec:sim_catalog}

\begin{figure*}
\begin{center}
 \includegraphics[width=0.99\textwidth]{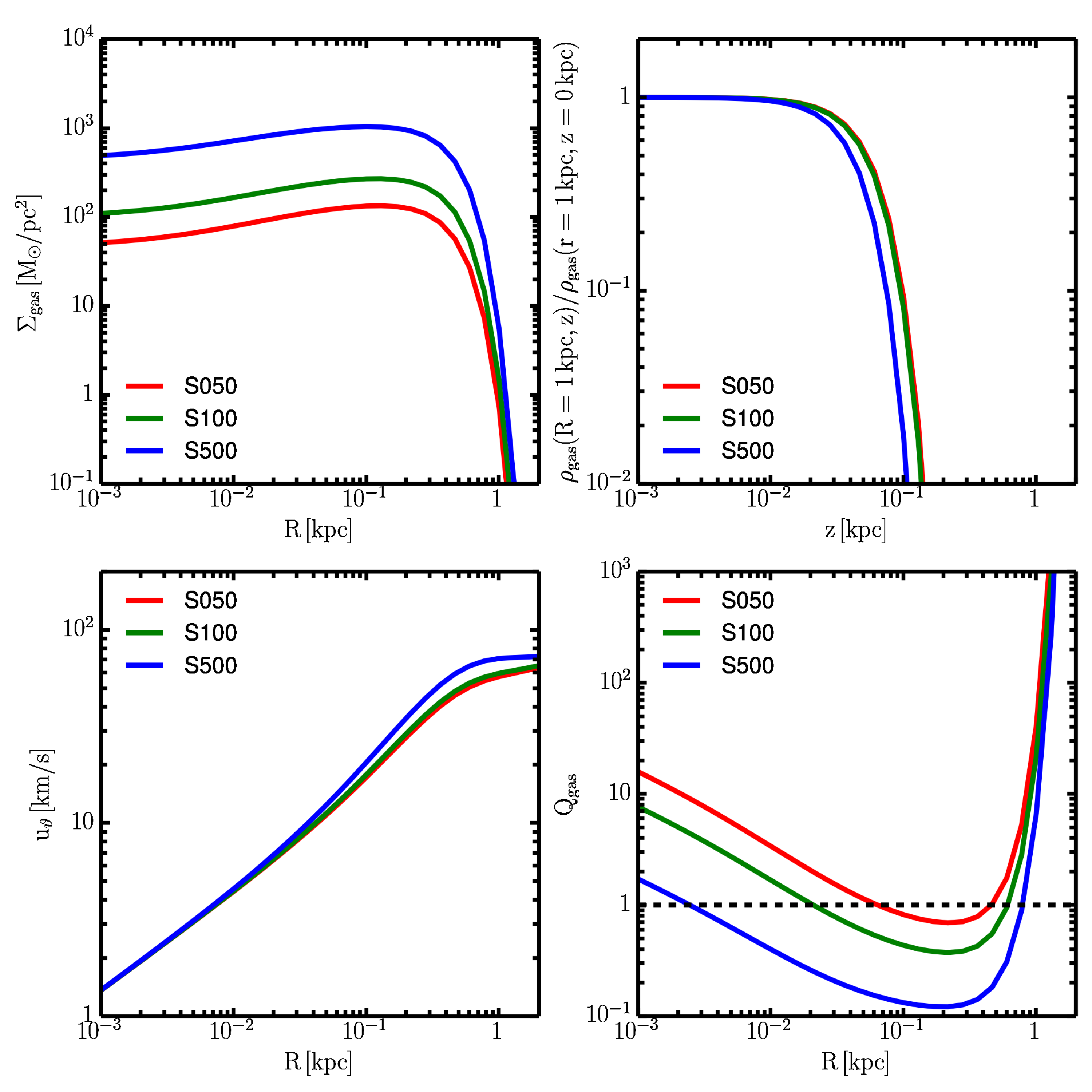}
\end{center}
\caption{The initial conditions considered in our simulations. The external potential generated by a dark matter halo, a stellar bulge and an old stellar disc is kept fixed. The amount of gas in the disc is varied in each model: S050, S100, S500 have respectively increasing gas surface density. Top left: the gas surface density as a function of radius. Top right: the normalised vertical gas density profile at radius ${R=1 \, kpc}$. Bottom left: the azimuthal velocity of the gas as a function of radius. Bottom right: the Toomre parameter $Q_{\rm gas}$ that characterise the disc stability against the gravitational instability. All the models we consider have regions prone to the development of the gravitational instability. When self-gravity is turned on, gravitational instability leads to highly clustered star formation and supernova feedback.}\label{fig:initial_conditions}
\end{figure*}

In summary, all simulations are evolved in a static external potential generated by an NFW dark matter halo of mass $M_{\rm 200} = 1.5\times 10^{11} \, M_{\odot}$,and old stellar disc of mass $M_{\rm star,old} = 5\times 10^8 \, M_{\odot}$. The disc potential scale height and radius are set to $h_{\rm R} = 0.4 \, {\rm kpc}$ and $h_{\rm z} = 0.2 \, {\rm kpc}$. The contribution from a stellar bulge was not included. 

\begin{figure*}
\begin{center}
 \includegraphics[width=0.245\textwidth]{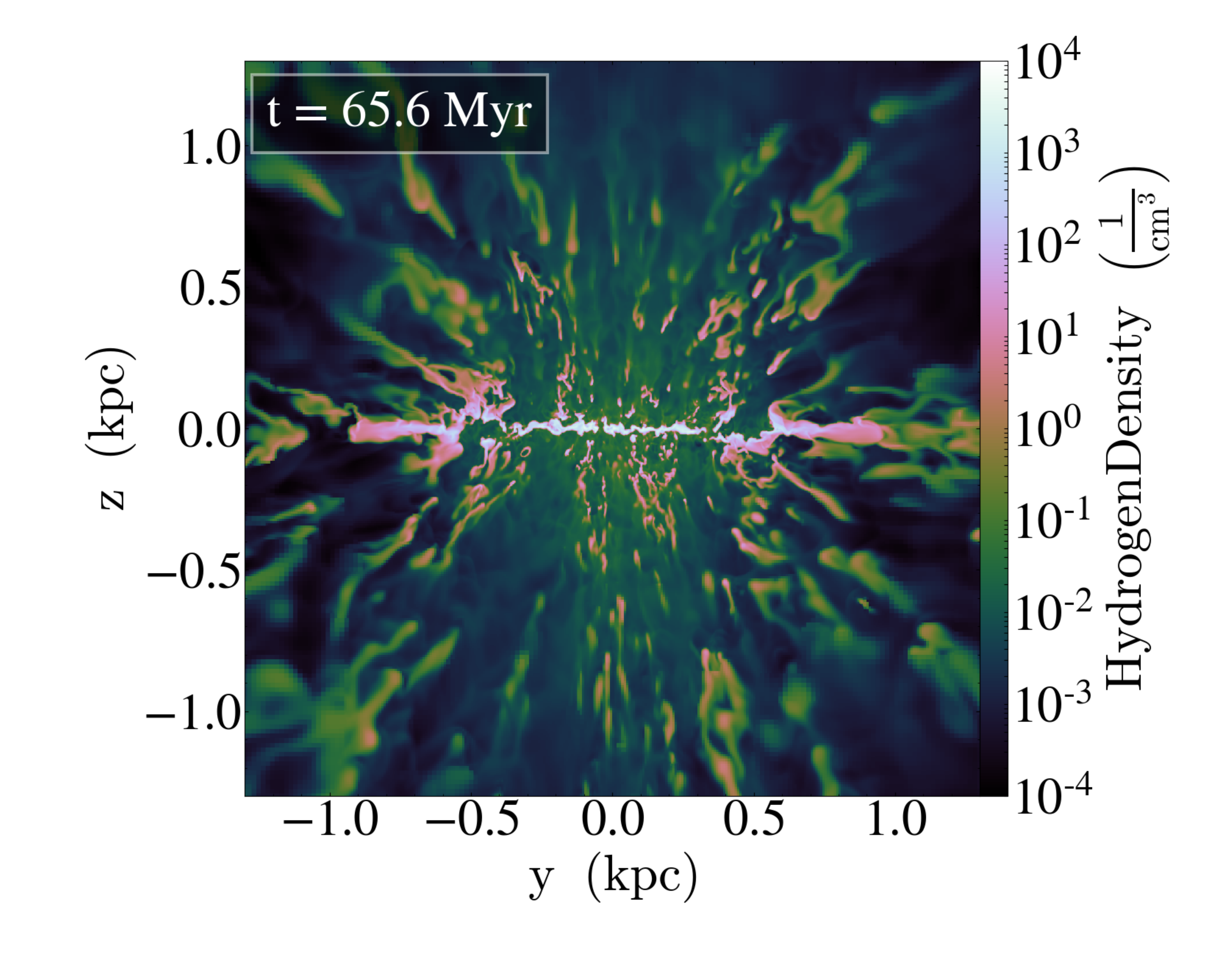}
 \includegraphics[width=0.245\textwidth]{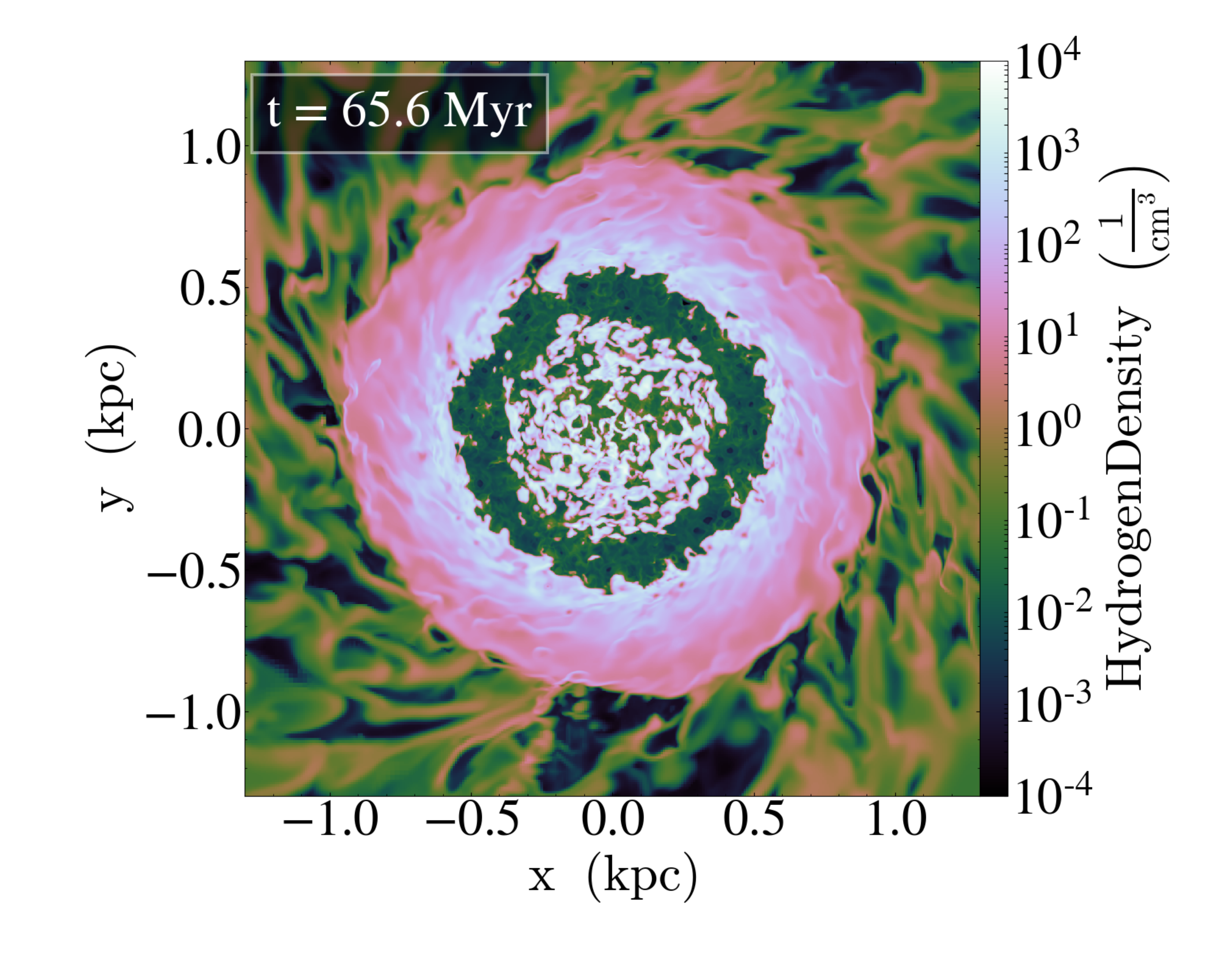}
 \includegraphics[width=0.245\textwidth]{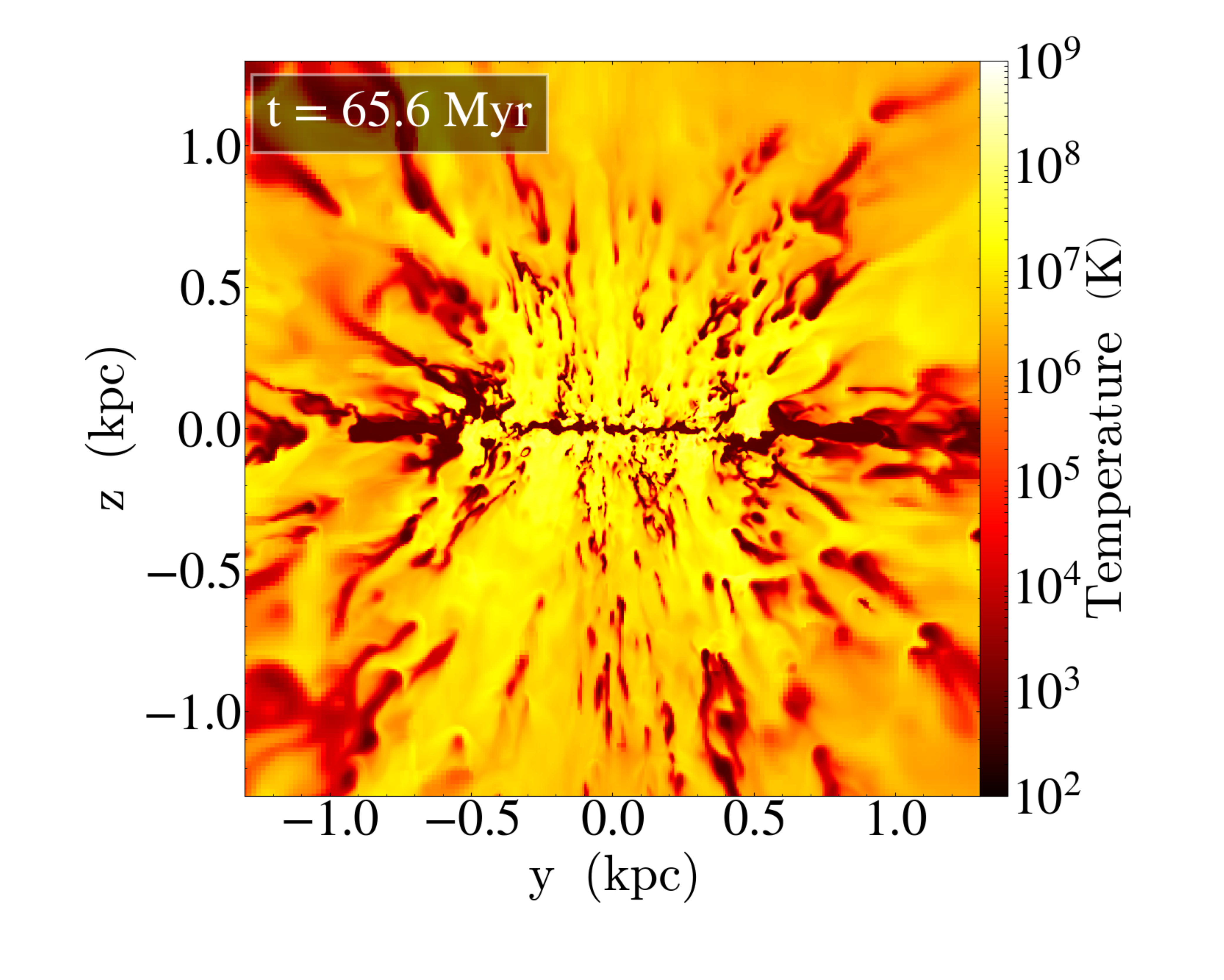}
 \includegraphics[width=0.245\textwidth]{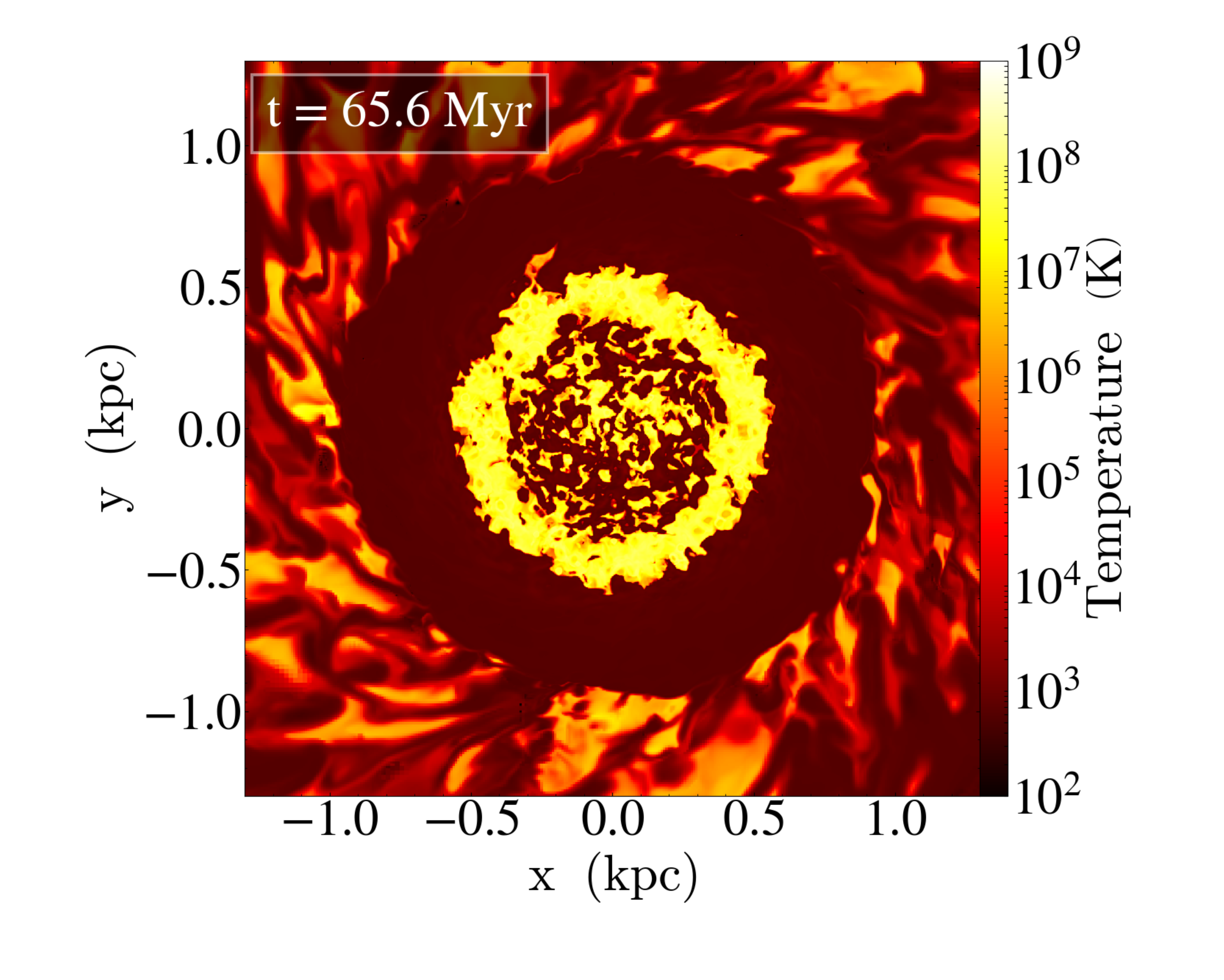}
 \includegraphics[width=0.245\textwidth]{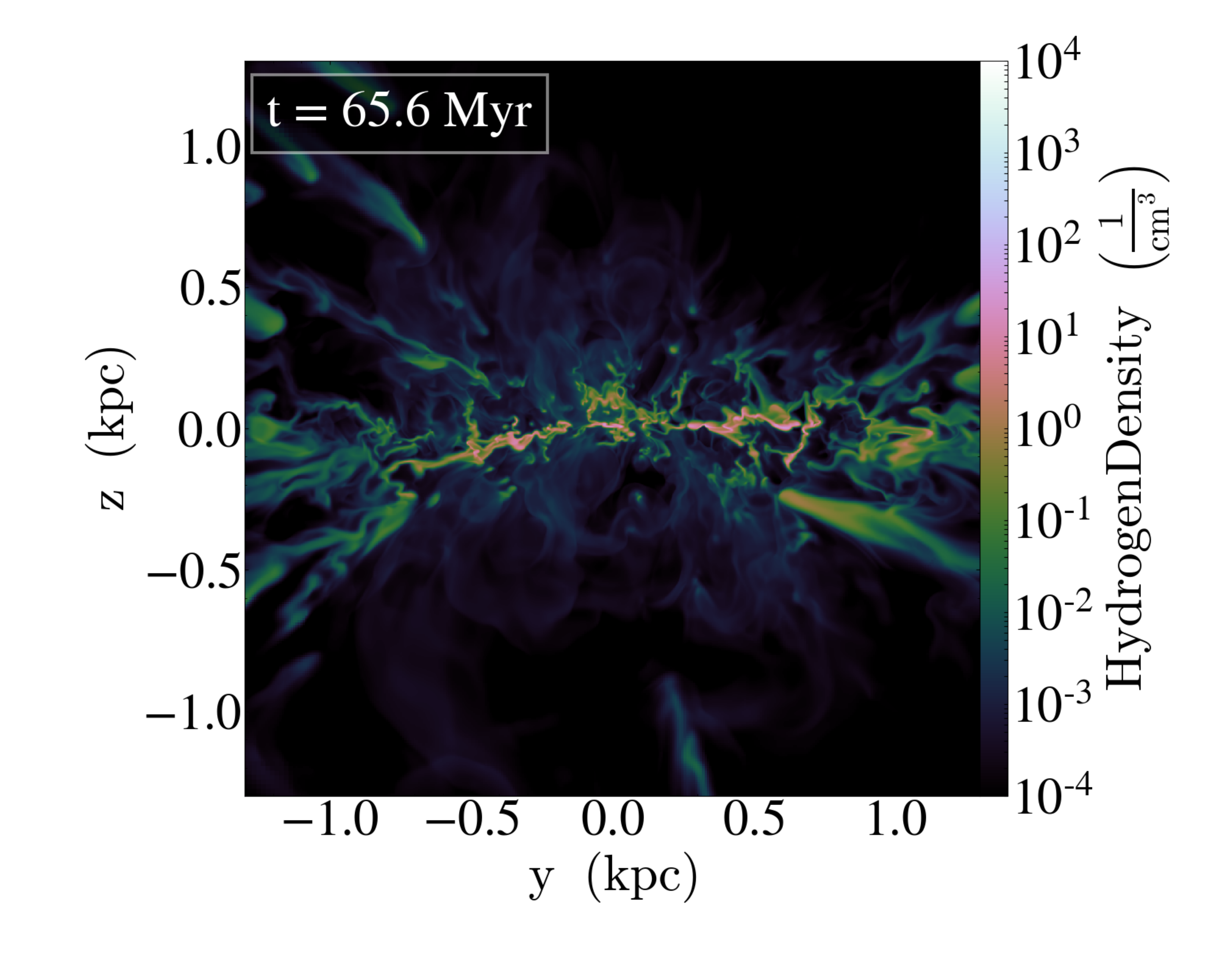}
 \includegraphics[width=0.245\textwidth]{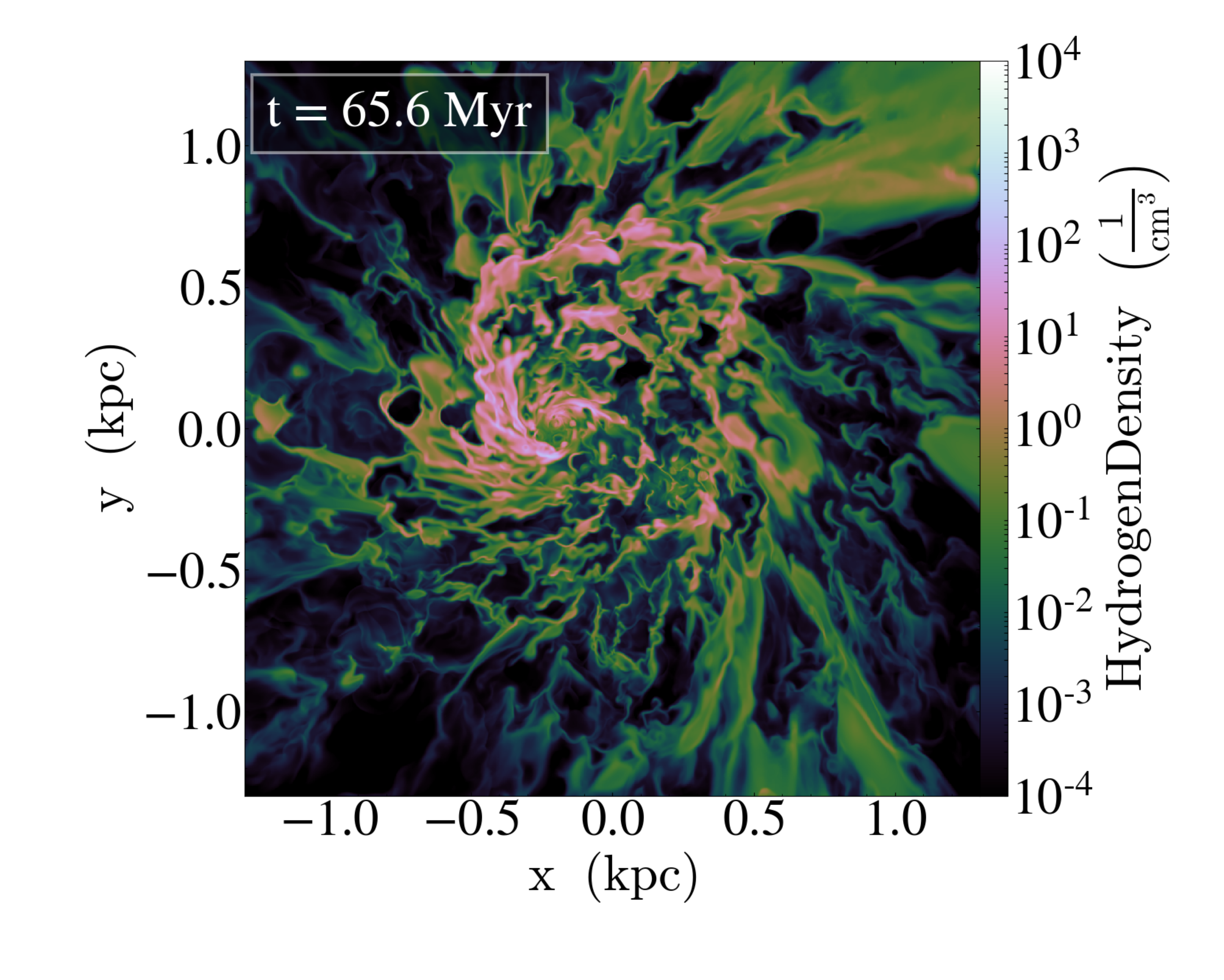}
 \includegraphics[width=0.245\textwidth]{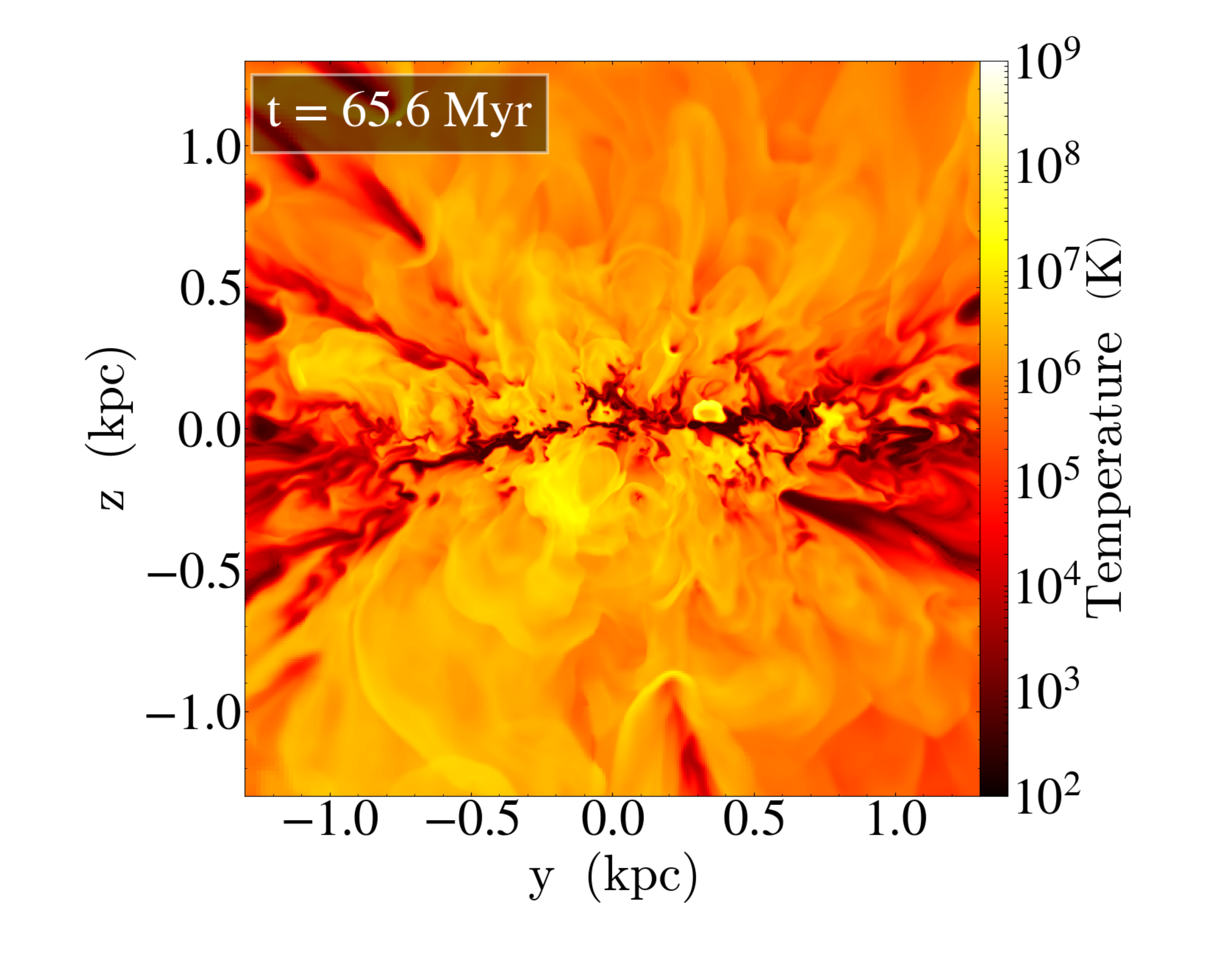}
 \includegraphics[width=0.245\textwidth]{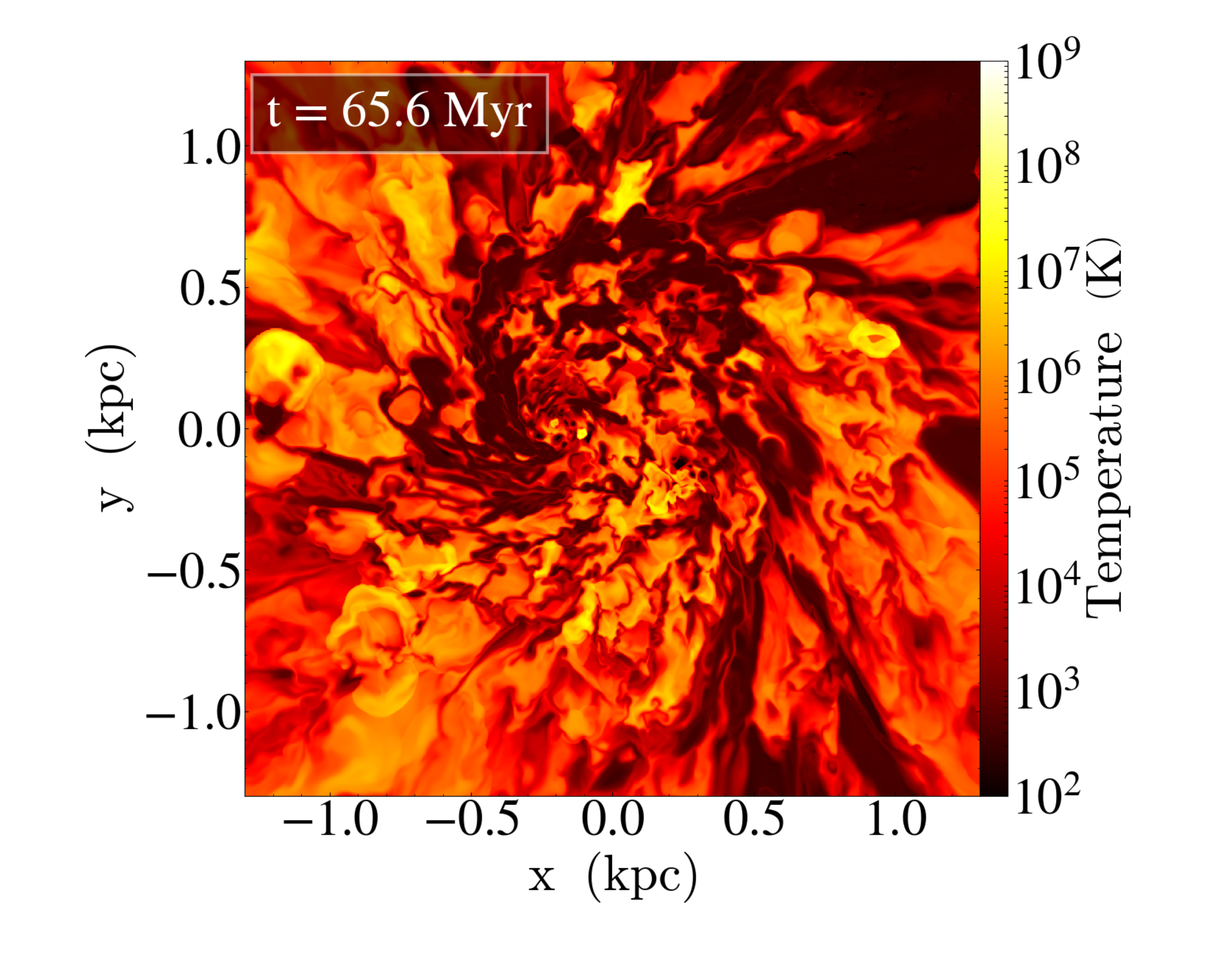}
\end{center}
\caption{Final snapshots of the S500\_NOSG (top row) and S500\_WSG simulations (bottom row). The plots represent gas density and temperature maps in razor-thin slices passing through the mid-plane of the disc and shown edge-on and face-on. Simulations with self-gravity (WSG) develop a highly turbulent disc structure as a result of highly clustered supernova feedback. Simulations without self-gravity have a smoother distribution of stellar particles and supernovae, which leads to a more regular disc structure. Galactic winds are driven by supernova feedback in both cases, but they have different properties and different temporal evolution. }\label{fig:maps_comp}
\end{figure*}

\begin{table*}
\centering
\caption{ This table summarises the parameters adopted for the  simulations considered in this paper. The columns show the simulation name, the box size $L$, the minimum cell size $\Delta x_{\rm fine}$, the dark matter halo mass $M_{\rm 200}$, the old stellar disc mass $M_{\rm star,old}$, the scale radius of the stellar disc potential $h_{\rm R}$, the scale height of the stellar disc potential $h_{\rm z}$, the mass of the gas disc $M_{\rm gas}$, whether the simulation includes self-gravity, and the dynamical time in the disc mid-plane in the initial conditions $t_{\rm dyn}$.}\label{tab:ic_params}
{\bfseries Parameters of the Simulations}
\makebox[\linewidth]{
\begin{tabular}{lccccccccc}
\hline
\hline
Simulation Name & $L \, [{\rm kpc}]$ & $\Delta x_{\rm fine} \, [{\rm pc}]$ & $M_{\rm 200} \, [M_{\odot}]$ & $M_{\rm star,old} \, [M_{\odot}]$ & $h_{\rm R} \, [kpc]$ & $h_{\rm z}$ & $M_{\rm gas}$ & Self-gravity & $t_{\rm dyn} [{\rm Myr}]$ \\
\hline
S050\_NOSG & 4.096 & 2 & $1.5\times 10^{11}$ & $5.0\times10^8$ & 0.4 & 0.2 & $1.0\times 10^8$ & No & 14.41 \\
S050\_WSG & 4.096 & 2 & $1.5\times 10^{11}$ & $5.0\times10^8$ & 0.4 & 0.2 & $1.0\times 10^8$ & Yes & 14.41 \\
S100\_NOSG & 4.096 & 2 & $1.5\times 10^{11}$ & $5.0\times10^8$ & 0.4 & 0.2 & $2.0\times 10^8$ & No & 12.12 \\
S100\_WSG & 4.096 & 2 & $1.5\times 10^{11}$ & $5.0\times10^8$ & 0.4 & 0.2 & $2.0\times 10^8$ & Yes & 12.12 \\
S100\_WSG\_LBOX & 8.192 & 2 & $1.5\times 10^{11}$ & $5.0\times10^8$ & 0.4 & 0.2 & $2.0\times 10^8$ & Yes & 12.12 \\
S100\_WSG\_LRES & 4.096 & 4 & $1.5\times 10^{11}$ & $5.0\times10^8$ & 0.4 & 0.2 & $2.0\times 10^8$ & Yes & 12.12 \\
S500\_NOSG & 4.096 & 2 & $1.5\times 10^{11}$ & $5.0\times10^8$ & 0.4 & 0.2 & $7.5\times 10^8$ & No & 7.50 \\
S500\_WSG & 4.096 & 2 & $1.5\times 10^{11}$ & $5.0\times10^8$ & 0.4 & 0.2 & $7.5\times 10^8$ & Yes & 7.50 \\
\hline
\hline
\end{tabular}
}
\end{table*}

Gas is initialised in a rotating disc of mass $M_{\rm gas}$. The external potential and self-gravity are included in half of the simulation suite (WSG), whereas the other half only includes the contribution from the external potential (NOSG). In order to study the response of the disc to supernova feedback and gravitational instability, simulations with multiple values of $M_{\rm gas}$ were performed. The S050\_WSG and S050\_NOSG simulations have $M_{\rm gas} = 10^8 \, M_{\odot}$, which corresponds to a central gas surface density $\Sigma_{\rm gas}\sim 50 \, M_{\odot}/{\rm pc}^2$. The S100\_WSG and S100\_NOSG simulations have $M_{\rm gas} = 2\times 10^8 \, M_{\odot}$, which corresponds to a central gas surface density $\Sigma_{\rm gas}\sim 100 \, M_{\odot}/{\rm pc}^2$. The S500\_WSG and S500\_NOSG simulations have $M_{\rm gas} = 7.5\times 10^8 \, M_{\odot}$, which corresponds to a central gas surface density $\Sigma_{\rm gas}\sim 500 \, M_{\odot}/{\rm pc}^2$. Figure~\ref{fig:initial_conditions} shows the properties of the gas distribution in the initial conditions: the radial surface density profile, the vertical gas density profile, the rotation velocity of the gas and Toomre's parameter $Q_{\rm gas}=c_s\kappa/(\pi G \Sigma)$. In particular, the $Q_{\rm gas}$ profile shows that the chosen initial conditions are all prone to gravitational instability at radii where $Q_{\rm gas}<1$ when gas-self gravity is turned on, as intended. 

To perform resolution and robustness tests, two alternative versions of the S100\_WSG simulation were run, one with a factor two lower spatial resolution (S100\_WSG\_LRES), and one with the same resolution, but with a twice bigger computational box (S100\_WSG\_LBOX). It was found that the predictions of the simulations do not vary significantly with spatial resolution (provided that it is at least 4 pc), and that varying the box size only introduces minor differences. For completeness, the results of the tests are shown in Appendix~\ref{app:tests}. The key parameters of all the simulations are summarised in Table~\ref{tab:ic_params}.

\section{Results} \label{sec:results}

All the global disc simulations considered in this paper have been evolved for $\gtrsim 7t_{\rm dyn}$, where $t_{\rm dyn}=(4\pi G\rho_{\rm 12})^{-1/2}$ is the dynamical time in the initial conditions, and $\rho_{12}$ is the average gas density measured in a slice of thickness 12 pc centered at the disc mid-plane (Table~\ref{tab:ic_params}). This time scale was found to be long enough to capture the response of the disc to star formation and SN feedback. The results of the analysis of the simulation suite are reported and discussed below. 

\begin{figure*}
\begin{center}
 \includegraphics[width=0.245\textwidth]{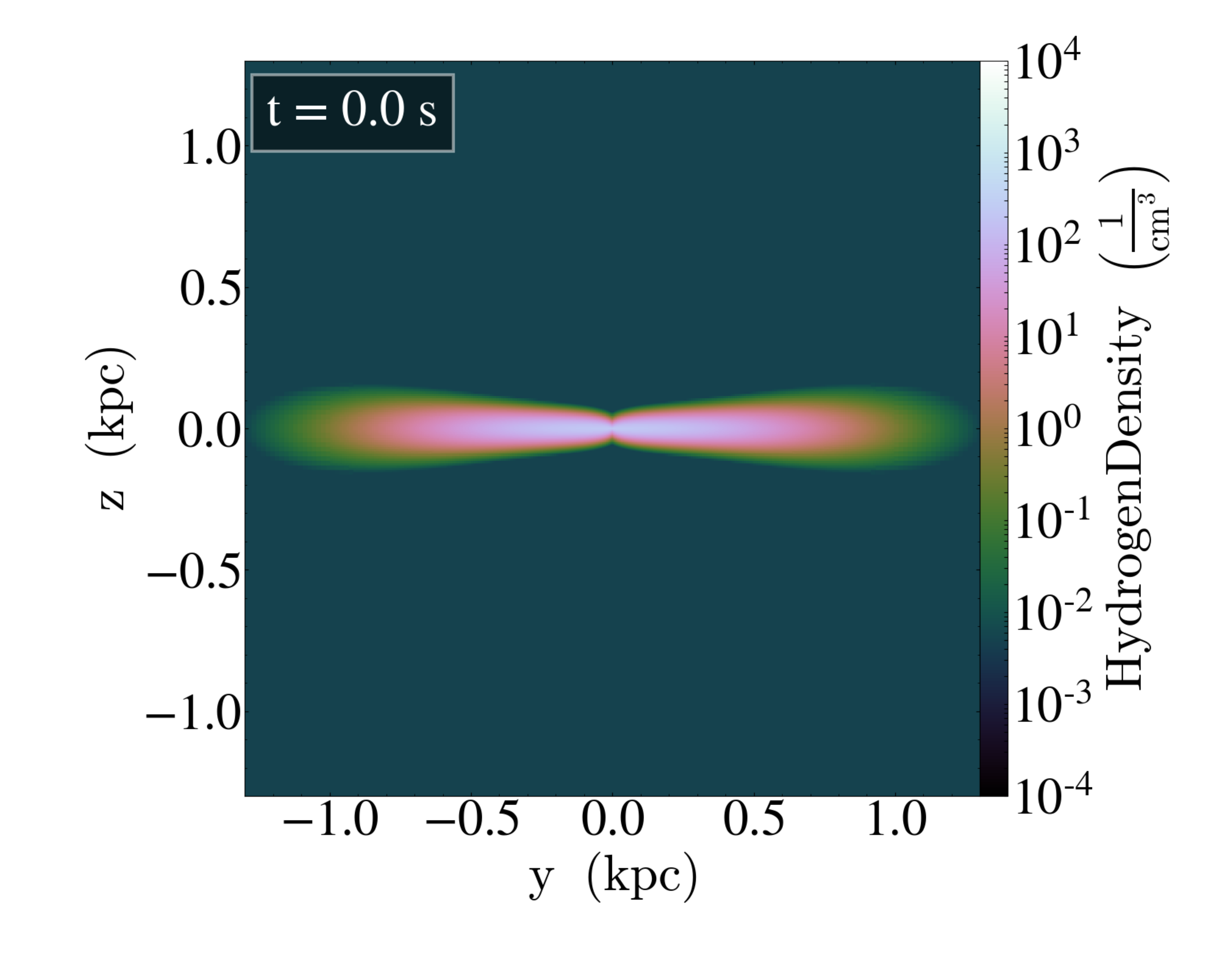}
 \includegraphics[width=0.245\textwidth]{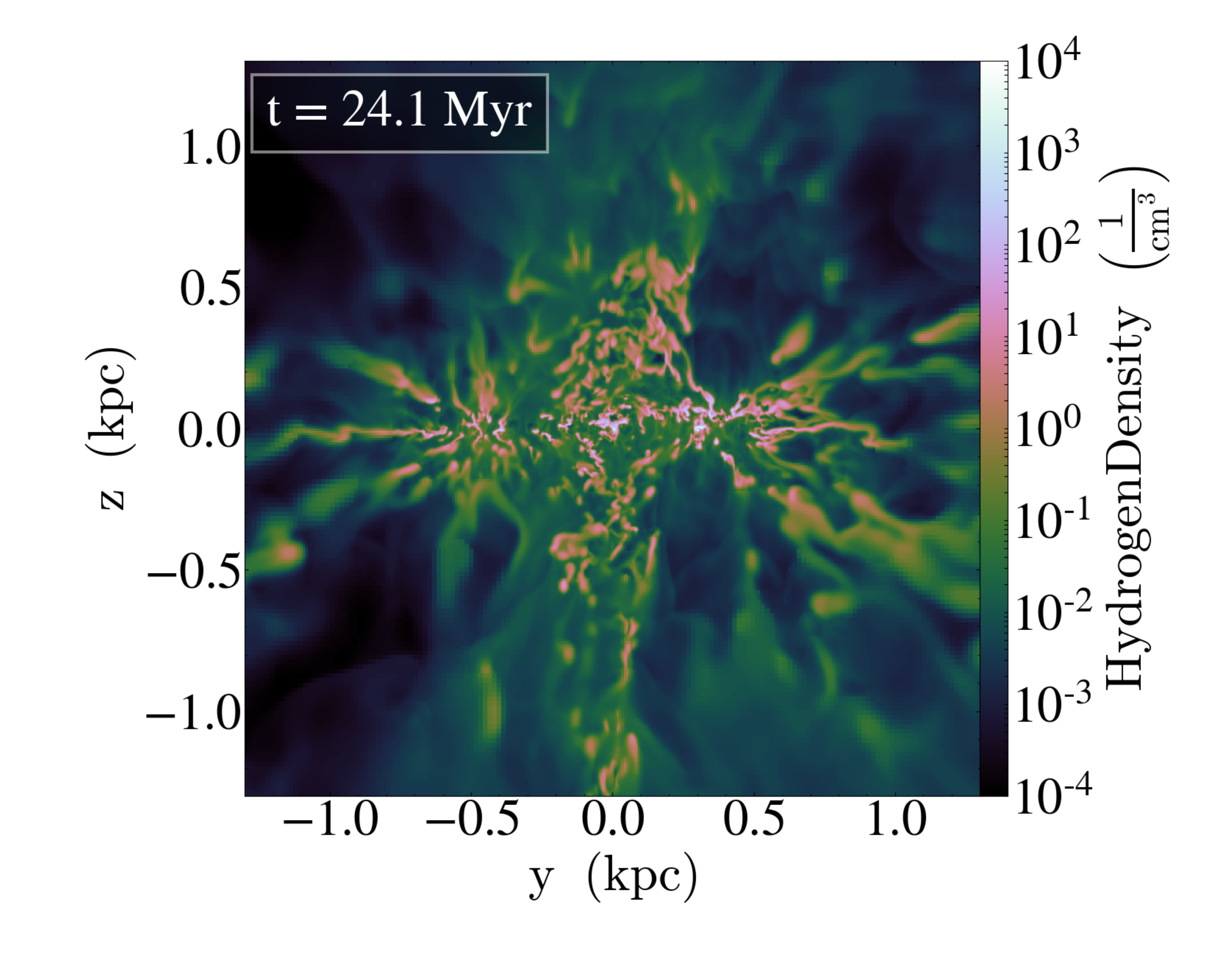}
 \includegraphics[width=0.245\textwidth]{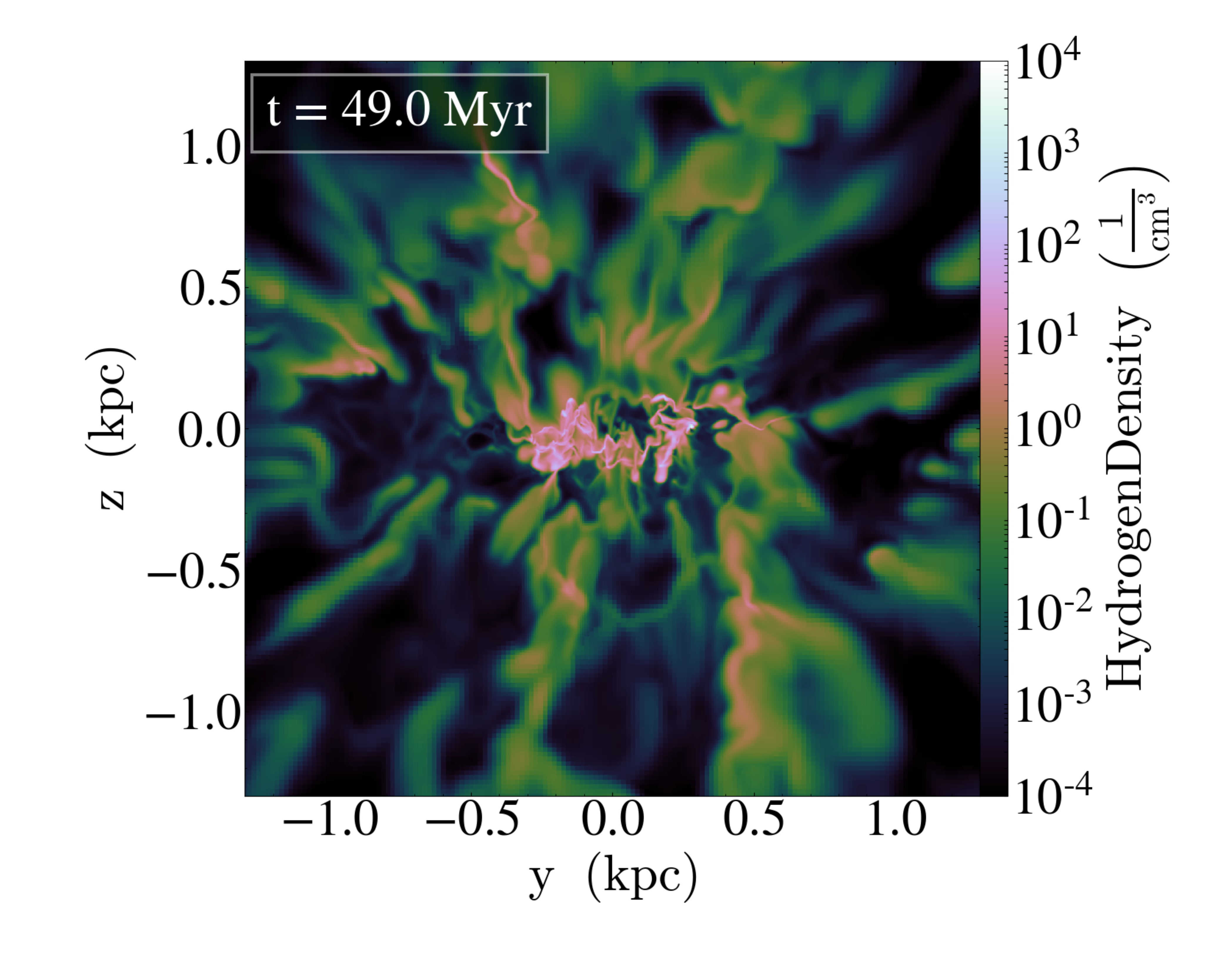}
 \includegraphics[width=0.245\textwidth]{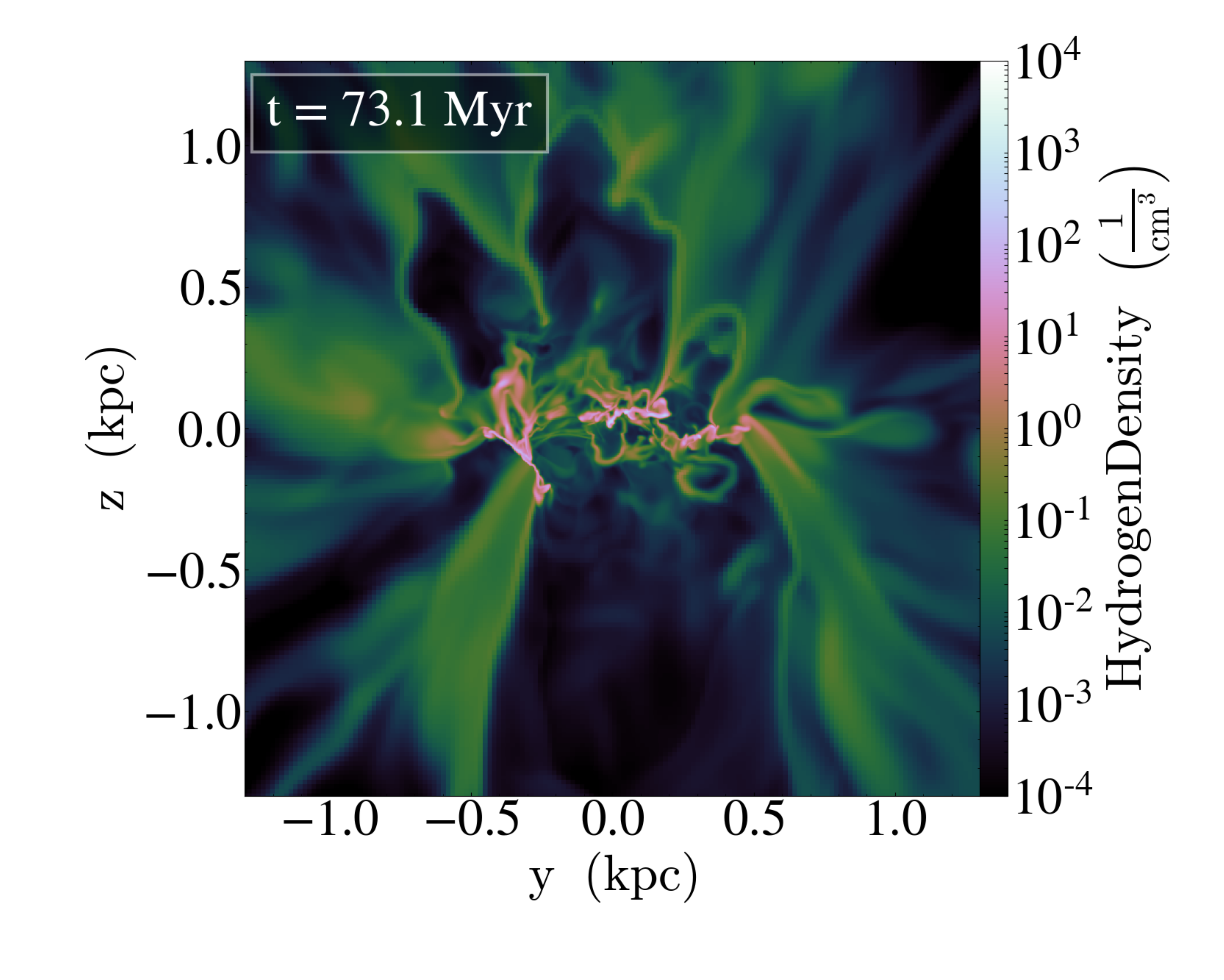}
 \includegraphics[width=0.245\textwidth]{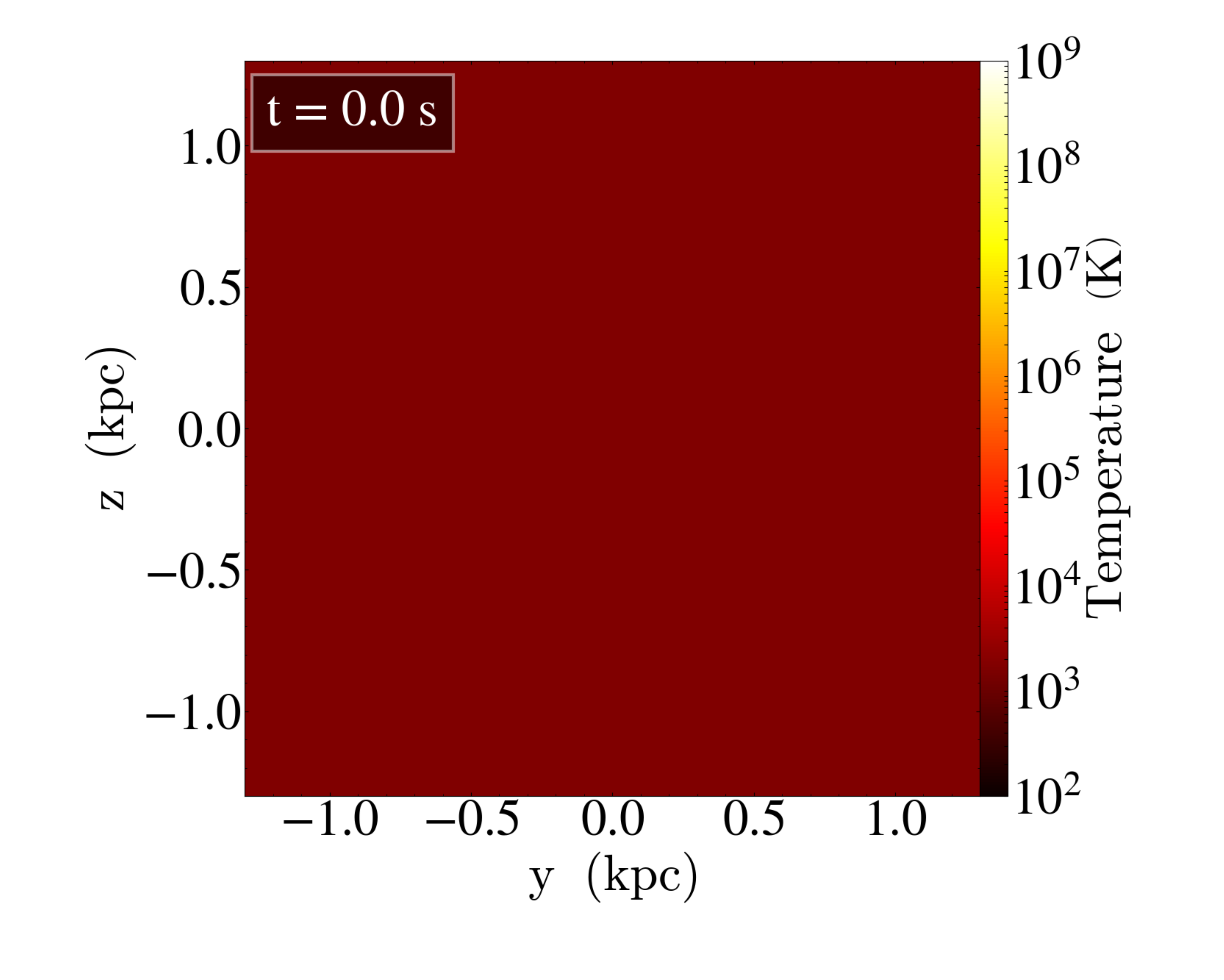}
 \includegraphics[width=0.245\textwidth]{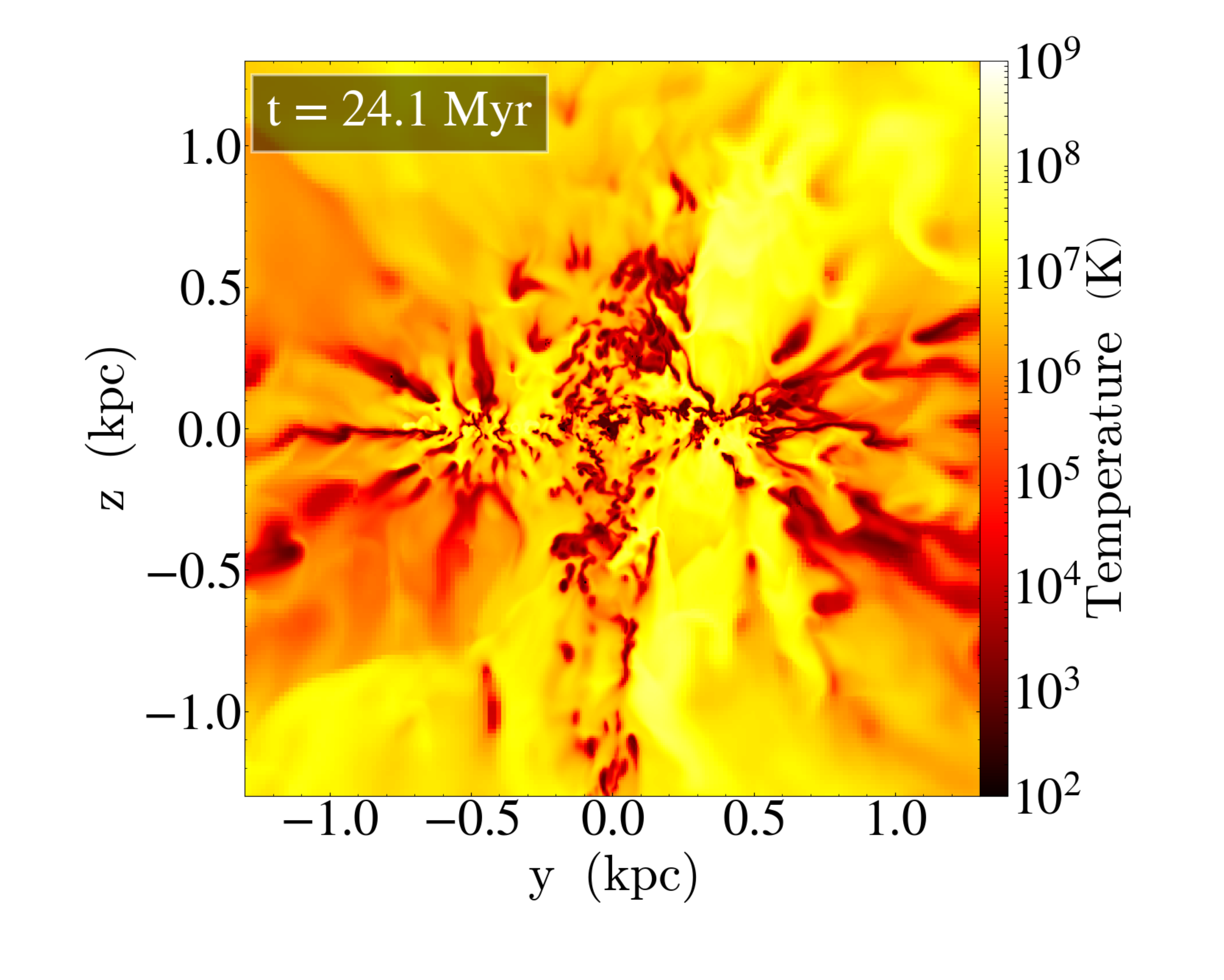}
 \includegraphics[width=0.245\textwidth]{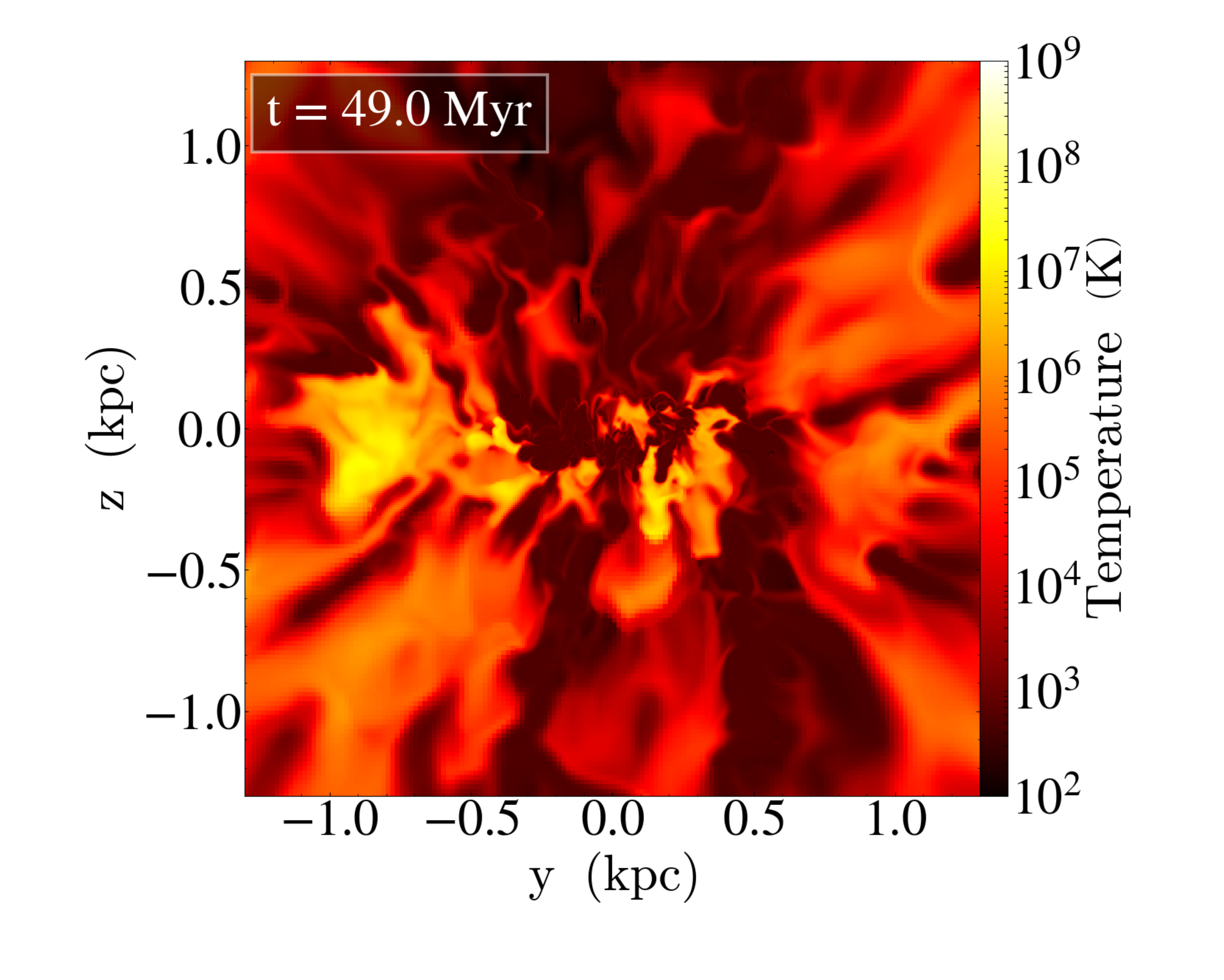}
 \includegraphics[width=0.245\textwidth]{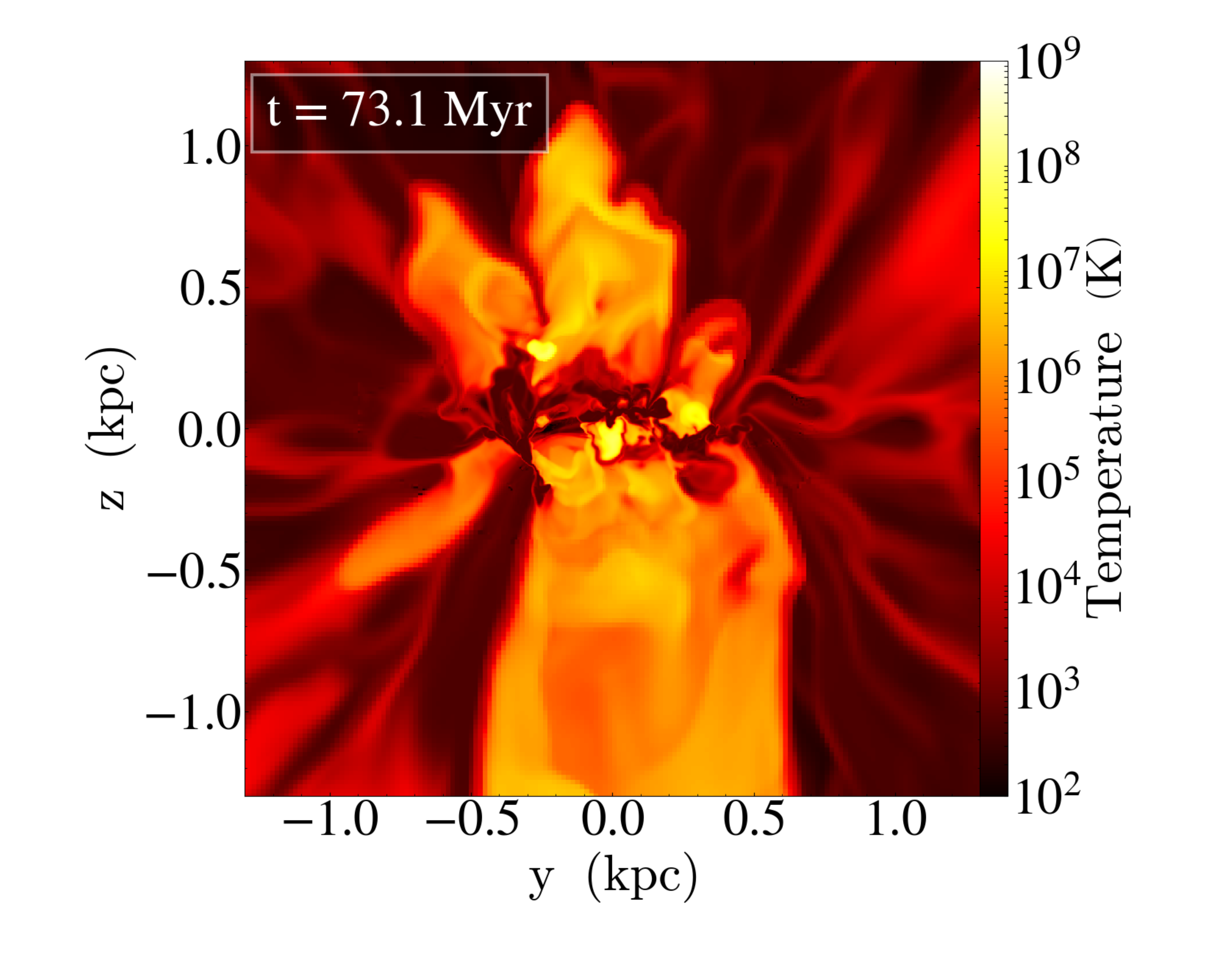}
 \includegraphics[width=0.245\textwidth]{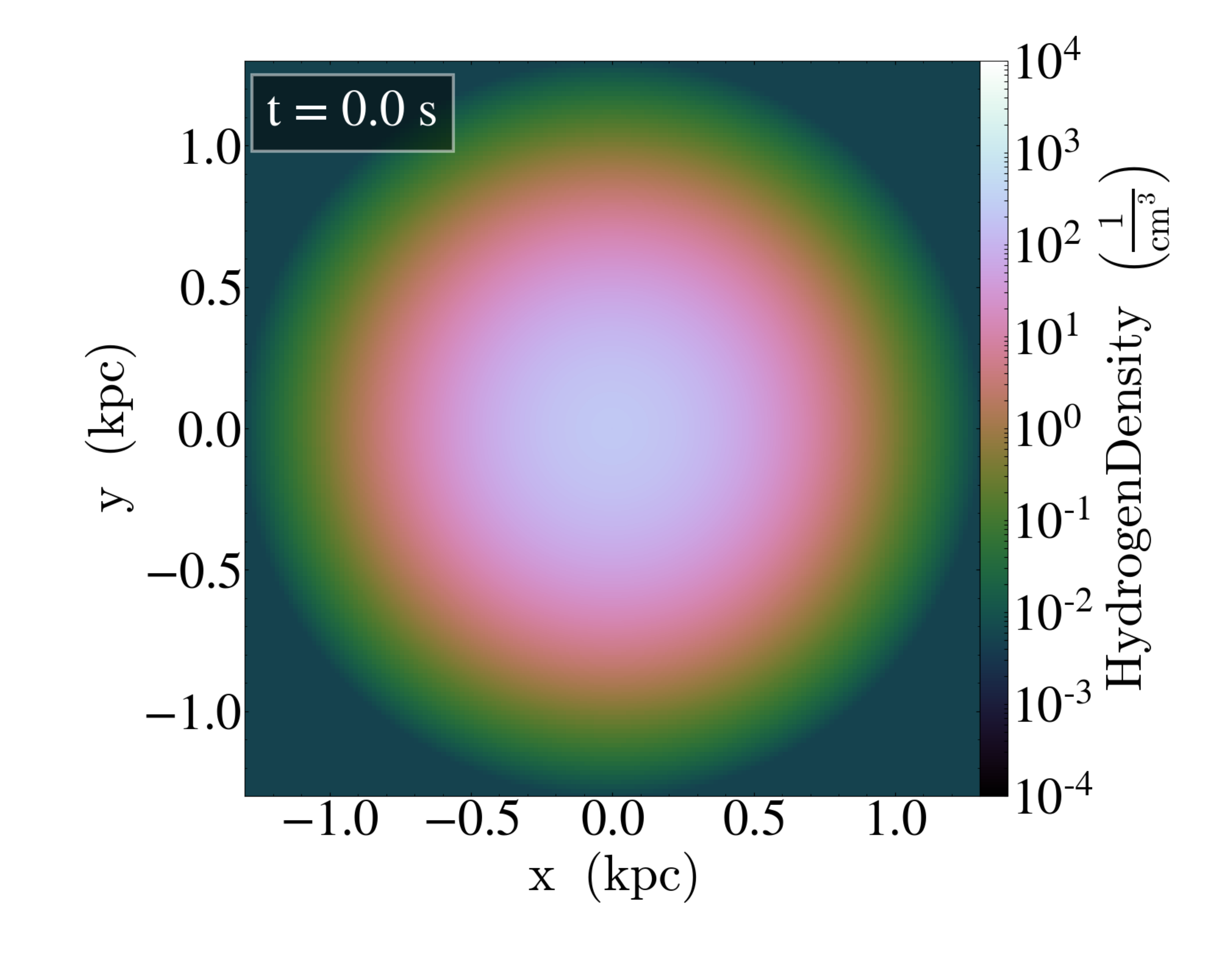}
 \includegraphics[width=0.245\textwidth]{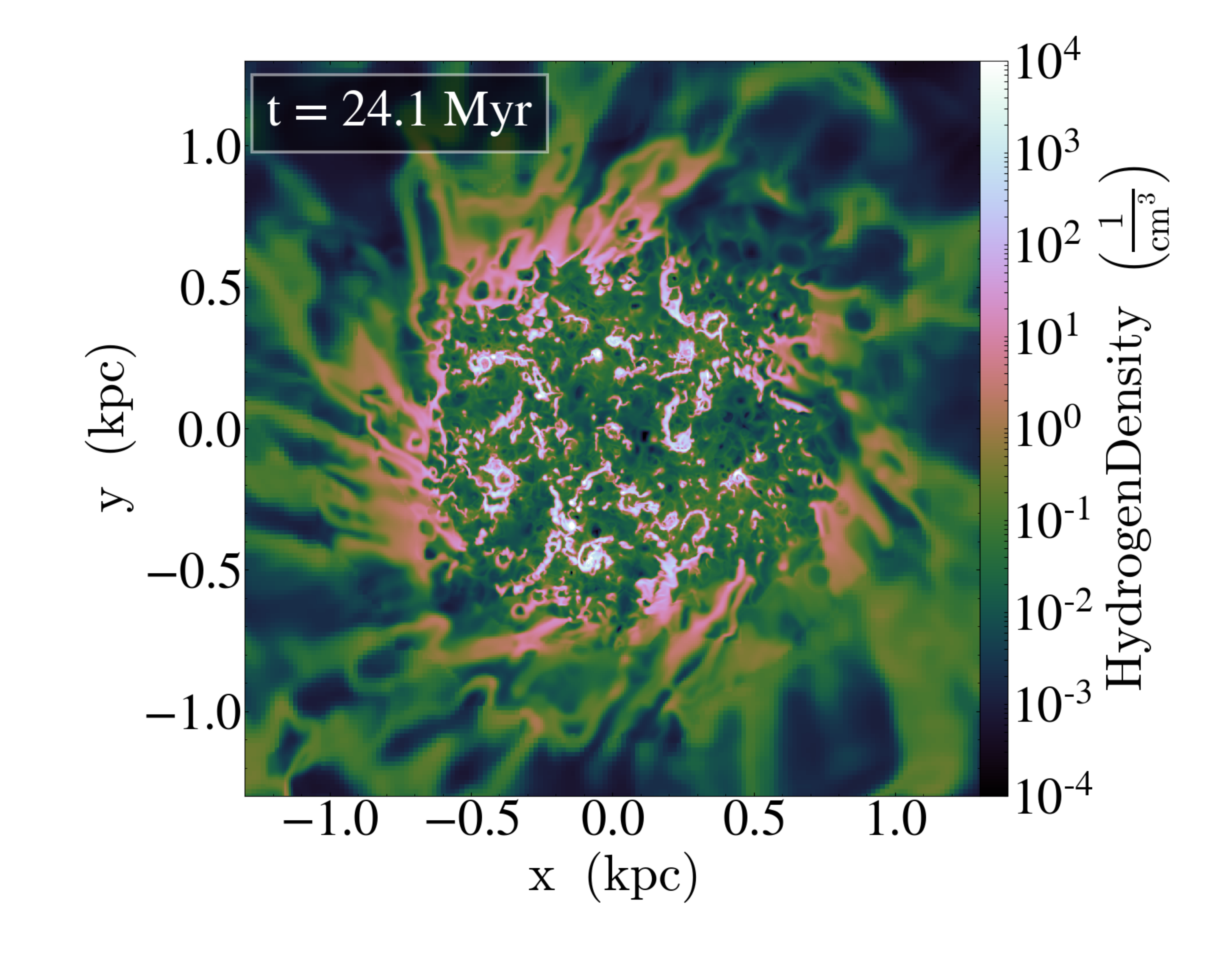}
 \includegraphics[width=0.245\textwidth]{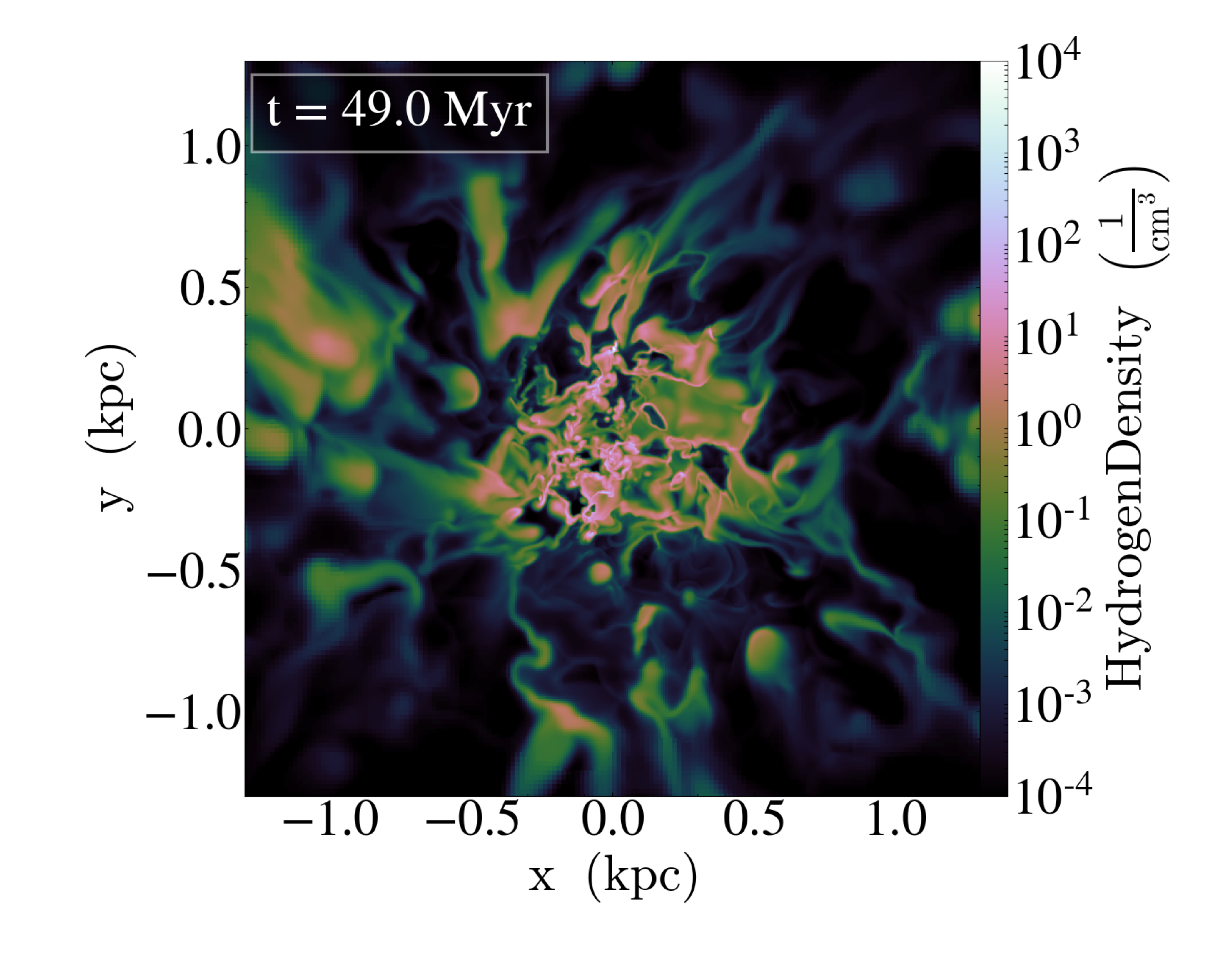}
 \includegraphics[width=0.245\textwidth]{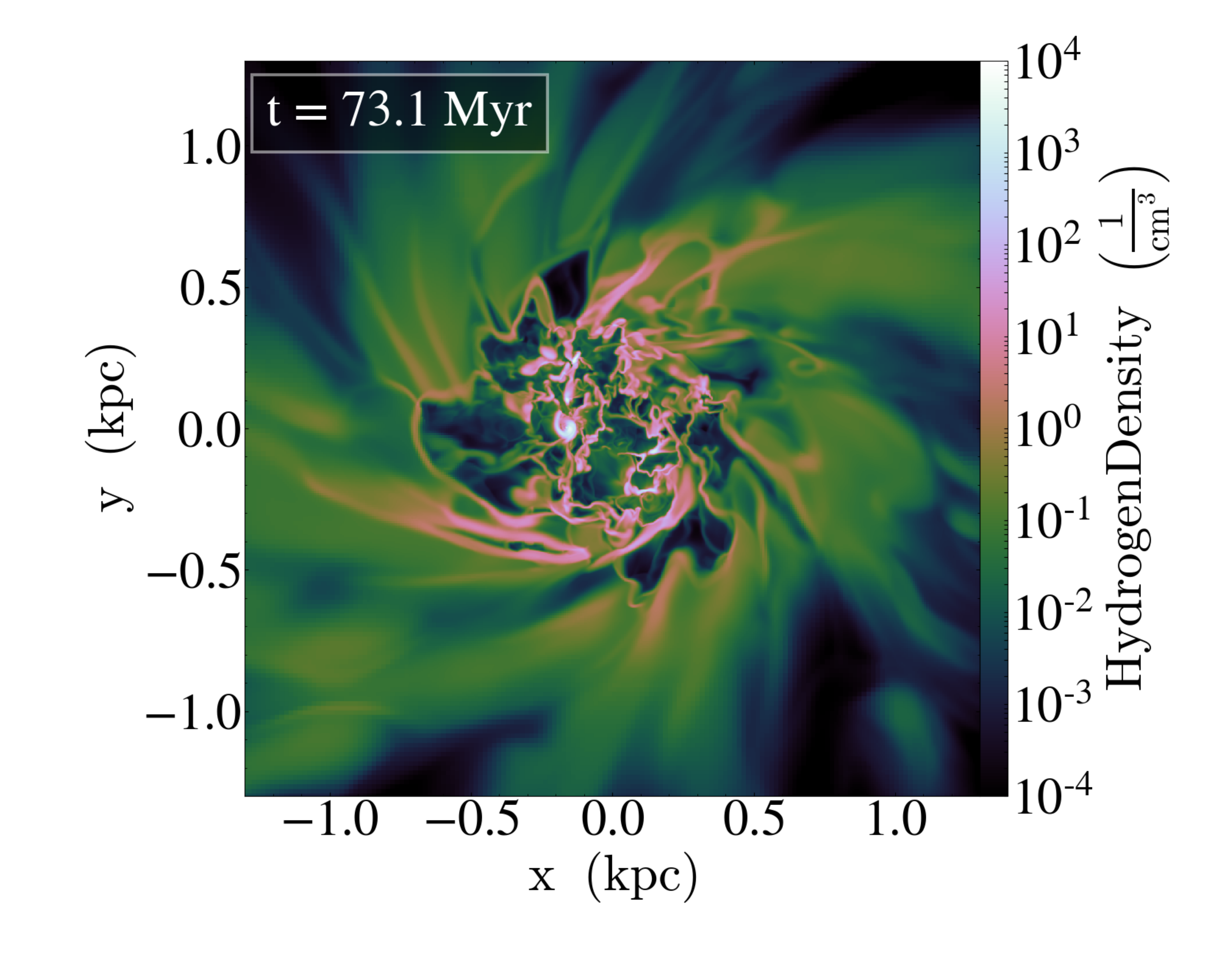}
 \includegraphics[width=0.245\textwidth]{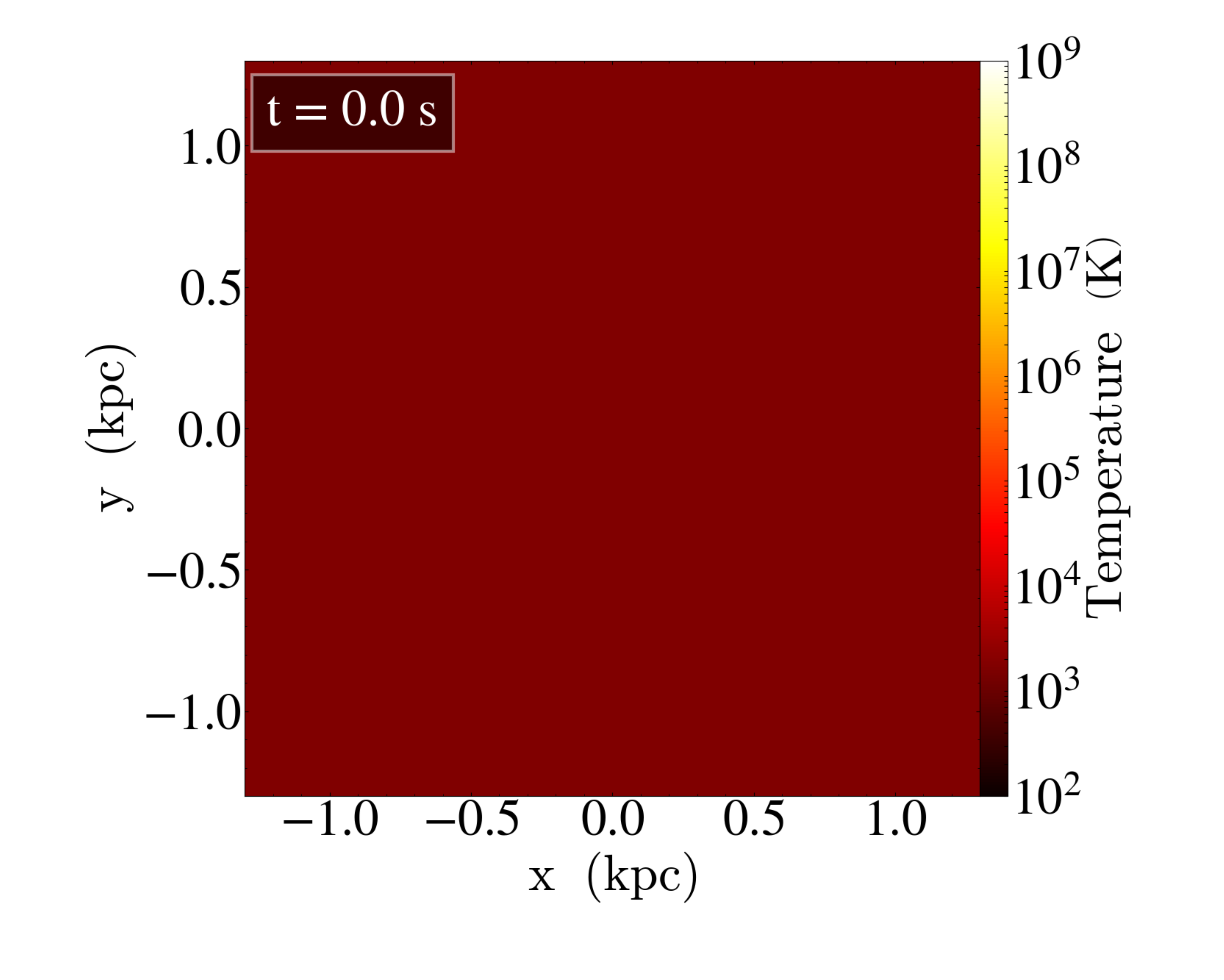}
 \includegraphics[width=0.245\textwidth]{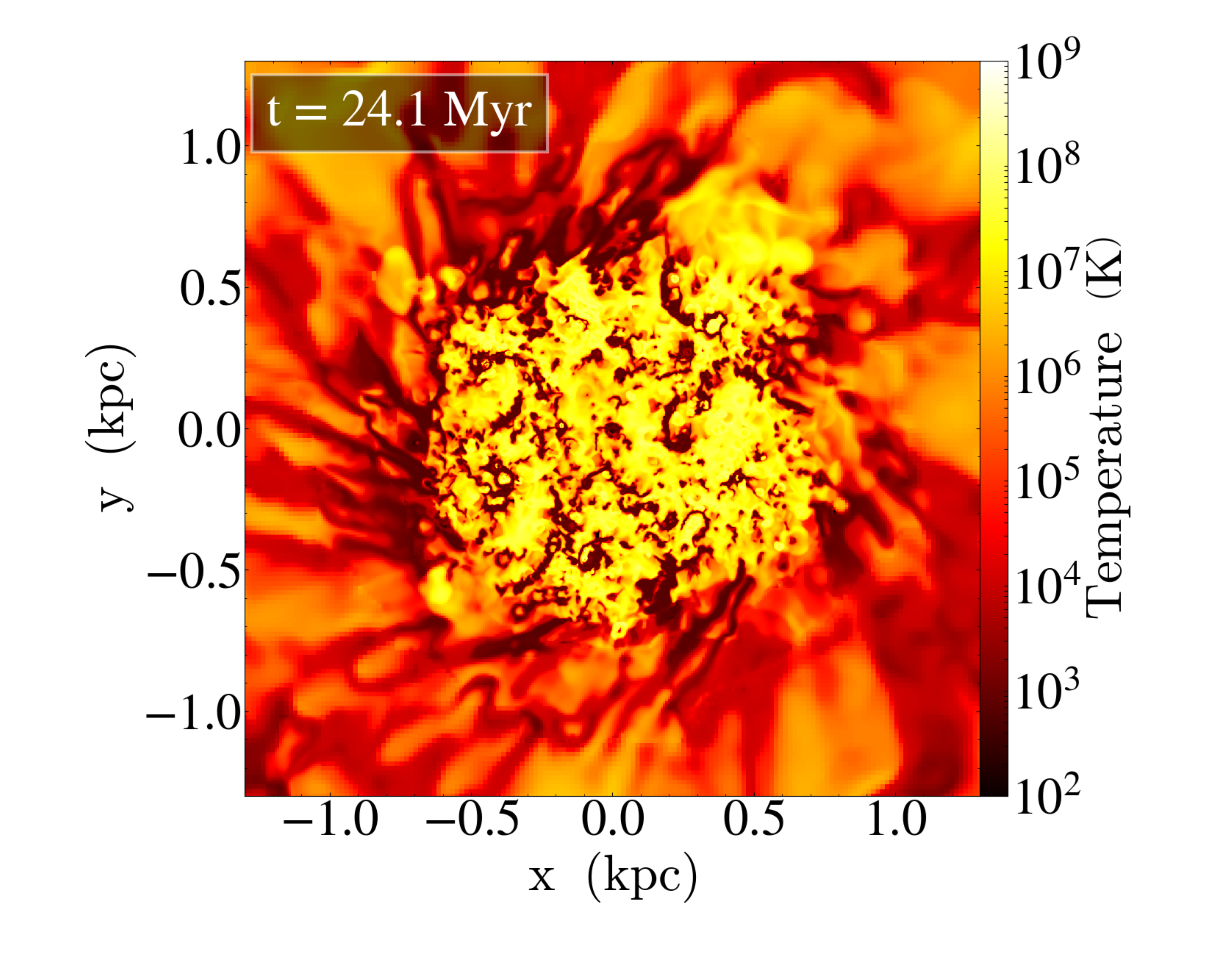}
 \includegraphics[width=0.245\textwidth]{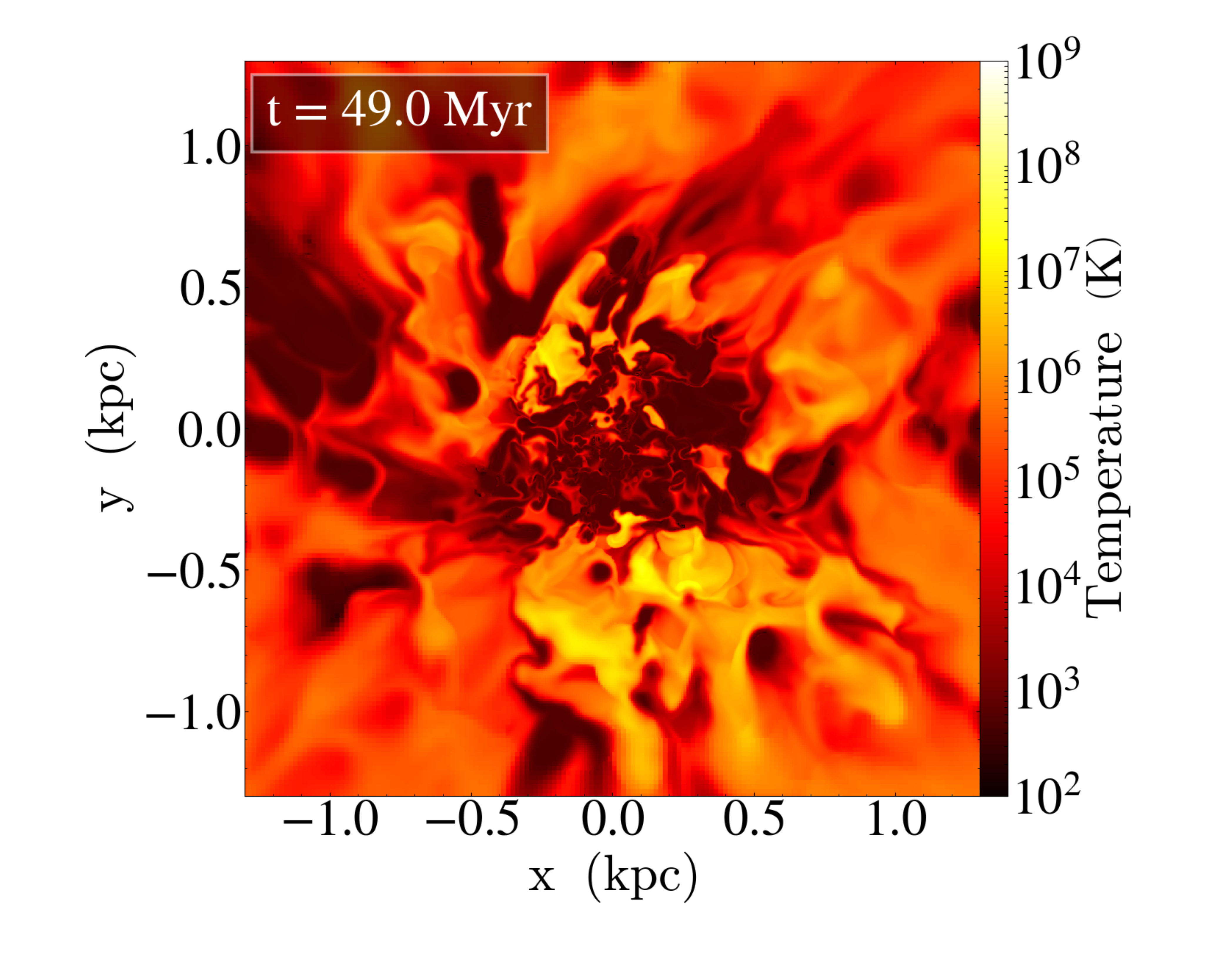}
 \includegraphics[width=0.245\textwidth]{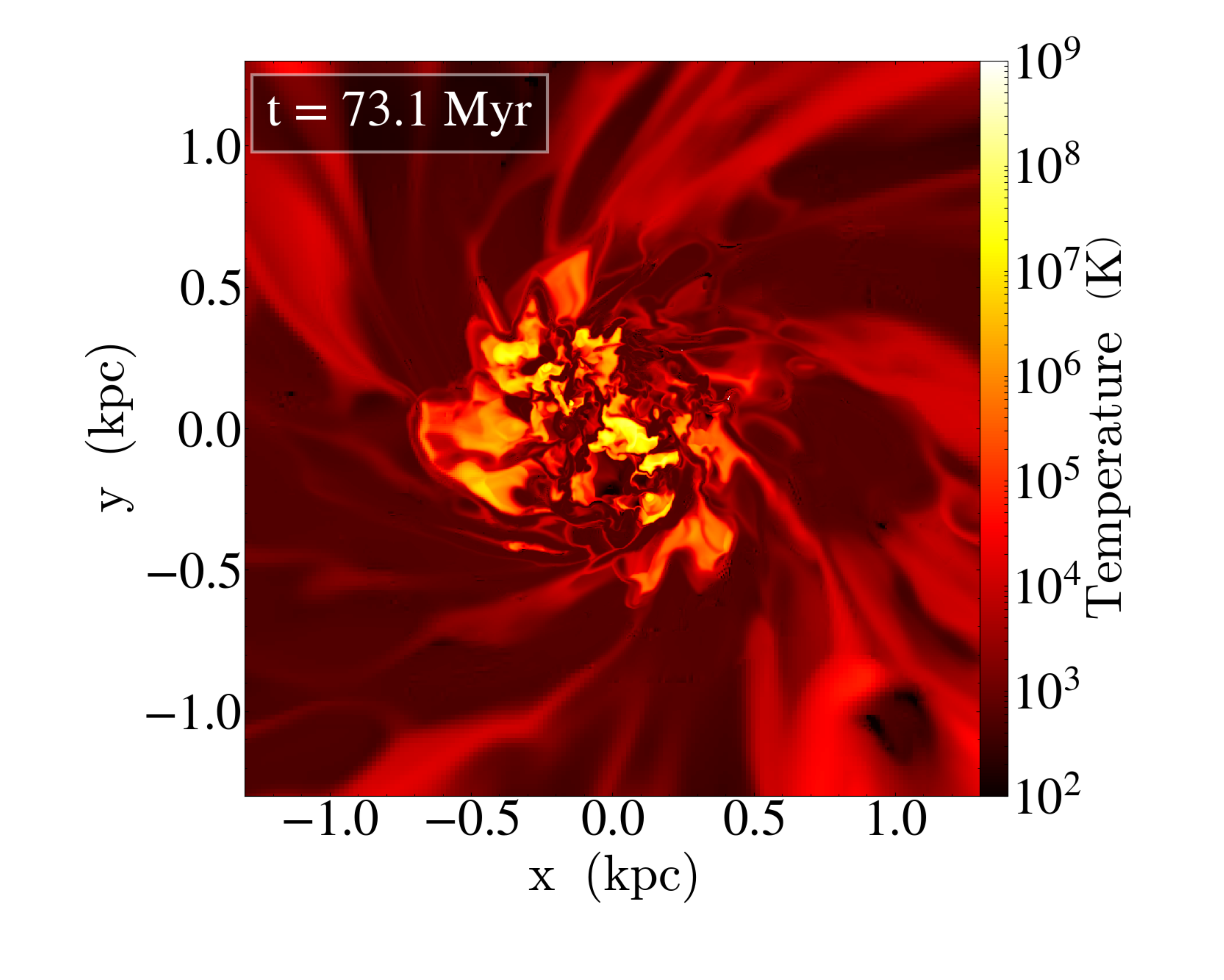}
\end{center}
\caption{Simulation S100\_WSG including self-gravity. From left to right the panels show density and temperature maps at different times taken from razor-thin slices passing through the mid-plane of the disc. Maps are shown edge-on and face-on. The far-left panels show the initial conditions. The mid-left panels show the moment when a violent supernova feedback event drives a strong galactic wind and partially destroys the structure of the disc. The mid-right panels show the time when some of the ejected gas is cooling and raining back towards the disc. The far-right panel is the final stage of the evolution when a highly turbulent disc forms. }\label{fig:maps_sg}
\end{figure*}

\begin{figure*}
\begin{center}
 \includegraphics[width=0.245\textwidth]{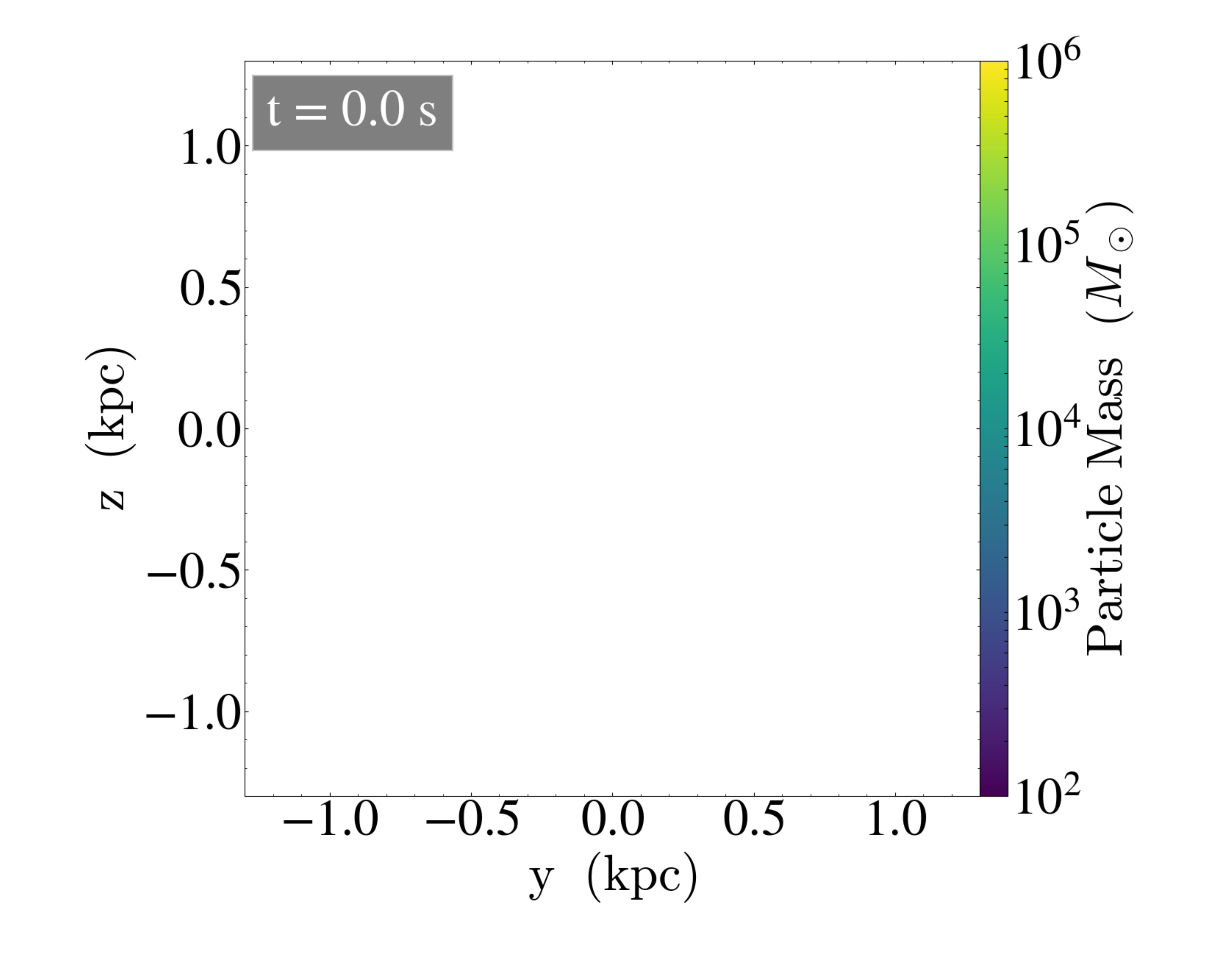}
 \includegraphics[width=0.245\textwidth]{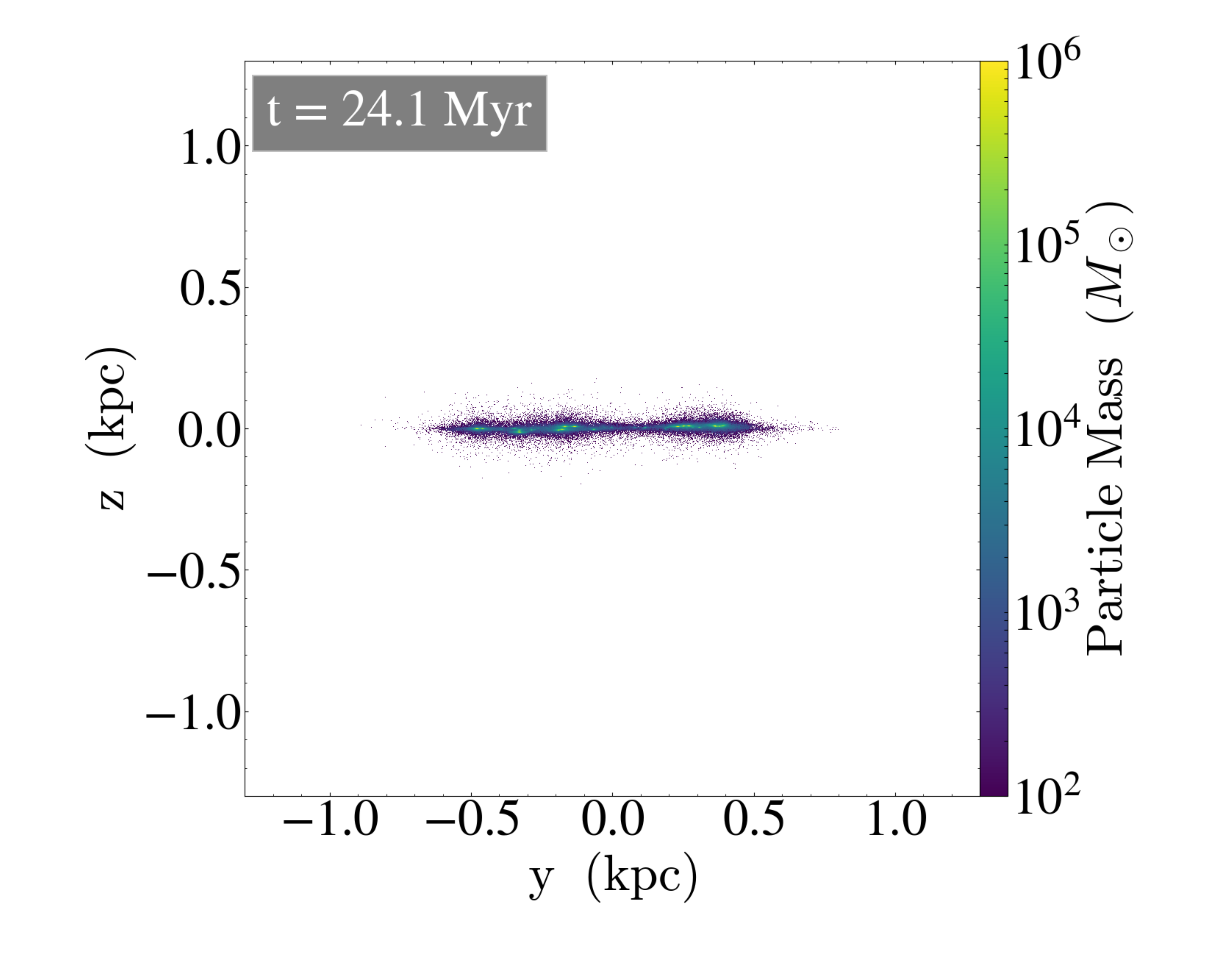}
 \includegraphics[width=0.245\textwidth]{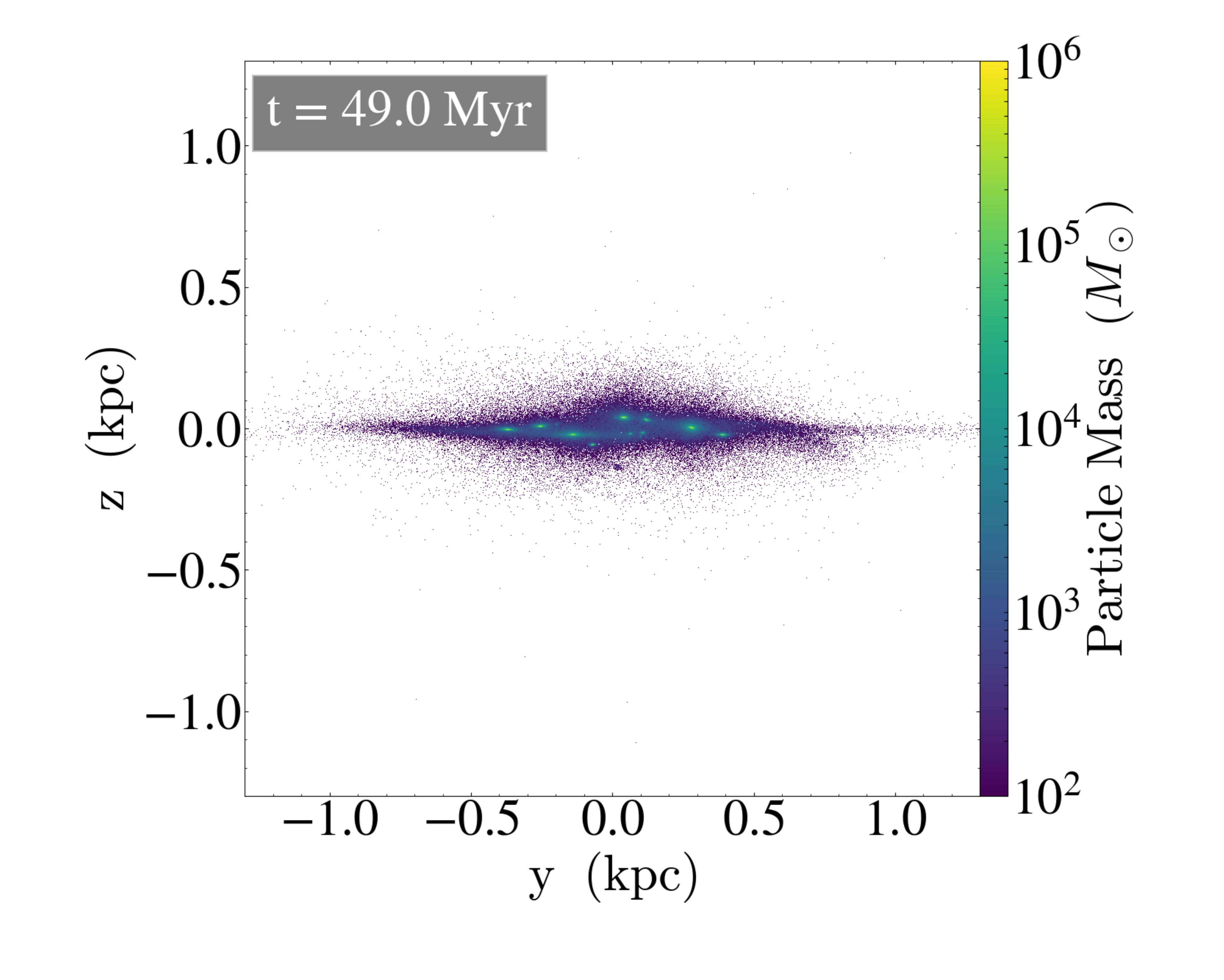}
 \includegraphics[width=0.245\textwidth]{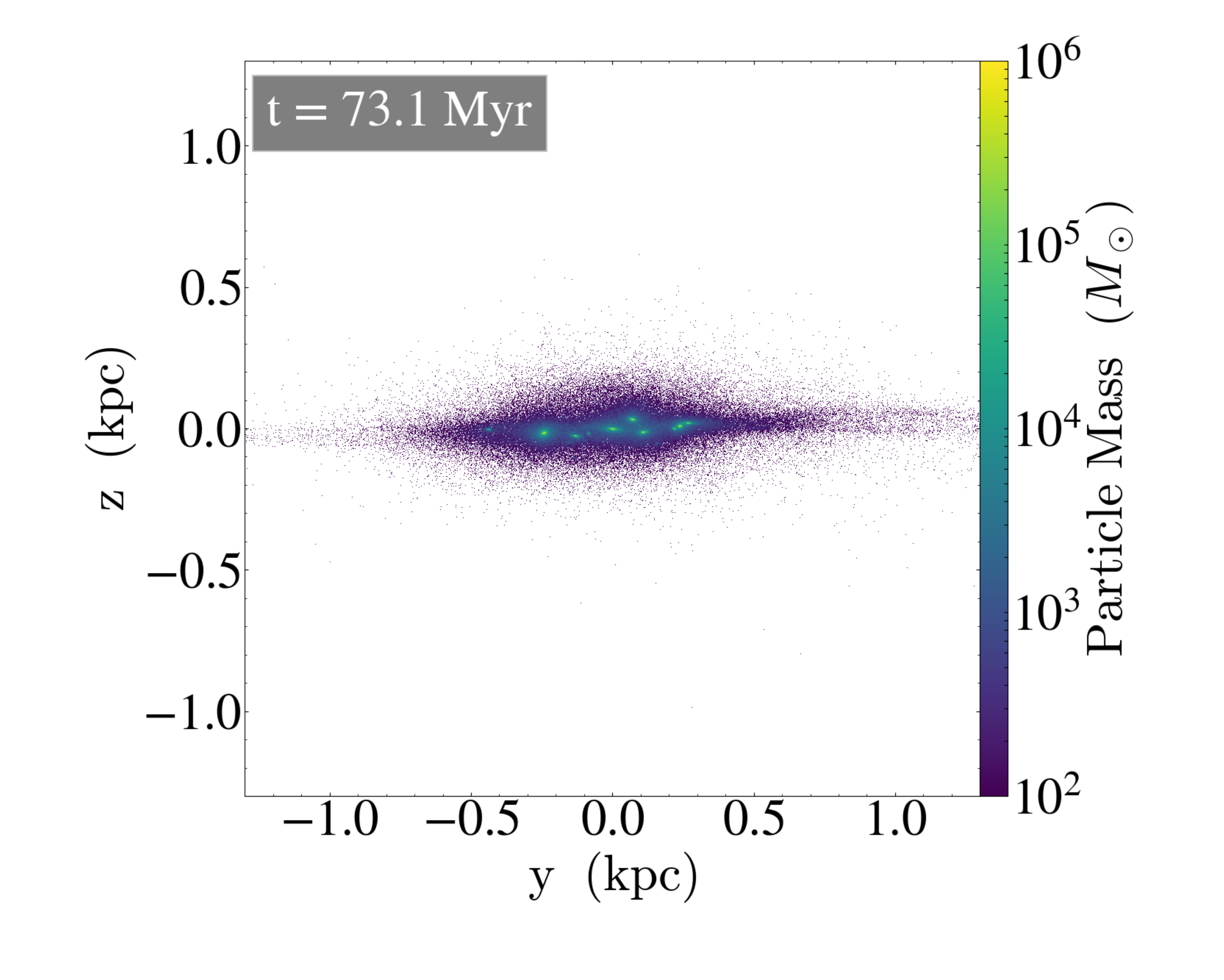}
 \includegraphics[width=0.245\textwidth]{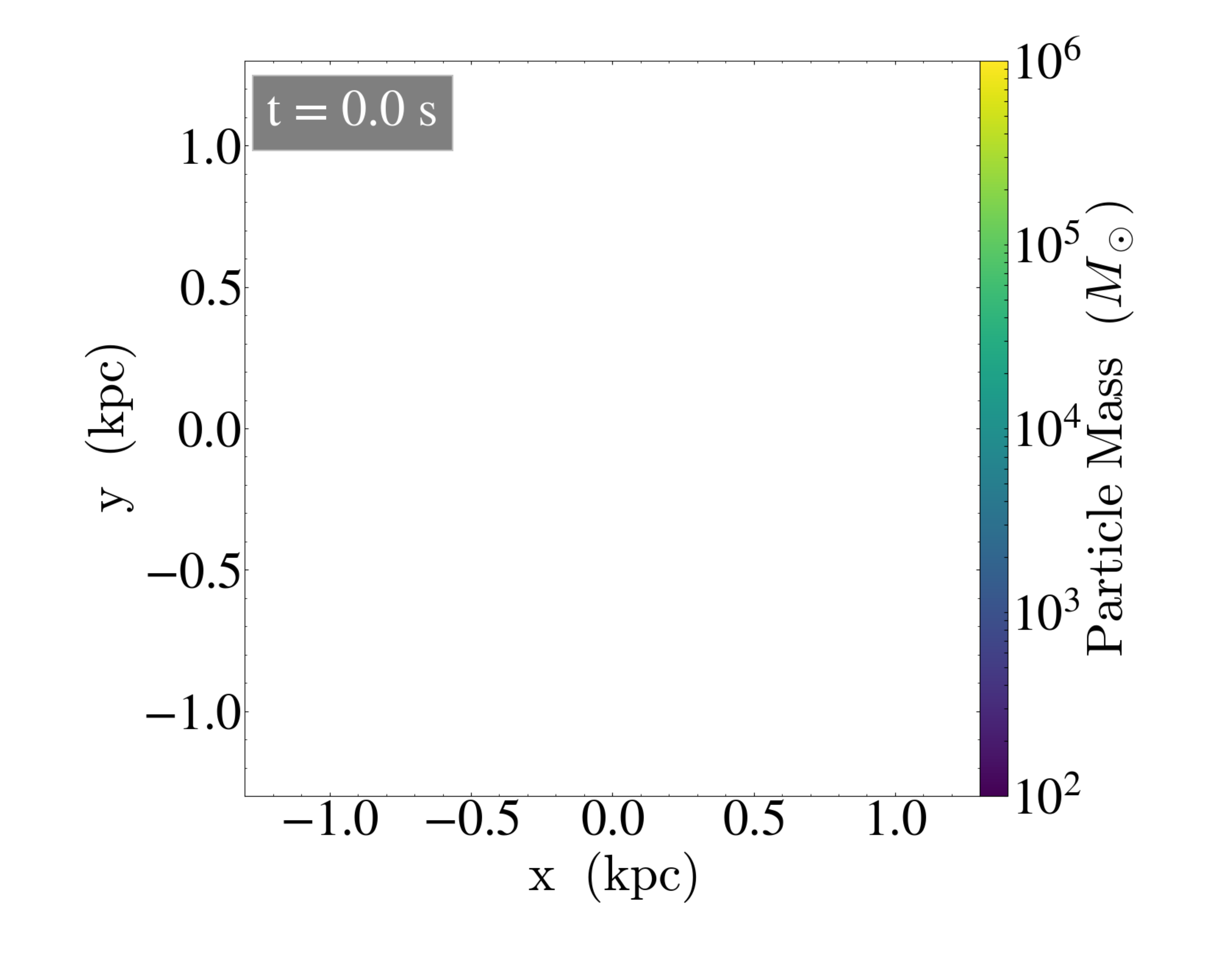}
 \includegraphics[width=0.245\textwidth]{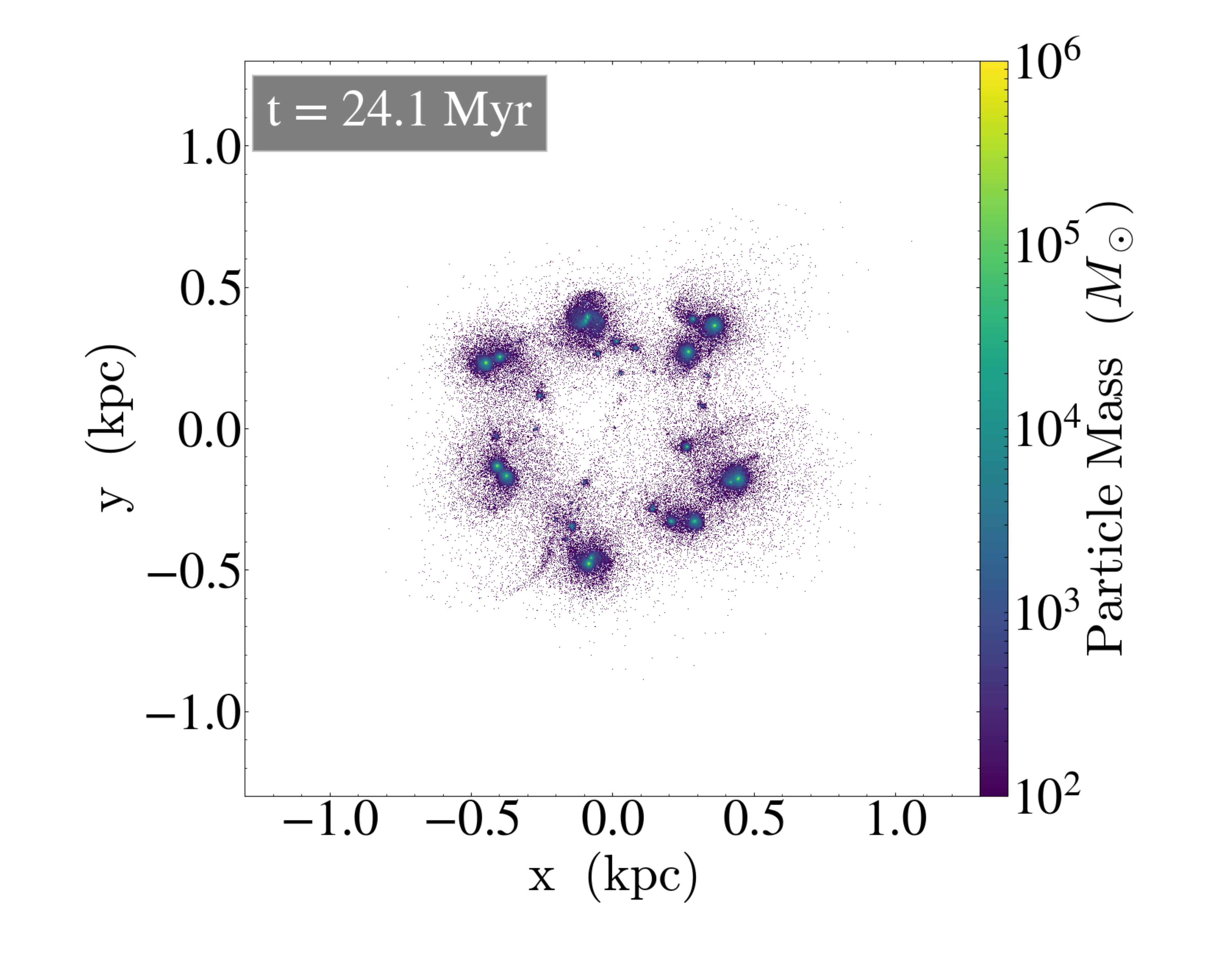}
 \includegraphics[width=0.245\textwidth]{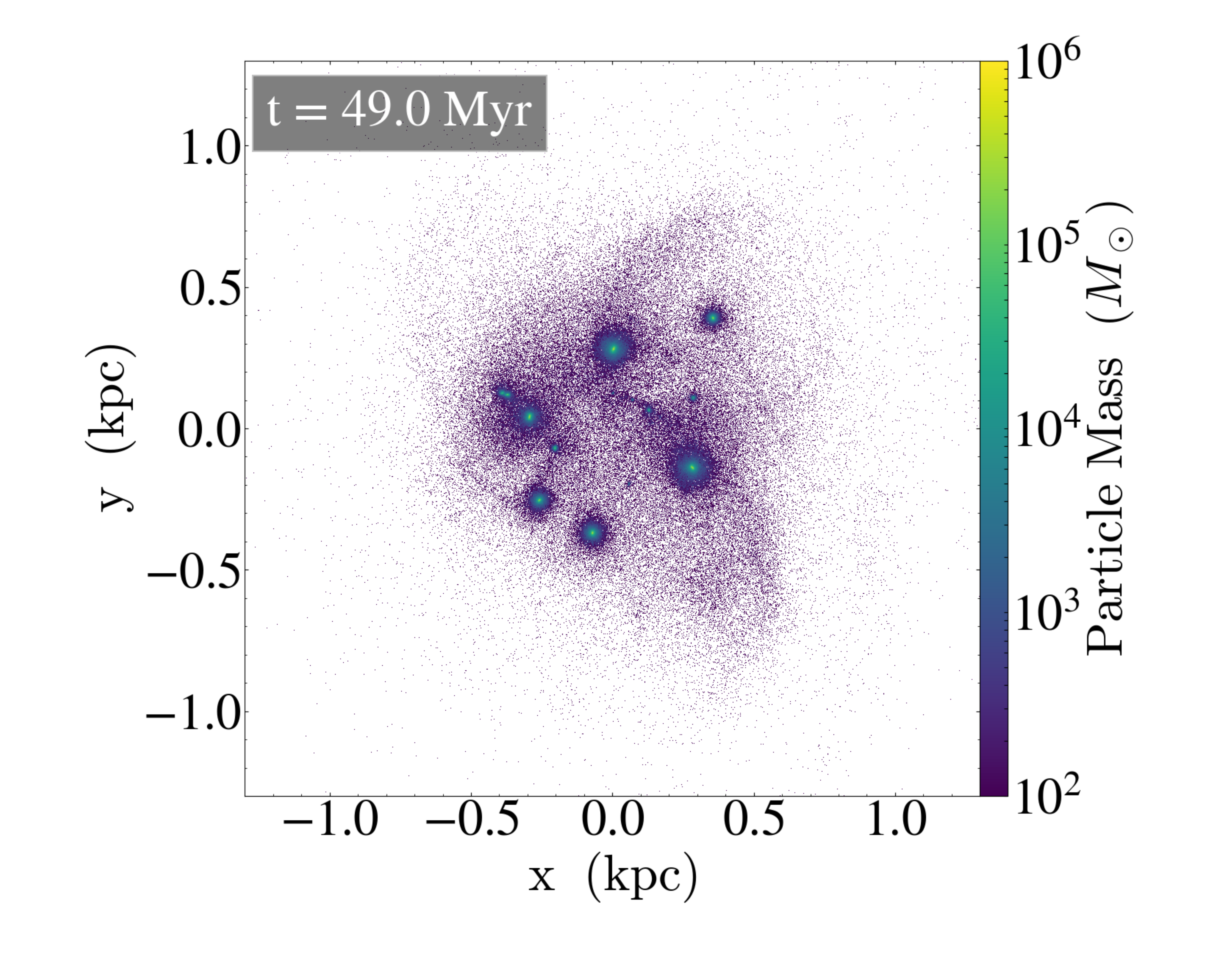}
 \includegraphics[width=0.245\textwidth]{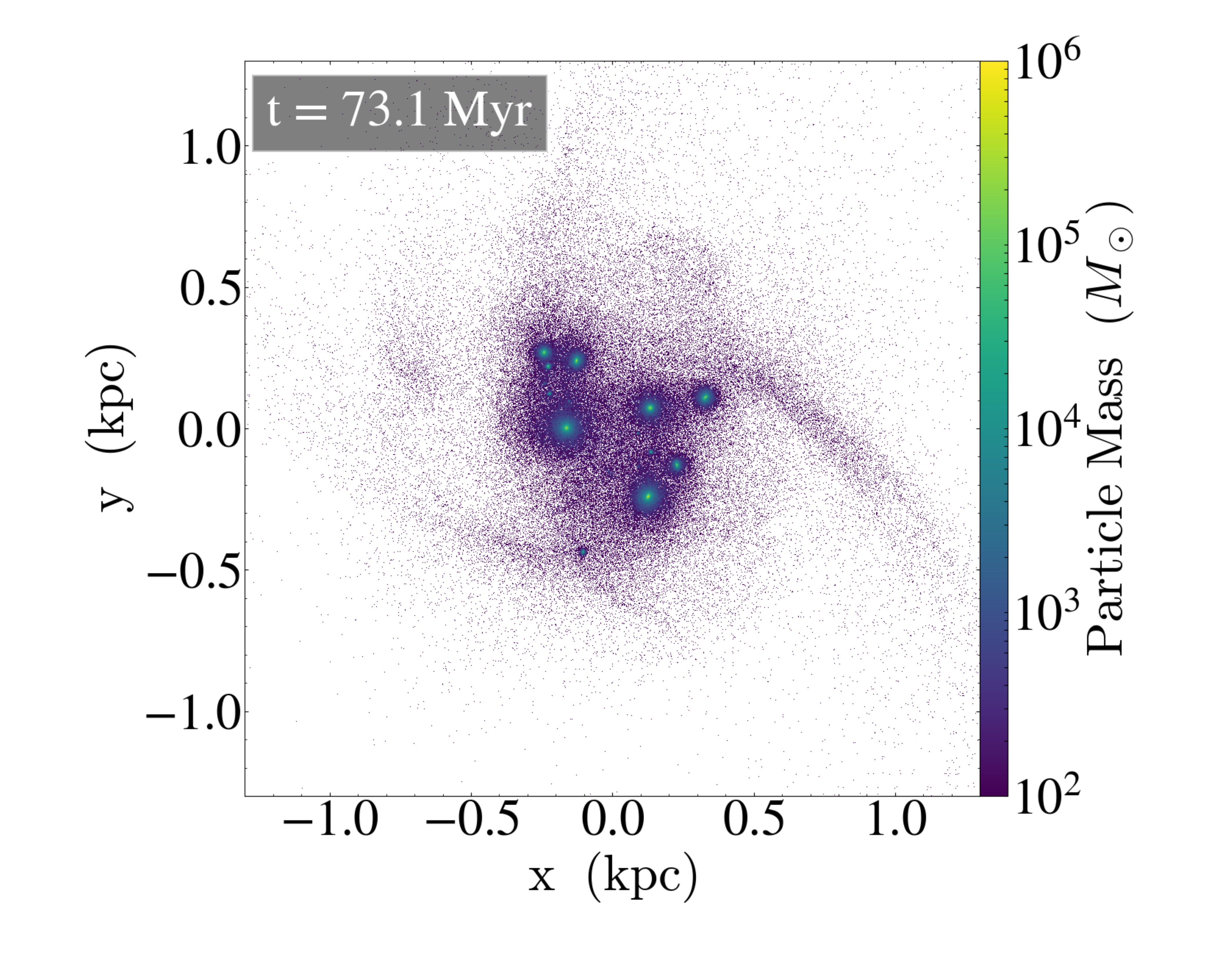}
\end{center}
\caption{Simulation S100\_WSG including self-gravity. From left to right the panels show maps of the stellar particle mass integrated along the line of sight. Only stars formed after the beninning of the simulation are included. The time sequence of Figure~\ref{fig:maps_sg} is shown for consistency. The result of gravitational instability is the formation of star clusters that give rise to clustered supernova feedback. }\label{fig:maps_sg_star}
\end{figure*}

\subsection{Qualitative Description of the Results}\label{sec:qual}

In the NOSG simulations in which self-gravity is not included, the gas responds to radiative cooling and to the external gravitational potential. These simulations behave like a toy model for supernova feedback and are suitable to isolate the response of the gas to a given supernova rate. Since star formation is triggered only when the gas density exceeds a threshold ($n_{\rm H} \geq 100 {\rm cm}^{-3}$), the distribution of stellar particles that can seed supernovae is smoother and confined to the regions where the star-formation density criterion is satisfied by the balance between pressure gradient decrease due to radiative cooling and the external gravitational force. It was found that for the disc with the lowest surface density and no self-gravity (S050\_NOSG), the physical restriction imposed by these choices artificially suppresses star formation in the disc; for this reason simulation S050\_NOSG is excluded from the rest of the analysis, because it does not allow to study supernova feedback. 

In the WSG simulations in which self-gravity is included, gas can become locally self-gravitating, leading to a more patchy and clustered star formation pattern in the disc. This also implies that a high fraction of the supernovae seeded by star particles are highly clustered, modifying the effects and efficiency of supernova feedback. 

Visual inspection of the NOSG and WSG simulations reveals qualitative differences between the two cases that reflect the considerations made above. A comparison between the final states of S500\_NOSG and S500\_WSG is shown in Figure~\ref{fig:maps_comp}. The figure highlights a general trend found in this suite of simulations: supernovae have a smooth spatial distribution in the NOSG simulations, leading to a well-defined disc structure and a quasi-steady configuration characterized by steady galactic wind; supernovae are highly clustered in the WSG simulations, which increases the efficiency of supernova feedback in regions with high density and star formation rate; this leads to violent, transient feedback events that drive more violent and time-varying galactic winds and highly turbulent discs.  

Figure~\ref{fig:maps_sg} shows gas density and temperature maps of the S100\_WSG simulation which includes self-gravity. This figure summarises the typical evolution of the WSG simulations. The initial evolution of the disc is dominated by the joint effect of radiative cooling and gravitational instability. These processes trigger gas fragmentation and compression, followed by a violent burst of star formation. Figure~\ref{fig:maps_sg_star} shows that star formation takes place in clusters of mass up to $\sim 10^6 \, M_{\odot}$ that can seed multiple clustered supernovae. The disc undergoes a violent change as a result of clustered supernova feedback. A strong, transient galactic wind is driven that heats and ejects a significant fraction of the disc gas. After this violent feedback event, part of the gas is irreversibly ejected (crossing the box boundary), part of the gas cools and rains back down towards the disc mid-plane, and the system settles into a highly turbulent disc configuration that drives a weaker, time-variable galactic wind. 

\subsection{Clustered Star Formation}\label{sec:clustered_sf}

\begin{figure}
\begin{center}
 \includegraphics[width=0.485\textwidth]{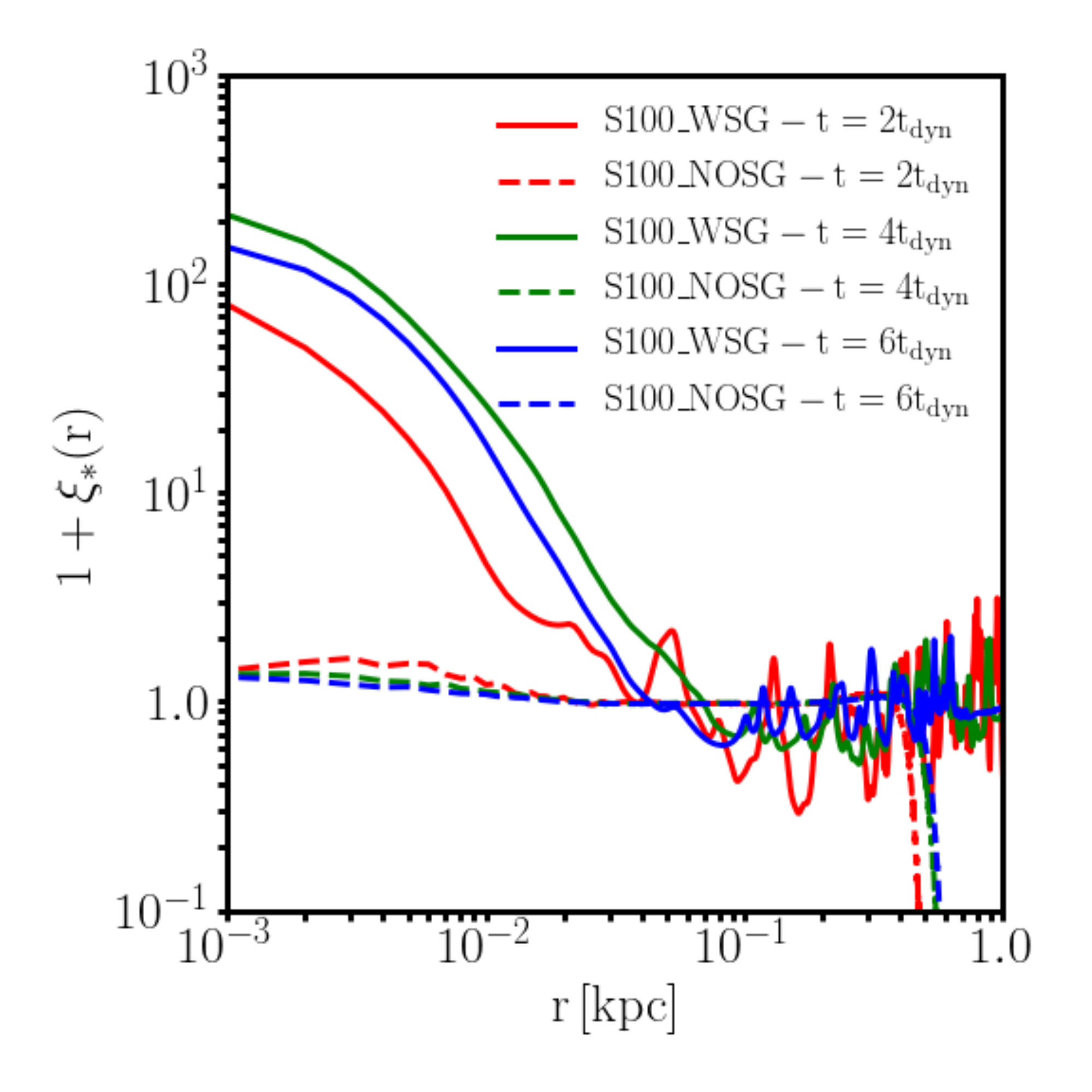}
\end{center}
\caption{ {The two-point correlation function of stellar particles $\xi_*(r)$ for the S100\_NOSG (dashed lines) and S100\_WSG (solid lines) simulations at multiple times. The simulation with gas self-gravity S100\_WSG exhibits highly clustered star formation and time variability, unlike the simulation without gas self-gravity S100\_NOSG.}  }\label{fig:two_point_stars}
\end{figure}

{Although visual inspection of the evolution of the NOSG and WSG suggests that star formation happens differently in the two scenarios, leading to different spatial distributions of stars and supernovae, a more quantitative analysis can demonstrate this effect explicitly. An effective way to quantify the clustering strength of star formation (and indirectly of supernova feedback) in each case is to compute the two-point correlation function $\xi_*(r)$ of the stellar particles, as shown in Figure~\ref{fig:two_point_stars}. The two-point correlation function of the NOSG runs exhibits weak clustering ($1+\xi_*(r)\approx 1$) at all radii $r$, and very weak time variability. On the other hand, the WSG runs have highly clustered star formation ($1+\xi_*(r)>2$) at radii $r<20 \, {\rm pc}$, as a consequence of the formation of locally self-gravitating regions. The WSG runs also have high time variability, with $1+\xi_*(r)$ varying more than a factor 2 from $t = 2t_{\rm dyn}$ to $t = 6 t_{\rm dyn}$. This is a consequence of the violent supernova feedback arising from clustered supernovae, that leads to the temporary destruction of star-forming clouds and, consequently, to variations in the local gravitational potential. }

\subsection{Bursts of Star Formation and Supernova-driven Galactic Winds}\label{sec:time_evo}

\begin{figure}
\begin{center}
 \includegraphics[width=0.485\textwidth]{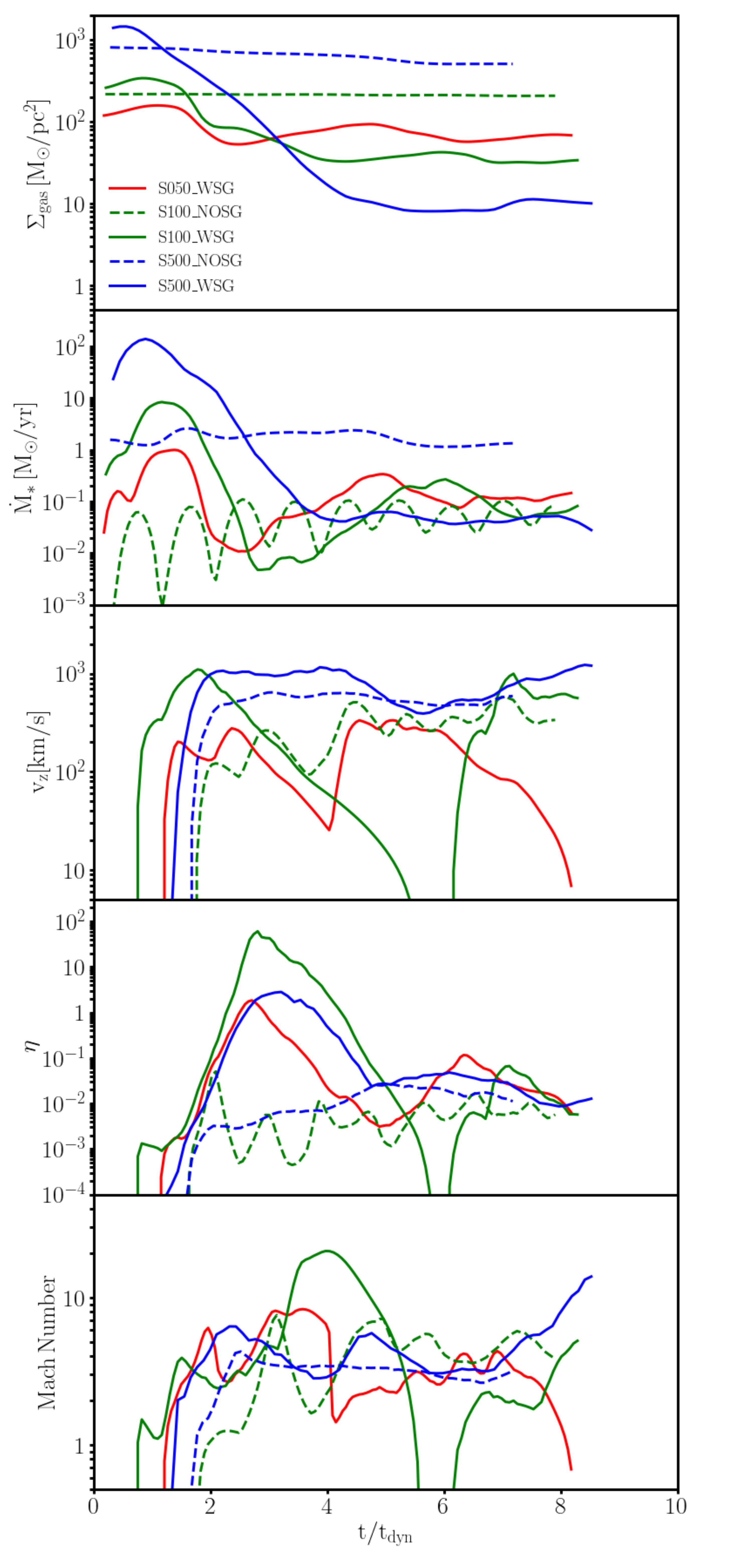}
\end{center}
\caption{Temporal evolution of properties of the system in a cylindrical region of radius ${\rm R = 400 \, pc}$. Time on the x-axis has been normalised to the value of the dynamical time in the disc mid-plane in each simulation, $t_{\rm dyn}$. First panel: gas surface density $\Sigma_{\rm gas}$ within ${\rm |z|<2 \, kpc}$. Second panel: star formation rate in the disc $\dot{M}_*$. Third panel: net velocity of the gas at height ${\rm |z|=2 \, kpc}$ from the disc mid-plane. Fourth panel: mass loading factor of the galactic wind at height ${\rm |z|=2 \, kpc}$ from the disc mid-plane; only outflowing material has been included in the calculation. Fifth panel: Mach number of the galactic outflow. All properties have been time-averaged on a time scale of 5 Myr. SN feedback drives supersonic galactic winds in all cases. Simulations without self-gravity (NOSG) reach a quasi-steady (or weakly oscillatory) solution for the star formation rate and the properties of the galactic winds. Simulations with self-gravity (WSG) show much higher time variability: an initial phase characterised by a strong star-formation burst that drives a highly mass-loaded galactic wind ($\eta>1$) is typically followed by a quasi-steady star-forming disc solution with weaker galactic winds ($\eta \sim 10^{-2}$). }\label{fig:galwind_evo}
\end{figure}

Figure~\ref{fig:galwind_evo} shows the temporal evolution of several quantities that characterise the state of the simulated galactic discs. All quantities in this figure have been time-averaged on a time scale of 5 Myr. The first panel of this figure shows the temporal evolution of the surface density of the gas measured in a cylindrical region defined by radius $R \leq 400 \, {\rm pc}$ and height $|z|\leq 2 \, {\rm kpc}$, $\Sigma_{\rm gas}$. The second panel show the star formation rate in the same region, $\dot{M}_*$. This Figure demonstrates how the NOSG simulations without self-gravity settle into a disc configuration with approximately constant gas surface density and star formation rate, whereas the WSG simulations with self-gravity are characterized by a violent burst of star formation followed by a quasi-steady configuration, with weak time-variability. The third panel of Figure~\ref{fig:galwind_evo} shows $v_{\rm z}$, the average $z$-component of the velocity within radius $R\leq 400 \, {\rm pc}$ and at height $z=2 \, {\rm kpc}$ from the mid-plane of the disc; this panel demonstrates that all simulations develop galactic winds with typical speeds $10^2 \, {\rm km/s} \lesssim v_{\rm z} \lesssim 10^3 \, {\rm km/s}$. The NOSG simulations develop a wind with approximately constant speed, whereas the WSG simulations exhibit high time variability. The same behavior is observed in the fourth panel of Figure~\ref{fig:galwind_evo}, which shows the mass-loading factor of the winds $\eta=\dot{M}_{\rm gas,out}/\dot{M}_* $, where $\dot{M}_{\rm gas,out}$ is the mass outflow rate at height $z=2 \, {\rm kpc}$ from the mid-plane of the disc. The NOSG simulations drive galactic winds with low mass loading factor $10^{-2}\lesssim \eta \lesssim 10^{-1}$, which is also the case for the WSG simulations in the late phase of their evolution. However, the WSG simulations drive winds with high mass loading factor $1 \lesssim \eta \lesssim 10^2$ during the star formation burst driven by gravitational instability in the disc at time $t\approx3t_{\rm dyn}$. The fifth panel of Figure~\ref{fig:galwind_evo} shows the Mach number of the galactic winds at height $z=2 \, {\rm kpc}$ and demonstrates that they are supersonic in all cases, in agreement with previous idealised global simulations of galactic discs \citep{2017MNRAS.470L..39F}. 

\subsection{Vertical Structure of the Discs}\label{sec:vert_struct}

\begin{figure*}
\begin{center}
 \includegraphics[width=0.49\textwidth]{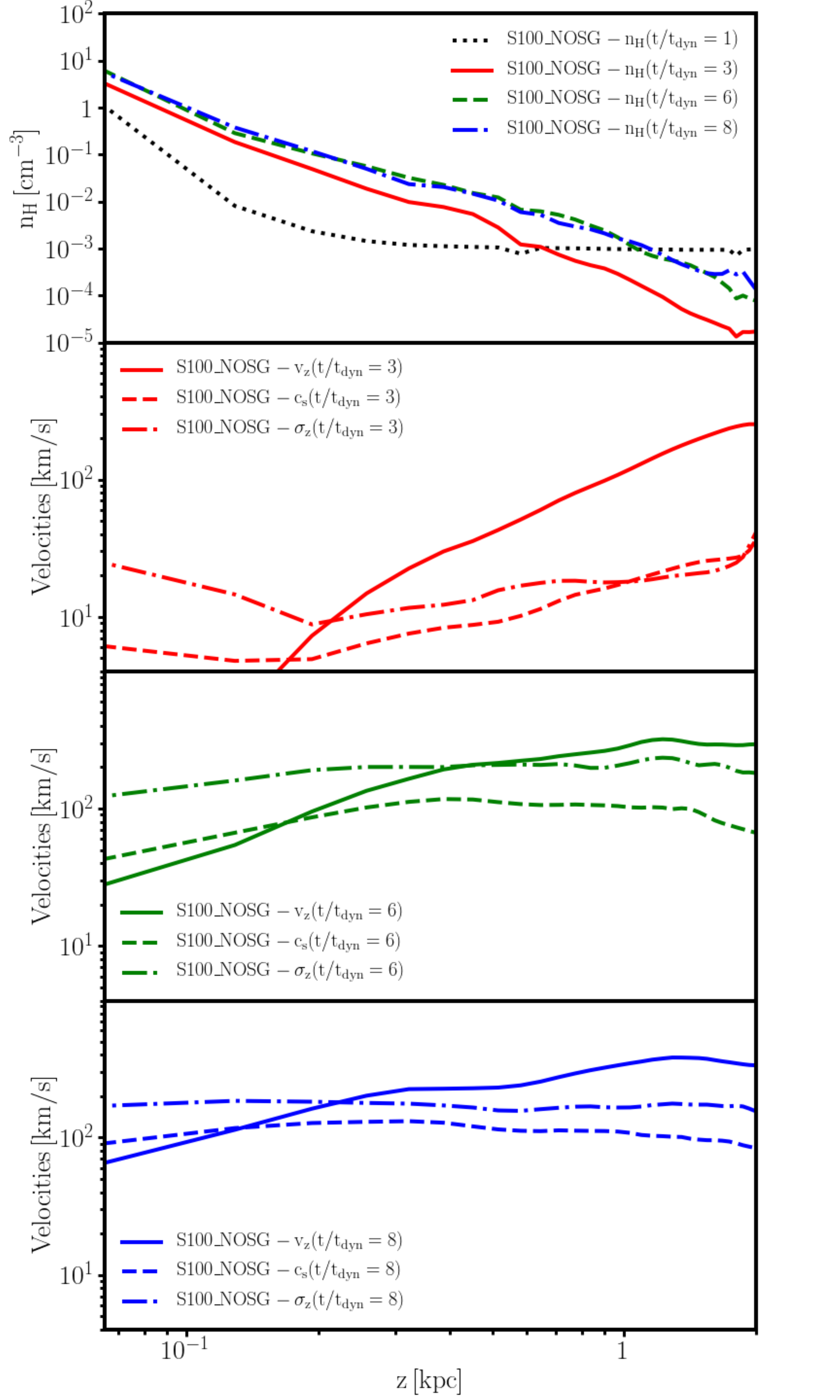}
 \includegraphics[width=0.49\textwidth]{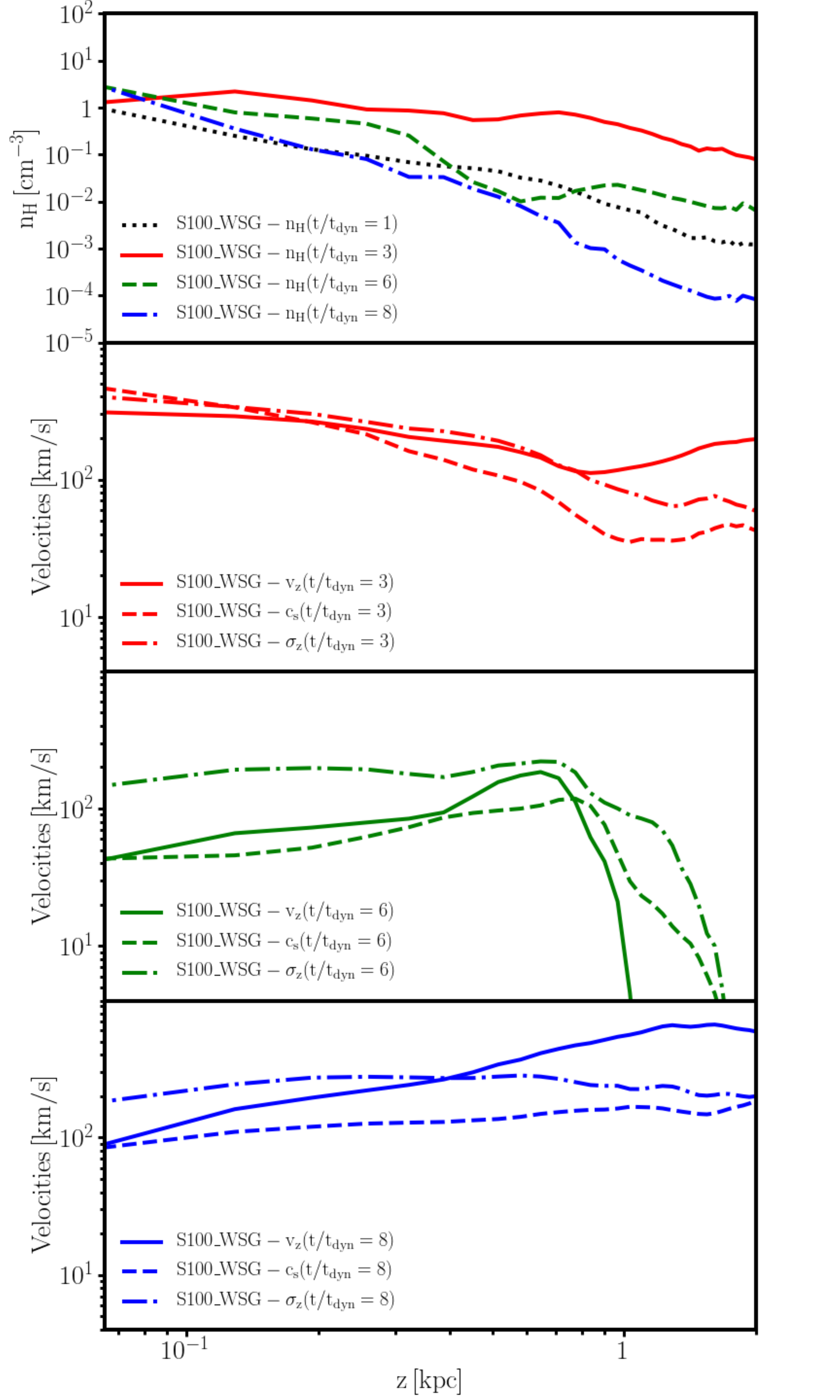}
\end{center}
\caption{Vertical profiles of of the hydrogen number density $n_{\rm H}$, the sound speed $c_{\rm s}$, the $z$-component of the gas velocity dispersion $\sigma_{\rm z}$, and the $z$-component of the gas velocity $v_{\rm z}$. All quantities are measures within cylindrical radius $R<400 \, {\rm pc}$, volume averaged in vertical bins of height $\Delta z = 64 \, {\rm pc}$, and time averaged over a time-scale of 5 Myr. The left panels show the S100\_NOSG simulation (no self-gravity), whereas the right panels show the S100\_WSG simulation (with self-gravity). First row: $n_{\rm H}$ at different times. Second row: $v_{\rm z},\sigma_{\rm z},c_{\rm s}$ at time $t=3t_{\rm dyn}$. Third row: $v_{\rm z},\sigma_{\rm z},c_{\rm s}$ at time $t=6t_{\rm dyn}$. Fourth row: $v_{\rm z},\sigma_{\rm z},c_{\rm s}$ at time $t=8t_{\rm dyn}$. Simulations without self-gravity settle into a quasi-steady state, whereas simulations with gas-self gravity have higher variability under the effect of gravitational instability triggering violent supernova feedback events. }\label{fig:vert_struct}
\end{figure*}

Important details of the response of galactic discs to supernova feedback can be appreciated by examining the vertical structure in the simulations presented in this paper. Figure~\ref{fig:vert_struct} shows an analysis of the vertical profiles of the hydrogen number density $n_{\rm H}$, the sound speed $c_{\rm s}$, the $z$-component of the gas velocity dispersion $\sigma_{\rm z}$, and the $z$-component of the gas velocity $v_{\rm z}$. The profiles are taken from a disc region defined by cylindrical radius $R\leq 400 \, {\rm pc}$, and time averaged over a time-scale of 5 Myr, whereas the quantities are volume averaged within vertical bins of size $\Delta z = 64 \, {\rm pc}$, .

The top panels of Figure~\ref{fig:vert_struct} show the density profile of the S100\_NOSG (left) and S100\_WSG (right) simulations at different time. In both simulations, the disc structure evolves significantly from the initial configuration, as a response of kinetic and thermal energy injection by supernovae. The disc thickens under the effect of thermal and turbulent pressure generated by supernova feedback, and the regions above the disc are filled with gas with higher density than before, as an effect of the galactic winds. In the case without self-gravity, S100\_NOSG, the system settles to a quasi-steady configuration at time $t>3t_{\rm dyn}$. This not observed in the case with self-gravity, S100\_WSG, which exhibits higher time variability. 

The second, third and fourth rows of Figure~\ref{fig:vert_struct} show the characteristic velocities of the system ($v_{\rm z},\sigma_{\rm z},c_{\rm s}$) at times $t=3t_{\rm dyn}$, $t=6t_{\rm dyn}$, and $t=8t_{\rm dyn}$, respectively. At time $t=3t_{\rm dyn}$, the mass loading factor of the galactic wind reaches its peak in the WSG simulations (see Figure~\ref{fig:galwind_evo}). At time $t=6t_{\rm dyn}$, the galactic wind of the S100\_WSG  simulation has stopped receiving enough input from supernovae to maintain its peak power and turns into a galactic fountain raining back towards the mid-plane of the disc. At time $t=8t_{\rm dyn}$ in the S100\_WSG  simulation, the disc has settled into another configuration that launches a galactic wind. The left panels of Figure~\ref{fig:vert_struct} demonstrate that the S100\_NOSG simulation settles into a quasi-steady configuration with a stable vertical structure at $t>3t_{\rm dyn}$, after an initial transient. The $z$-component of the velocity dispersion $\sigma_{\rm z}$ appears to be larger than the sound speed $c_{\rm s}$ at all heights $z$, but their values are comparable. This fact implies that the thermal pressure is always comparable to, but weaker than the turbulent pressure in these discs. The galactic wind speed $v_{\rm z}$ becomes larger than both the sound speed and the velocity dispersion at $z\gtrsim 100 \, {\rm pc}$, implying that this is the region where the galactic wind has its largest dynamical influence. The right panels of Figure~\ref{fig:vert_struct} show the vertical profiles of the same quantities for the S100\_WSG simulation, in which similar conclusions can be drawn, although more time variability is observed.

Finally, although the results for the S050 and S500 simulations are not shown in this Subsection, the temporal evolution of the vertical profiles in these simulated discs is qualitative similar to that of the S100 simulations. 

\subsection{Maximum Mass Loading Factor from Clustered Supernova Feedback}\label{sec:eta_peak}

It is possible to estimate the mass loading factor of a galactic wind from analytical arguments under some simplifying assumptions. In this Subsection, this prediction will be compared to the results of the suite of simulations. 

Let us assume that galactic winds are driven by supernova remnants whose cooling radius $R_{\rm cool}$ is comparable to the scale height of the galactic disc \citep[e.g.][]{2017MNRAS.470L..39F}. This condition is necessary to prevent radiative losses in the gas in the supernova remnant bubbles, so that they can expand until they blow material out of the galactic disc. In quantitative terms, this condition on the cooling radius is: 
\begin{equation}\label{eq:rcool_condition}
    R_{\rm cool}(n_{\rm H},Z) \geq h,
\end{equation}
where the dependence of the supernova remnant cooling radius on the ISM density $n_{\rm H}$ and metallicity $Z$ has been explicitly introduced. Let us assume that the cooling radius follows the scaling of \cite{2015MNRAS.450..504M}, which was calibrated on numerical simulations of isolated supernova remnants:
\begin{equation}\label{eq:martizzi15}
    R_{\rm cool}(n_{\rm H},Z) = 3.0 \, {\rm pc}\left(\frac{Z}{Z_{\odot}} \right)^{-0.082}\left(\frac{n_{\rm H}}{100 \, {\rm cm}^{-3}} \right)^{-0.42},
\end{equation}
where $n_{\rm H}$ is the local ISM hydrogen number density and $Z$ is the local ISM metallicity. Putting together equations~\ref{eq:rcool_condition} and \ref{eq:martizzi15}, it is possible to verify that a galactic wind will be driven from regions where:
\begin{equation}\label{eq:nHlim}
    n_{\rm H} \leq n_{\rm H, lim} = 100 \, {\rm cm}^{-3} \left( \frac{h}{3.0 \, {\rm pc}}\right)^{-2.38}\left(\frac{Z}{Z_{\odot}}\right)^{-0.195}
\end{equation}

\begin{figure*}
\begin{center}
 \includegraphics[width=0.99\textwidth]{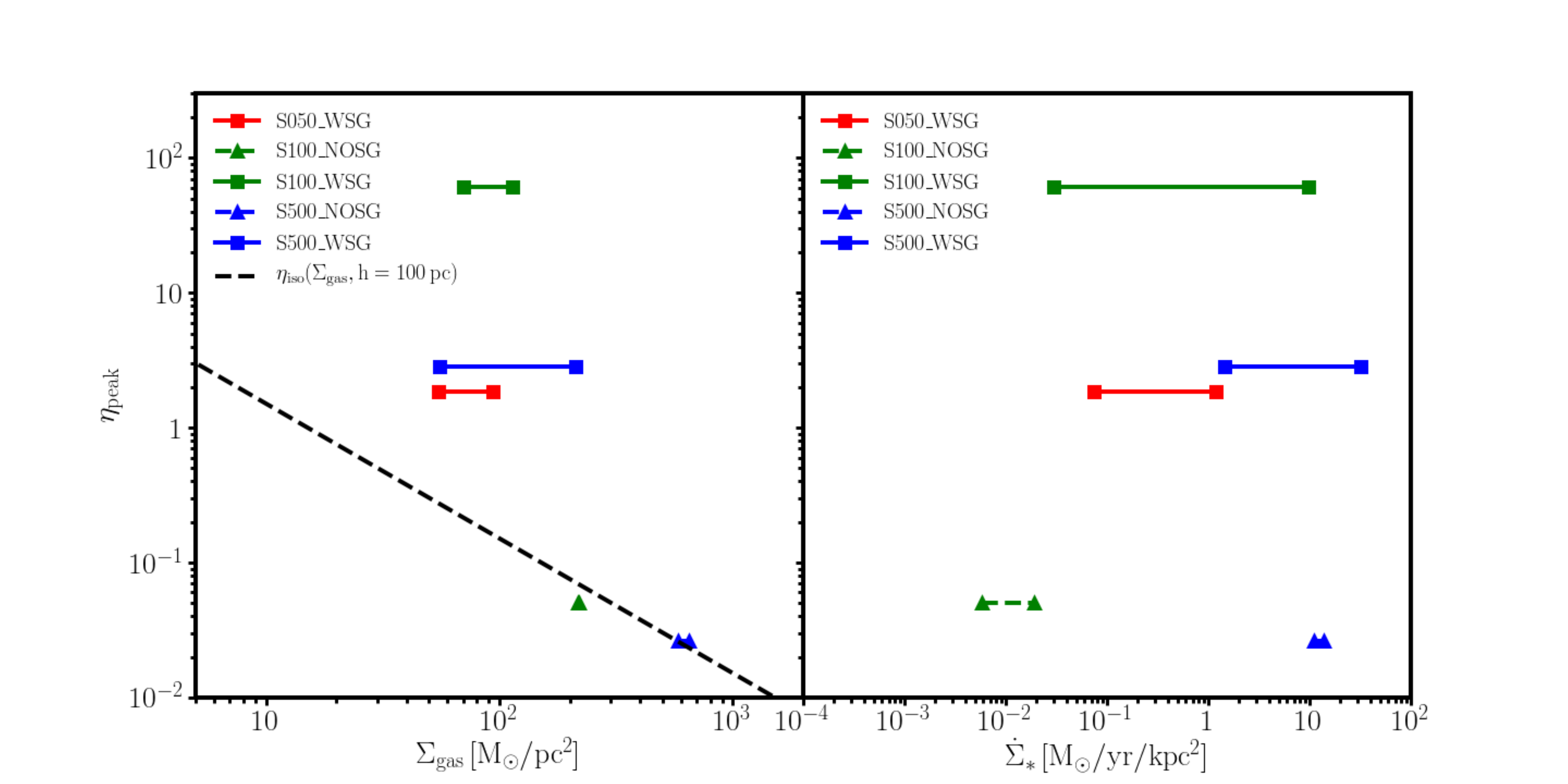}
\end{center}
\caption{The maximum value of the galactic wind mass loading factor at height ${\rm |z|=2 \, kpc}$ as a function of gas surface density $\Sigma_{\rm gas}$ (left panel) and star formation rate density $\dot{\Sigma}_*$. Each simulation is plotted as a bar, indicating the variation of $\Sigma_{\rm gas}$ and $\dot{\Sigma}_*$ from $t_{\rm peak}-t_{\rm dyn}$ to $t_{\rm peak}$, where $t_{\rm peak}$ is the time at which the mass loading factor peaks and $t_{\rm dyn}$ is the dynamical time in the disc mid-plane. Simulations without self-gravity (NOSG, triangles) develop galactic winds with mass loading factor  $10^{-2}<\eta<10^{-1}$. The mass loading factors of these winds follow very closely the analytical prediction expected for galactic winds driven by isolated SNRs, $\eta_{\rm iso}$ (see Subsection~\ref{sec:eta_peak}). The peak mass loading factor in simulations with self-gravity is much higher than this prediction, as a consequence of the effects of non-linear supernova clustering. In particular, the mass loading factor is not a monotonic function of neither $\Sigma_{\rm gas}$ nor $\dot{\Sigma}_*$.}\label{fig:eta_peak}
\end{figure*}

Let us further assume that vertical structure of a galactic disc can be approximated by an exponential distribution:
\begin{equation}
 n_{\rm H}(z) = n_{\rm H,0}\exp\left(-\frac{|z|}{h}\right),
\end{equation}
where $z$ is the height above the disc mid-plane, $n_{\rm H,0}$ is the mid-plane density, $h = 100 \, {\rm pc}$ (approximately the gas scale-height of the simulated discs). By using this assumption on the vertical structure, the condition of equation~\ref{eq:nHlim} can be expressed as a condition on the distance from the mid-plane $z$:
\begin{equation}\label{eq:zlim}
    |z|\geq z_{\rm lim} = h\log\left(\frac{n_{\rm H,0}}{n_{\rm H,lim}}\right).
\end{equation}

Finally, the mass loading factor $\eta_{\rm iso}$ can be estimated as the ratio between the mass in the regions where the condition in equation~\ref{eq:zlim} is satisfied and the total gas mass that is converted in stars:
\begin{equation}\label{eq:eta_iso}
    \eta_{\rm iso} \approx \frac{\Sigma_{\rm wind}}{\epsilon\Sigma_{\rm gas}}=\frac{\int_{z_{\rm lim}}^{\infty}n_{\rm H,0}\exp\left(-\frac{z}{h}\right)dz}{\epsilon \int_{0}^{\infty}n_{\rm H,0}\exp\left(-\frac{z}{h}\right)dz}=\frac{1}{\epsilon}\exp\left(-\frac{z_{\rm lim}}{h}\right),
\end{equation}
where $\epsilon$ is the star formation efficiency. 

The expression for $\eta_{\rm iso}$ in equation~\ref{eq:eta_iso} represents an approximation of the mass loading factor expected for a galactic wind driven by non clustered supernovae. Figure~\ref{fig:eta_peak} shows a comparison of this analytical model to the results of found in the simulated galaxies of this paper. $\eta_{\rm iso}$ is compared to the peak mass loading factor measured throughout the whole simulated history, $\eta_{\rm peak}$. Figure~\ref{fig:eta_peak} shows that for a scale height $h=100 \, {\rm pc}$ the analytical model for $\eta_{\rm iso}$ is in excellent agreement with the simulations without self-gravity (NOSG). In practice, this can be interpreted as a consequence of the fact that the supernova spatial distribution is smooth in the NOSG simulations, leading to rare clustering of supernovae. On the other hand, the simulations with self-gravity (WSG) naturally produce stronger feedback caused by supernova clustering. For this reason, the peak mass loading factor $\eta_{\rm peak}$ in the WSG simulations can be  orders of magnitude larger than $\eta_{\rm iso}$. 

The horizontal bars in Figure~\ref{fig:eta_peak} show the variation of gas surface density (left panel) and star formation rate density (right panel) from time $t_{\rm dyn}-t_{\rm peak}$ to time $t_{\rm peak}$, where $t_{\rm peak}$ is the time at which the mass loading factor peaks and $t_{\rm dyn}$ is the dynamical time in the disc mid-plane. At time $t_{\rm peak}$ the gas surface density and star formation rate density are at peak value and they decrease steadily over the following dynamical time. These variations appear to be up to $\sim 0.5$ dex for the case with self-gravity and are quite modest for the case without self-gravity. 

In summary, the effect of supernova clustering produces winds that are much more time varying and violent compared to naive expectations based on the physics of individual supernova remnants. Supernova feedback is a highly non-linear process. 

\subsection{Mass Fluxes in Galactic Winds, Fountains and their Phase Structure}\label{sec:mass_flux}

\begin{figure*}
\begin{center}
 \includegraphics[width=0.49\textwidth]{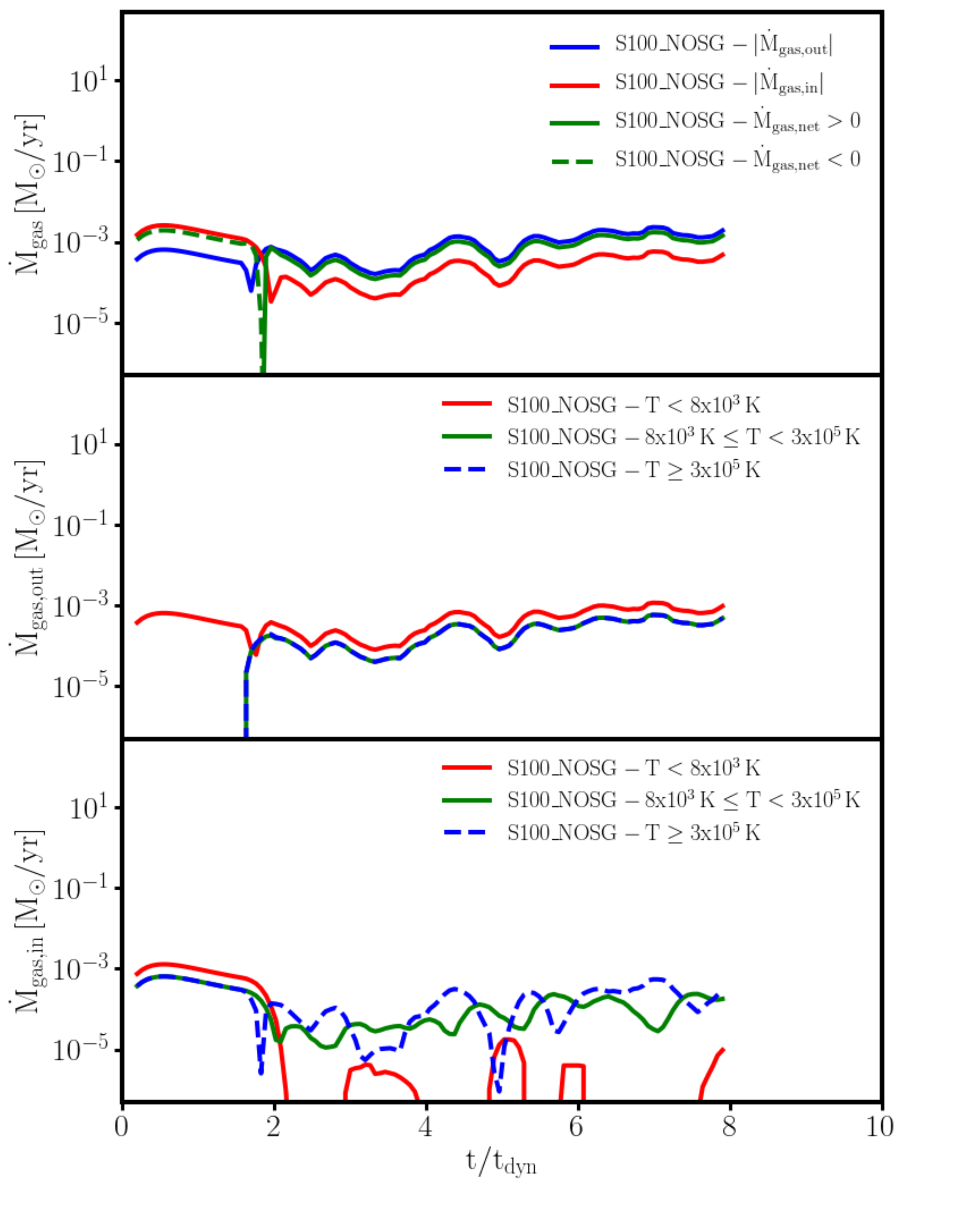}
 \includegraphics[width=0.49\textwidth]{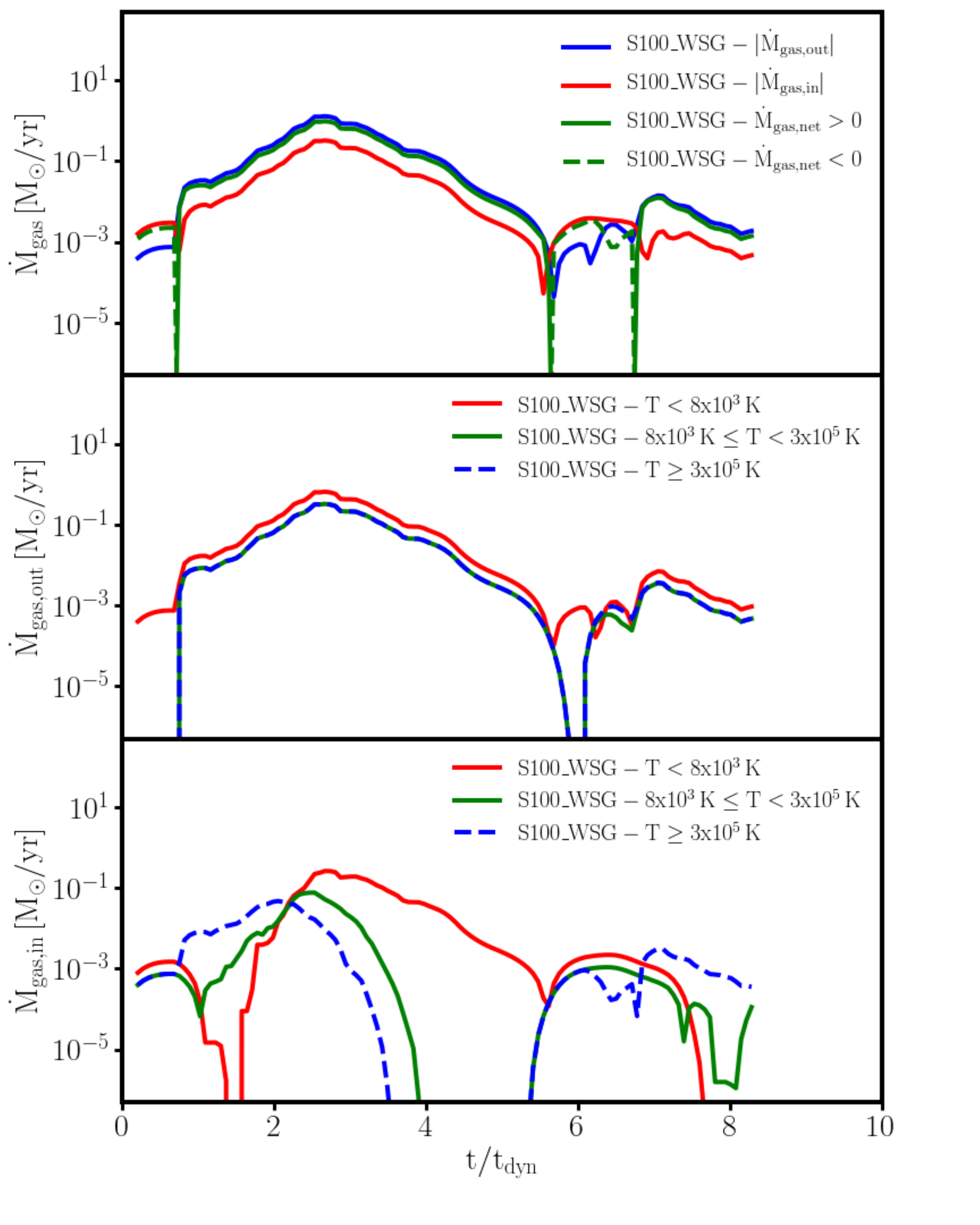}
\end{center}
\caption{Inflowing ($\dot{M}_{\rm gas,in}$), outflowing ($\dot{M}_{\rm gas,out}$) and net ($\dot{M}_{\rm gas,net}=\dot{M}_{\rm gas,out}-\dot{M}_{\rm gas,in}$) mass fluxes plotted as a function of time. The mass fluxes were measured at height $|z|=2 \, {\rm kpc}$ from the disc mid-plane and within a cylindrical radius $R<400 \, {\rm pc}$. The left panels show the results of the S100\_NOSG simulation (no self-gravity), whereas the right panels show the results of the S100\_WSG simulation (with self-gravity). First row: total fluxes $\dot{M}_{\rm gas,in}$, $\dot{M}_{\rm gas,out}$ and $\dot{M}_{\rm gas,net}$. The net mass flux is plotted as a dashed line when it is negative (net inflow) and as a solid line when it is positive (net outflow). Second row: $\dot{M}_{\rm gas,out}$ for gas in multiple temperature ranges. Third row: $\dot{M}_{\rm gas,in}$ for gas in multiple temperature ranges. }\label{fig:mdot_evo}
\end{figure*}

In order to complete the analysis of the evolution of the simulated galactic discs, the gas inflow and outflow rates measured at height $|z|=2 \, {\rm kpc}$ from the disc mid-plane were examined. The results are reported in Figure~\ref{fig:mdot_evo}. The top panels of this figure show the gas outflow rate $\dot{M}_{\rm gas,out}$, the gas inflow rate $\dot{M}_{\rm gas, in}$, and the net outflow rate $\dot{M}_{\rm gas,net}=\dot{M}_{\rm gas,out}-\dot{M}_{\rm gas, in}$ as a function of time. The S100\_NOSG (no self-gravity, left panels) experiences an initial net inflow phase caused by the combination of gravity and radiative cooling. When the galactic wind turns on, the inflow is partially suppressed and a steady outflow with $\dot{M}_{\rm gas,net} \approx \dot{M}_{\rm gas,out}\sim 10^{-3} \, M_{\odot}/{\rm yr}$ develops. In the S100\_WSG case (with self-gravity, right panels) there are large variations in the net outflow rate in the range $10^{-3} \, M_{\odot}/{\rm yr} \lesssim \dot{M}_{\rm gas,net}\lesssim 10 \, M_{\odot}/{\rm yr}$. However, as discussed in Subsection~\ref{sec:vert_struct}, at times $6t_{\rm dyn}\lesssim t \lesssim 7t_{\rm dyn}$ the galactic wind loses power and turns into a galactic fountain that rains back towards the disc; during this phase the net outflow rate $\dot{M}_{\rm gas,net}$ is negative, i.e. the outflow turns into an inflow. This galactic fountain phase lasts for approximately a dynamical time $t_{\rm dyn}$. 

An analysis of the temperature structure of the mass inflow and outflow rates reveals the physics of galactic winds and fountains. In order to expose these details, the second row of Figure~\ref{fig:mdot_evo} shows the mass outflow in three temperature ranges: cold gas at temperature $T \leq 8\times 10^3 \, {\rm K}$, warm gas at temperature $8\times 10^3 \, {\rm K}<T<3\times 10^5 \, {\rm K}$ and hot gas at temperature $T\geq 3\times 10^5 \, {\rm K}$. The third row of Figure~\ref{fig:mdot_evo} shows a similar plot for the inflow rates. These plots demonstrate that the outflows are composed of cold and warm gas entrained in hot gas, in both the NOSG and WSG cases. On the other hand, inflows are observed in all the three phases, but are dominated by warm and cold gas. The existence of prominent cold and warm phases in the inflows is a strong indication that a fraction of the inflowing gas is cooling from the hot phase to the warm phase, and then to the cold phase. A comparison between the NOSG and WSG cases also shows that cold inflows are much more prominent in simulations with self-gravity. This effect is caused by the fact that the WSG galactic winds produce much higher densities above the disc than in the NOSG case (see Figure~\ref{fig:vert_struct}), which accelerates radiative cooling from the hot phase into the cooler inflowing phases. 

\section{Discussion and Conclusions} \label{sec:conclusions}

In this paper, idealised adaptive mesh refinement simulations of the response of  galactic discs to supernova feedback were presented and analysed. Initial conditions prone to gravitational instability where intentionally chosen so that the effects of supernova feedback during strong bursts of star formation could be isolated. Simulations were run for discs with several gas surface densities in two versions: (I) in presence of an external, static halo + old stellar disc gravitational potential, and (II) in presence of the external gravitational potential and self-gravity from the gas and newly formed stars. The simulations include radiative cooling, photoelectric heating, star formation from the gas, and the seeding of thermal energy and momentum from individual supernova remnants with the analytic formulae of \cite{2015MNRAS.450..504M}. This setup assumes that the simulated galactic discs are hosted by a dark matter halo of mass $M_{\rm 200}=1.5\times 10^{11} \, M_{\odot}$, which is appropriate for dwarf galaxies. However, due to the idealised nature of these simulations, which simplifies their interpretation, the predictions can be generalised to larger gravitationally unstable galactic discs (possibly at high-redshift), and to unstable galaxy core configurations during galaxy mergers.

Simulations without self-gravity produce a smooth spatial distribution of supernova remnants, with relatively low probability of overlap. These discs rapidly settle into a quasi-steady state with a turbulent disc with regulated star formation that launches steady supersonic galactic winds. Similar quasi-steady states were observed in local 2D simulations of galactic disc patches  \citep{2011ApJ...731...41O,2012ApJ...754....2S}, idealised 3D simulations of galactic disc patches \citep{2006ApJ...653.1266J,2009ApJ...704..137J,2013MNRAS.429.1922C,2016MNRAS.459.2311M,2016MNRAS.456.3432G}, global simulations of galactic discs \citep{2017MNRAS.470L..39F,2018ApJ...862...56S}, and invoked in analytical models that fit the properties of main-sequence star-forming galaxies  \citep{2010ApJ...721..975O,2013MNRAS.433.1970F,2017MNRAS.465.1682H,2019arXiv190300962D}. In this paper, the galactic winds in the simulations without self-gravity have speeds $100 \, {\rm km/s}\lesssim v \sim 500 \, {\rm km/s}$ and mass loading factors $0.01\lesssim \eta \lesssim 0.1$ in good agreement with the recent galactic wind measurements at redshift $0.6\leq z\leq 2.7$ \citep{2019ApJ...875...21F}. 

When self-gravity is turned on the evolution of these discs changes drastically: as the discs become gravitationally unstable, clustered regions of star formation develop which produce a distributions of supernovae correlated in space and time. Clustered supernova feedback induced by gravitational instability is much more violent than in the discs without self-gravity and drives highly mass loaded, supersonic winds  ($0.5\lesssim \eta \lesssim 50$, $100 \, {\rm km/s}\lesssim v \lesssim 500 \, {\rm km/s}$). This result is in qualitative agreement with recent numerical work on supernova clustering  \citep{2017MNRAS.465.2471G,2017ApJ...834...25K,2018MNRAS.481.3325F,2019MNRAS.483.3647G,2019arXiv190209547E}. The wind is composed by cold gas entrained in hot gas, and it eventually cools giving rise to a galactic fountain. During the fountain phase, some of the gas cools from the hot phase and rains towards the mid-plane of the disc. Eventually, the disc settles into a new quasi-steady, turbulent state with lower surface density, lower star formation rate and a weaker wind. Nonetheless, this quasi-steady state exhibits more time variability than in the simulations without self-gravity. 

In the final states of the disc evolution of all the simulations presented in this paper, the vertical structure within the scale height of the disc appears to be jointly determined by turbulent and thermal pressure, with the former playing a slightly more predominant role. This result is in agreement with those of recent local 3D simulations of stratified patches of galactic discs \citep{2015ApJ...815...67K}. Most notably, the global simulations presented in this paper produce random (turbulent) motions with velocity dispersion $\sigma\sim 100 \, {\rm km/s}$ , which are observed in star-forming galaxies at redshift $z\sim 2$ \citep[e.g.]{2011ApJ...733..101G}, but that were not reproduced by previous simulations that did not include self-consistent star formation and clustered supernovae \citep{2016MNRAS.459.2311M}. 

One notable aspect of the simulations with self-gravity is that the time variability in disc properties allows them to produce both a highly mass loaded wind and a weaker wind during a time scale of a $\sim 5-10$ disc dynamical times ($\sim 100 \, {\rm Myr}$). The results of the simulations presented in this paper suggest that the gas surface density, star formation rate and mass loading factor of the wind can vary significantly over a short time scale. This might imply that timing information is important to properly interpret measurements of galactic wind properties from observations. The outflow rates measured/inferred from observations are instantaneous rates, whereas measured star formation rates are averaged over a time scale that depends on the chosen indicator. If time variability is important, then the mass loading factor computed from an outflow rate and a star formation rate measured on two different time scales may lead to erroneous interpretation of the observations. Timing information might partially reconcile differences in the mass loading factors measured in recent literature (e.g. compare \citealt{2019ApJ...875...21F} with \citealt{2017MNRAS.469.4831C} and \citealt{2017ApJ...850...51S}). \cite{2019arXiv190601890L} recently showed that the ISM of post-starburst galaxies evolves as a function of time after the starburst, but the implications on galactic winds have not been explored yet. 

An important caveat that complicates a direct comparison of the predicted wind properties from the presented simulations to observational data (and to the result of cosmological simulations) is that the simulations in this paper are performed in boxes of a few kpc, which limits the distance at which wind properties are measured. Conversely, measurements of galactic wind properties are typically obtained at larger distances from the galaxy ($\sim 10-100 \, {\rm kpc}$; \citealt{2005ApJ...621..227M,2005ARA&A..43..769V,2009ApJ...697.2030S}). One way to address this issue with currently feasible numerical setups would be to perform idealised global disc simulations in much larger boxes, with an approach similar to that used by \citep{2017MNRAS.466.3810F} for the circumgalactic medium, but with explicit modeling of supernova feedback at scales $\lesssim  1 \, {\rm pc}$ \citep[see ][for setups at slightly lower resolution]{2017MNRAS.466.1093G,2019arXiv190508806M}. The problem can also be studied with high-resolution cosmological large-box simulations \citep{2019arXiv190205554N}, zoom-in simulations \citep{2015ApJ...804...18A, 2015MNRAS.454.2691M, 2016MNRAS.460.2731C}, but the currently achievable resolution does not allow to explicitly resolve the emergence of supernova feedback on an individual supernova basis. 

Although supernova feedback is the (energetically) dominant regulating process in star forming galactic discs, the role of stellar winds, radiative feedback (heating and pressure) and cosmic rays (diffusion and streaming) in galaxies might further boost the efficiency of feedback, and recent work has shown huge progress in the modeling of these physical processes \citep{2013ApJ...770...25A, 2014MNRAS.445..581H, 2014MNRAS.437.2882K, 2015MNRAS.451...34R, 2017MNRAS.465.4500P, 2018ApJ...854....5J, 2019arXiv190508806M, 2019arXiv190504321H}. Such physics is not included in the presented simulations, and its effect needs to be accurately explored in future work. A final caveat is the fact that entrainment of cold clouds and filaments in a hotter medium, which regularly happens in the presented simulations, is also object of ongoing studies \citep{2015MNRAS.449....2M, 2018ApJ...862...56S, 2017MNRAS.470..114A, 2019arXiv190704771G}. Whether a spatial resolution between 2 pc and 16 pc like the one considered in this paper is sufficient to properly capture entrainment and mixing of cold and hot gas in galactic winds is still unclear, and the full theoretical scenario might depend on the presence of magnetic fields and the role of thermal conduction. Although the predictions of this paper are robust against variation of the spatial resolution within a factor two (Appendix~\ref{app:tests}), they will probably need to be updated with new simulations once the physics of cold gas entrainment in galactic winds is better understood.

\section*{Acknowledgments}
DM was supported by the DARK Fellowship. DM also acknowledges contribution from the Danish council for independent research project DFF - 6108-00470, and from the Danish National Research Foundation project DNRF132. DM thanks the Reviewer for their useful comments that greatly improved the quality of the paper.

\bibliography{main}

\appendix

\section{Resolution and Robustness Tests}\label{app:tests}

\begin{figure}
\begin{center}
 \includegraphics[width=0.485\textwidth]{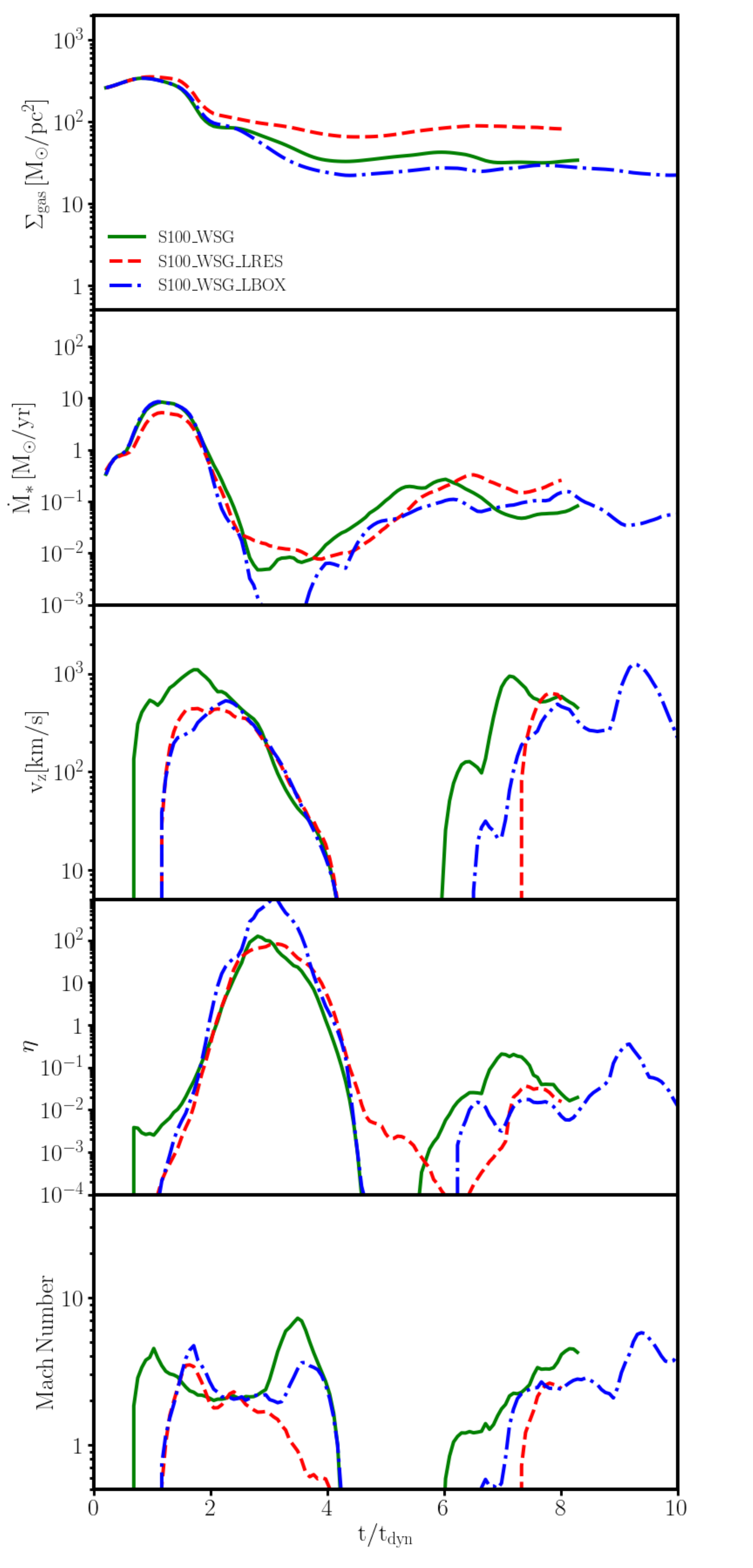}
\end{center}
\caption{Temporal evolution of properties of the system in a cylindrical region of radius ${\rm R = 400 \, pc}$. Time on the x-axis has been normalised to the value of the dynamical time in the disc mid-plane in each simulation, $t_{\rm dyn}$. First panel: gas surface density $\Sigma_{\rm gas}$ within ${\rm |z|<1 \, kpc}$. Second panel: star formation rate in the disc $\dot{M}_*$. Third panel: net velocity of the gas at height ${\rm |z|=1 \, kpc}$ from the disc mid-plane. Fourth panel: mass loading factor of the galactic wind at height ${\rm |z|=1 \, kpc}$ from the disc mid-plane; only outflowing material has been included in the calculation. Fifth panel: Mach number of the galactic outflow. All properties have been time-averaged on a time scale of 5 Myr. The predictions of the fiducial simulation S100\_WSG are in qualitative and quantitative agreement with the lower resolution version S100\_WSG\_LRES. The simulation with the same initial conditions and resolution in a larger box, S100\_WSG\_LBOX provides similar results too, but with a slight offset in time. }\label{fig:galwind_evo_tests}
\end{figure}

\begin{figure}
\begin{center}
 \includegraphics[width=0.485\textwidth]{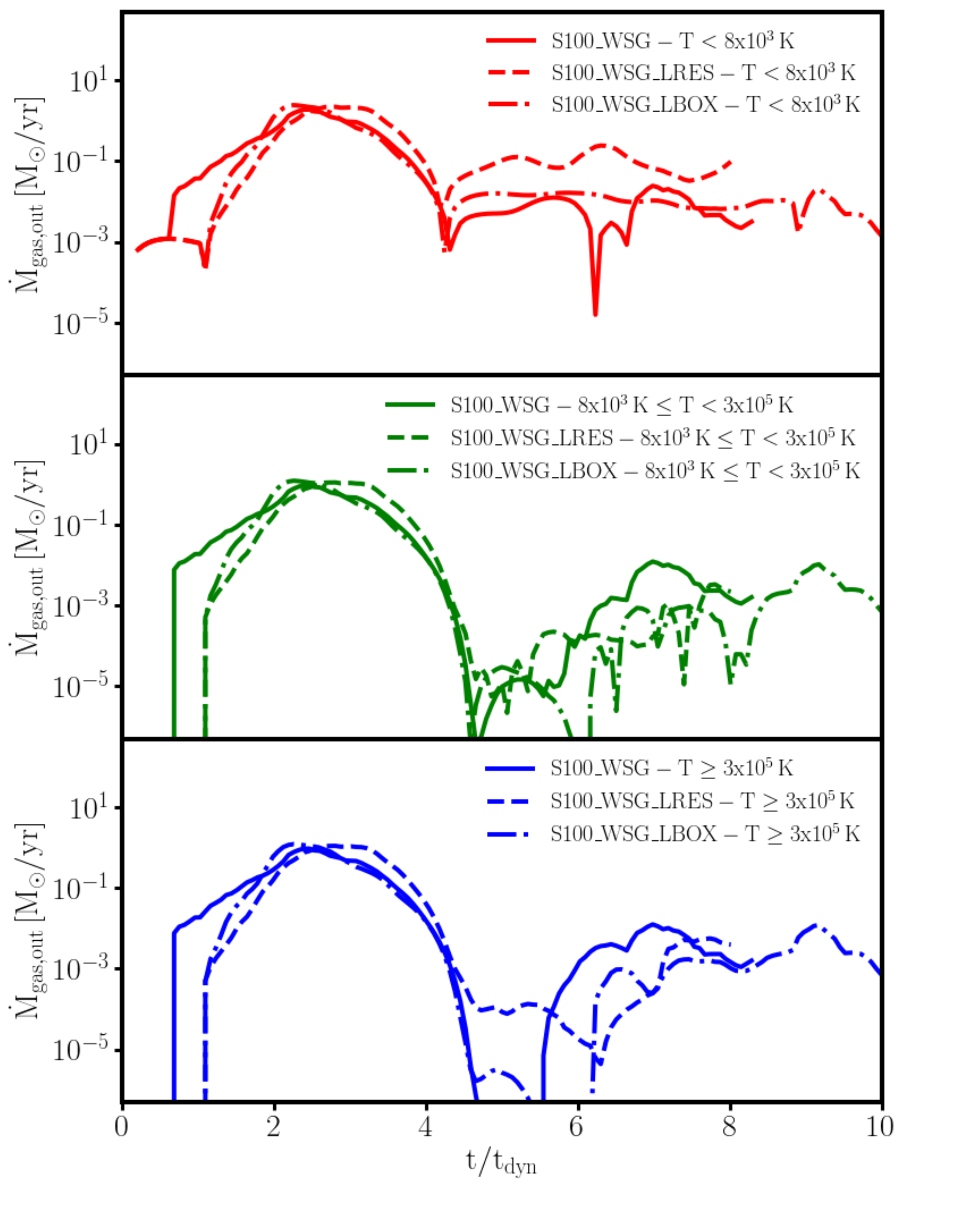}
\end{center}
\caption{ Temporal evolution of the mass outflow rate measured within a cylindrical region of radius ${\rm R = 400 \, pc}$ and at height $|z|=1 \, {\rm kpc}$ from the disc mid-plane. Time on the x-axis has been normalised to the value of the dynamical time in the disc mid-plane in each simulation, $t_{\rm dyn}$. Top panel: cold phase. Mid panel: warm phase. Bottom panel: hot phase. Variations of the numerical resolution and box size have a quantitative impact on the phase structure of the wind (especially on the cold phase), but the results are qualitatively robust. }\label{fig:mdotout_evo_tests}
\end{figure}

The robustness of the numerical predictions of the simulations presented in this paper has been assessed by performing resolution tests and variations of the box size. For the resolution tests, the fiducial simulations have been performed with the same base AMR level, corresponding to a minimum spatial resolution $\Delta x_{\rm coarse}=16 \, {\rm pc}$, but with a maximum spatial resolution $\Delta x_{\rm fine} = 4 \, {\rm pc}$, which is worse than the fiducial one by a factor two. The stellar particle mass has not been varied. For the box size test, alternative simulations were performed in a computational domain with twice the fiducial box size ($L=8.192 \, {\rm kpc}$ instead of $L=4.096 \, {\rm kpc}$), with coarse spatial resolution $\Delta x_{\rm coarse} = 32 \, {\rm pc}$ and maximum spatial resolution $\Delta x_{\rm fine} = 2 \, {\rm pc}$, the same as the fiducial case.

Figure~\ref{fig:galwind_evo_tests} shows the result of one of the tests for the S100\_WSG disc. Similar results are found for the other setups discussed in the paper. The S100\_WSG and its lower resolution version S100\_WSG\_LRES are in qualitative and quantitative agreement when the evolution of star formation rates, wind speed, wind mass loading factor and wind Mach number are taken into account. The evolution of the disc surface densities in the two simulations are also in qualitative agreement, but with slightly different final values. This discrepancy is probably due to the fact that the vertical structure of the disc is slightly `puffed up' when the resolution is lower, resulting in the slightly higher final surface density. Varying the box size also produces very similar results as the fiducial run, with minor differences related to the timing of the events related to the galactic wind/fountain. With a larger box the galactic wind is triggered earlier. This effect is caused by the increased weight of the column of gas above the disc, which will flow towards the disc mid-plane as soon as the simulation starts and cooling/heating turns on. This causes the initial dynamics of the wind to be slightly different. Timing differences in the late evolution of the wind/fountain are caused by the fact that with a smaller box some of the material in the wind is irreversibly lost when it passes through the box boundary, whereas when the box is larger, the same material will travel further, cool and rain down to the disc mid-pane later than in the fiducial case. 

{Figure~\ref{fig:mdotout_evo_tests} shows the temporal evolution of the wind outflow rate measured at height $|z|=1 \, {\rm kpc}$ for the cold, warm and hot gas phases and for each of the test runs, respectively. There are quantitative differences in the outflow rates of the different phases, but the qualitative picture is robust. The phase that is most influenced by changes in numerical resolution is the cold phase, whose mass outflow rate is higher at low resolution than at high resolution. This result suggests that formal convergence is yet not achieved for this phase in $\sim {\rm pc}$ resolution simulations. This is likely due to (I) over-cooling of under-resolved clouds and filaments, and (II) under-resolved hydrodynamical processes at the interface between the cold and warm phase. Nonetheless, the differences among the test runs are not large enough to qualitatively change the conclusions of the paper. }

\end{document}